\documentclass[12pt, a4paper]{article}
\pdfoutput=1

\usepackage{amsmath}
\usepackage{amsfonts}
\usepackage{amssymb}
\usepackage{bm}
\usepackage{bbm}
\usepackage{verbatim}
\usepackage{dsfont}
\usepackage{booktabs}
\usepackage{slashed}
\usepackage{nicefrac}
\usepackage{mathtools}
\usepackage{physics}
\usepackage{array}
\usepackage{mathrsfs}
\usepackage{tikz-cd}

\usepackage{epsfig}
\usepackage[dvipsnames]{xcolor}
\usepackage{graphicx}
\usepackage[font=small]{caption}
\usepackage[listofformat=empty,subrefformat=empty]{subfig}

\usepackage{float}
\usepackage{overpic}
\usepackage{tikz}
\usepackage{tabularx}

\usepackage{enumerate}
\usepackage{hhline}
\usepackage{multirow}

\usepackage[nosort]{cite}
\usepackage{xspace}
\usepackage{setspace}

\usepackage[pdftitle={},
  pdfauthor={},
  pdfsubject={},
  bookmarksopen, bookmarksnumbered, bookmarksopenlevel=2, colorlinks=false, linkcolor=blue, citecolor=blue, urlcolor=blue]{hyperref}

\usepackage[nameinlink,capitalize]{cleveref}

\makeatletter{}
\g@addto@macro\bfseries{\boldmath}
\makeatother

\usepackage[normalem]{ulem}

\newcommand{\be}{\begin{equation}}
\newcommand{\ee}{\end{equation}}
\newcommand{\ba}{\begin{aligned}}
\newcommand{\ea}{\end{aligned}}

\def\Z{{\mathbb{Z}}}
\def\R{{\mathbb{R}}}

\begin{document}

% format
\baselineskip=18pt  % a la harvmac
\numberwithin{equation}{section}  % make eq labels (sec.num)
\allowdisplaybreaks  % allow page breaks in displayed eqs

\thispagestyle{empty}

\vspace*{-2cm}

\vspace*{2.0cm}
\begin{center}
{\Large{\textbf{Notes on $(-2)$-form symmetries
}}} 
\\
\vspace*{1.0cm}
Pinak Banerjee$^1$, Alonso Perez-Lona$^1$, Daniel Robbins$^2$, Subham Roy$^2$, Eric Sharpe$^1$, Xingyang Yu$^1$ \\
\vspace*{0.5cm}
{\it $^1$Department of Physics MC 0435, 850 West Campus Drive, Virginia Tech, \\ Blacksburg, VA  24061, USA} \\
\vspace*{0.5cm}
{\it $^2$ Department of Physics, 1400 Washington Ave,
University at Albany, SUNY\\
 Albany, NY 12222, USA}
\vspace*{0.8cm}
\end{center}
\vspace*{.5cm}
We study $(-2)$-form symmetries of a $d$-dimensional quantum field theory, via a $(-1)$-form symmetry of its $(d+1)$-dimensional Symmetry Topological Field Theory (SymTFT), realized by a non-genuine codimension-one defect in the SymTFT bulk attached to a spacetime-filling topological operator. Unlike a $(-1)$-form symmetry of a $d$-dimensional theory, which merely shifts a parameter of the absolute theory, a $(-2)$-form symmetry modifies the SymTFT action, and thereby relates theories whose ordinary global symmetries differ by anomaly data or by the associator data of a non-invertible symmetry. We illustrate the construction in two-dimensional toy models, three-dimensional ABJM-type theories, four-dimensional generalized Yang--Mills theory, and in a fusion-categorical example relating the non-invertible symmetries $\operatorname{Rep}(D_4)$ and $\operatorname{Rep}(Q_8)$. We then develop a club-sandwich realization, in which a quarter-gauging operation interfaces between IR phases of distinct RG flows of a common UV theory, and an alternative realization via nested discrete gauging. Finally, we present a holographic, top-down realization in which the type IIA Romans mass plays the role of a $(-2)$-form background for a three-dimensional Chern--Simons-matter theory, with shifts of the Romans mass realizing shifts of the boundary anomaly coefficients. We also discuss related constructions for coupled bulk--boundary systems.
\noindent

\begin{flushleft}
    June 2026
\end{flushleft}

\newpage
%%%%%%%%%%%%%%%%%%%%%%%%%%%%%%%%%%%%%%%%%%%
%%%           TITLE ENDS HERE
%%%%%%%%%%%%%%%%%%%%%%%%%%%%%%%%%%%%%%%%%%%
\setcounter{tocdepth}{2}
\tableofcontents
%\printindex

\newpage

%%%%%%%%%%%%%%%%%%%%%%%%%%%%%%%%%%%%%%%%%%%
%%%        MAIN TEXT BEGINS HERE
%%%%%%%%%%%%%%%%%%%%%%%%%%%%%%%%%%%%%%%%%%%
\section{Introduction}
Quantum field theories are naturally organized into families, or more generally
into \emph{spaces} of theories. The coordinates on these spaces are not just
the local couplings entering the Lagrangian: they also include a wealth of
discrete data, such as theta-like parameters and other quantized couplings,
discrete torsion in orbifolds, global forms of gauge groups, 't~Hooft anomaly
coefficients, boundary conditions, and defect labels. A unifying viewpoint on
this collection of data is provided by the language of \emph{generalized
global symmetries}~\cite{Gaiotto:2014kfa}, whose modern
incarnation organizes higher-form, higher-group, and non-invertible symmetries
into a coherent framework; for reviews see~\cite{Sharpe:2015mja,Cordova:2022ruw,Schafer-Nameki:2023jdn,Brennan:2023mmt,Bhardwaj:2023ayw,Freed:2022iao, Bah:2022xfv,Bhardwaj:2023kri,Shao:2023gho,Gomes:2023ahz,McGreevy:2022oyu,Chang:2018iay,Bhardwaj:2022yxj,Iqbal:2024pee,Costa:2024wks}.

A particularly clean topological structure on such theory spaces is provided
by $(-1)$-\emph{form symmetries}. In this picture, a parameter is viewed as
the background field for a $(-1)$-form symmetry, and anomalies in the space
of coupling constants~\cite{Cordova:2019jnf,Cordova:2019uob} are reinterpreted
as genuine 't~Hooft anomalies involving $(-1)$-form backgrounds. Gauging a
finite part of such a $(-1)$-form symmetry promotes the parameter to a
partially dynamical topological field, and the resulting theory exhibits
\emph{decomposition} \cite{Hellerman:2006zs}
into a collection of universes, accompanied by a quantum dual $(d-1)$-form
symmetry. This perspective has been developed in a number of works, see
for example \cite{Caldararu:2010ljp,Sharpe:2014tca,Sharpe:2019ddn,Tanizaki:2019rbk,Nguyen:2021yld,Nguyen:2021naa,Komargodski:2020mxz,Cherman:2020cvw,Robbins:2020msp,Robbins:2021xce,Sharpe:2021srf,Pantev:2022kpl,Pantev:2022pbf,Lin:2022xod,Perez-Lona:2023llv,Pantev:2024kva,Pantev:2023dim,Santilli:2024dyz,Yu:2024jtk,Najjar:2024vmm,Robbins:2025urk,Lin:2025oml,Perez-Lona:2025add} (see \cite{Sharpe:2022ene} for a recent review)
where invertible and non-invertible variants of $(-1)$-form symmetries are
shown to govern shifts of $\theta$-like parameters, rescalings of complexified
couplings, and other natural operations relating points of theory space.

A higher-dimensional reorganization of these structures is provided by the
SymTFT framework. There a
$d$-dimensional QFT is realized as the boundary of a $(d+1)$-dimensional
topological theory, and its ordinary generalized symmetries are encoded by
topological defects of the SymTFT bulk, with their fusion and linking
structure capturing higher-categorical and anomaly data. Within this
dictionary, ordinary $(-1)$-form symmetries of the absolute theory correspond
to certain genuine codimension-one defects of the SymTFT, whose linking with
bulk point operators implements the shift of background parameters of the
boundary theory \cite{Yu:2024jtk,Najjar:2024vmm,Robbins:2025urk,Lin:2025oml,Aloni:2024jpb}.

The main observation of this paper is that this dictionary admits a further
layer. Not every codimension-one operation in the SymTFT is a genuine defect:
some are \emph{non-genuine}, in the sense that they must be attached to a
spacetime-filling topological defect in the bulk in order to be well defined \cite{Kapustin:2014gua}.
Such non-genuine operations do not merely shift a parameter of the boundary
QFT; they modify the SymTFT action itself, for instance by shifting anomaly
inflow terms. From the absolute QFT perspective, this is the SymTFT
realization of phenomena such as shifting an anomaly inflow phase, where the
shift moves between theories that differ by an anomaly coefficient rather
than by a QFT parameter.

Such an operation has been described \cite{Heckman:2024oot} as a
$(-2)$-\emph{form symmetry}
of the
absolute QFT. Equivalently, a $(-2)$-form symmetry of a $d$-dimensional QFT
is a $(-1)$-form symmetry of its $(d+1)$-dimensional SymTFT. In the same way
that a $(-1)$-form symmetry can be visualized as a spacetime-filling object
in the physical theory, a $(-2)$-form symmetry corresponds to a
spacetime-filling object in the SymTFT bulk. The construction is naturally
iterative: passing to a SymTFT of the SymTFT, or more generally including
the physical theory in a larger descent structure
\cite{Antinucci:2024zjp,Cvetic:2024dzu,Borsten:2025pbx}, suggests an extension to
$(-n)$-form symmetries for $n \geq 3$, which is suggested in \cite{Heckman:2024oot}.

This notion of $(-2)$-form symmetry naturally connects QFTs whose generalized
symmetry data differ by an anomaly. We illustrate this through three families
of examples spanning different dimensions. In a simple two-dimensional toy
model, the operation relates a theory with a non-anomalous $\Z_2$ symmetry
to one with an anomalous $\Z_2$ symmetry. Analogous structures appear in
three-dimensional ABJM-type theories, where the operation interpolates
between distinct Chern--Simons levels, and in four-dimensional generalized
Yang--Mills theory, where it captures the interplay between $\theta$-shifts
and the 't~Hooft anomalies of one-form symmetries.

From the absolute QFT point of view, all of these examples share a common
mechanism: the interplay between decomposition and a higher-group or
mixed-anomaly structure involving a quantum $(d-1)$-form symmetry.
Gauging or undoing the decomposition in the SymTFT, controlled by suitable
discrete torsion, moves between universes whose ordinary global symmetries
carry different anomaly data. This is the mechanism by which a $(-2)$-form
operation materializes as a topological morphism between absolute theories.

These observations suggest an enlarged theory-space viewpoint. The $(-2)$-form
symmetries need not act within a single ordinary deformation class; instead,
they are most naturally understood as topological morphisms between anomaly
sectors, between SymTFTs, or between IR phases, typically realized through
interfaces or common UV embeddings.\footnote{One would attempt to build a
groupoid or even a higher-category structure to encode $(-2)$-form symmetry.
Namely, for spaces of QFTs which are labeled by different anomalies, treat
them as $0$-morphisms/objects in the groupoid/category, while the $(-2)$-form
symmetry transformations connecting different spaces of QFTs are treated as
$1$-morphisms.}
The higher-categorical and functorial language developed in
\cite{Stockall:2025ngz,Heckman:2026beg,Bergman:2026lnz,Perez-Lona:2025xxm,Teixeira:2025qsg,PerezLona:2026tba}
provides a natural setting in which to phrase these morphisms, and we take a baby step in this work to
exploit this language, and leave a complete formal definition of $(-2)$-form symmetries as a future work.

We motivate this morphism picture illustratively using the \emph{club-sandwich} construction
\cite{Baume:2023kkf,Bhardwaj:2023bbf}. By performing a \emph{quarter gauging} operation in
the SymTFT, we obtain interfaces between distinct IR phases arising from
different RG flows of a common UV theory. In examples such as the
$\Z_4 \to \Z_2/\Z_2^\omega$ construction, the resulting interface realizes
the effect of a $(-2)$-form operation relating phases with different anomaly
data. This realization sits comfortably alongside related constructions of
interfaces and corners between symmetry topological theories
\cite{Cvetic:2024dzu,KNBalasubramanian:2025vum}.

We also discuss an alternative realization through \emph{nested discrete
gauging}, in which a sequence of topological gaugings changes the twist or
anomaly term of a three-dimensional Dijkgraaf--Witten theory. This provides
another concrete realization of a $(-1)$-form symmetry of the SymTFT, and
hence of a $(-2)$-form symmetry of the associated absolute theory.

A more general source of motivation comes from QFTs realized as boundaries or
defects of higher-dimensional systems, which are not limited to bulk TFTs. Such defect QFTs have their own
characteristic spaces of boundary conditions\footnote{One familiar case is the boundary conformal manifold for a CFT, e.g., space of all possible A- and B-brane configurations for a 2D SCFT.}, edge modes, and topological
boundary data. The notion of a boundary SymTFT makes it possible to study
topological operations acting on these spaces, including operations, which
change the boundary data while leaving the bulk theory fixed. We find that
many of the structures we identify in the absolute setting have natural
counterparts in this boundary setting.

Finally, we describe a holographic realization in which the type IIA \emph{Romans mass}~\cite{Romans:1985tz} plays the role of a $(-2)$-form background for a three-dimensional boundary theory along the lines of the AdS$_4$/CFT$_3$. The specific Romans-mass identification was first suggested in \cite{Yu:2024jtk}, while the general dictionary mapping a bulk gravity $(-1)$-form symmetry to a $(-2)$-form symmetry of the boundary CFT was proposed in \cite{Heckman:2024oot}; we develop this realization in detail in Section~\ref{sec:romans-minus-two-topdown}, where changing the Romans mass shifts the Chern--Simons levels of the dual theory, and hence shifts the anomaly coefficients of its ordinary generalized symmetries, providing a string-theoretic incarnation of a $(-2)$-form operation.

A comment on scope is in order. The construction presented in this paper by
no means gives the most general definition of $(-2)$-form symmetries.
Rather, it provides an interesting starting point for their investigation,
and a possible angle for characterizing defect and boundary QFTs.

\subsubsection*{Organization of the paper}
The rest of the paper is organized as follows. In
Section~\ref{sec:codim-1-defects-symtft} we review the codimension-one
defects of the SymTFT and their relation to $(-1)$-form symmetries of the
absolute theory, including invertible, higher-charged, and non-invertible
variants. In Section~\ref{sec:-2form-general} we formulate our proposal for
$(-2)$-form symmetries, both from a category-theoretic perspective and via a
concrete SymTFT prescription, and we work through the two-, three-, and
four-dimensional examples summarized above. Section~\ref{sec:rgflowand-2fs}
develops the club-sandwich realization, including the quarter-gauging
construction, several examples, and the alternative nested-discrete-gauging
construction. Section~\ref{sec:romans-minus-two-topdown} presents the
holographic realization based on Romans mass.
Section~\ref{sec:bulk/boundary_system} discusses related examples involving
coupled bulk--boundary systems and the boundary SymTFT. We close in
Section~\ref{sec:conclusion} with conclusions and an outlook, including
possible $(-n)$-form analogues and the role of $(-2)$-form symmetries in
the broader context of defect and boundary QFTs.

\section{Codimension one defects of SymTFT}\label{sec:codim-1-defects-symtft}
In this section, we would like to review the codimension one defects -- $0$-form symmetries -- of the SymTFT and their implications on the absolute theory. We can characterize these defects via their fusion, which correspond to invertible and non-invertible $0$-form symmetries of the SymTFT. Amongst the invertible objects, we might also have defects with higher generalized charges \cite{Bhardwaj:2023wzd}, that acts on extended operators instead of point operators. This characterization only captures genuine codimension one defects. 
\subsubsection*{Invertible $0$-form symmetries of SymTFT}

To begin with, we will talk about $0$-form symmetries of the SymTFT that can be associated with genuine codimension one defects. One of the obvious ways to have such extended defects is to require the presence of $\pi \int_{\mathcal{M}_{d+1}} b_0 \cup \delta c_d$ type of coupling in the SymTFT action. The full SymTFT takes the form,  
\begin{equation}
    S_{\text{SymTFT}} \sim \pi\int_{\mathcal{M}_{d+1}}  \Big( b_0 \cup \delta c_d + S' \Big).
\end{equation}
where $S'$ denotes additional couplings required to probe other global symmetries of the $d$-dimensional absolute theory. Due to the presence of the BF pair between $b_0$ and $c_d$, the $c_d$ codimension one surfaces link with topological point operators living in the bulk of the SymTFT. Depending on the choice of boundary condition on the symmetry boundary, such codimension one defects eventually boil down to traditional notion of $(-1)$-form symmetries of the absolute theory, which essentially captures the shift of the $\theta$-angle type parameters of the absolute theory,
see for example \cite{Yu:2024jtk, Aloni:2024jpb, Najjar:2024vmm, Robbins:2025urk, Lin:2025oml, Perez-Lona:2025add}. 

Gauging such $(-1)$-form symmetries from the absolute theory or imposing Dirichlet boundary condition on $c_d$ on the symmetry boundary of the SymTFT leads to a quantum field theory with $(d-1)$-form symmetry\cite{Hellerman:2006zs, Sharpe:2014tca, Sharpe:2019ddn, Tanizaki:2019rbk, Komargodski:2020mxz, Cherman:2020cvw, Robbins:2020msp, Nguyen:2021naa, Sharpe:2021srf, Cherman:2021nox, Sharpe:2022ene,Lin:2022xod, Bhardwaj:2023bbf}, where the universes will be identical.

\subsubsection*{Invertible $0$-form symmetries of SymTFT with higher charges}

Codimension one defects can also be found in SymTFTs devoid of point operators. These are usually constructed via higher gauging techniques - commonly known as condensation defects \cite{Lin:2022xod,Roumpedakis:2022aik, Kong:2013aya, Kong:2014qka, Else:2017yqj, Gaiotto:2019xmp}. These defects are porous to point operators by construction but they can act on extended operators of the SymTFT. Moreover, these defects can often have non-invertible fusion but right now, we are going to focus on the ones with invertible fusion. Examples of such defects can be found in three dimensional $\Z_2$ BF theory \cite{Roumpedakis:2022aik}, which can be treated as the SymTFT of a $2d$ quantum field theory with a $\Z^{(0)}_2$ symmetry. 
One of the most well known examples of such surface operator is the invertible \textit{electro-magnetic duality defect} of $3d$ $\Z_2$ gauge theory with a $\Z_2$ fusion, constructed by $1$-gauging a non-anomalous subgroup of the $1$-form symmetries of the theory \cite{Gaiotto:2020iye, Roumpedakis:2022aik, Kaidi:2022cpf}. In the following, $\mathcal{D}$ is the aforementioned surface operator,  
\begin{equation}
    \mathcal{D}(\mathcal{M}_2) \sim \sum_{\gamma_1 \in H_1(\mathcal{M}_2, \Z_2)} L_{(1,-1)} (\gamma)
\end{equation}
and $L_{(1,-1)}$ are anyons of the theory\footnote{We are using the nomenclature of \cite{Kaidi:2022cpf}. In their notation, the action of the gauge theory was given by $S = \pi \int a\delta \hat{a}$ and anyons are labeled as the following, 
\begin{equation}
    L_{(e,m)} = \exp\left(i\pi \oint e a + \oint m \hat{a} \right), \qquad (e,m) \in \Z_2 \times \Z_2.
\end{equation}
}. 
In general, these correspond to certain discrete gauging operation in the absolute theory \cite{ Robbins:2025urk, Kaidi:2022cpf, Hasan:2024aow, Arbalestrier:2025jsg, Niro:2022ctq}. In some theories like the two dimensional free boson or the four dimensional Maxwell theory, certain discrete gauging operation correspond to rescaling of parameters of the theory itself. To be specific, for the free boson theory, the radius is rescaled under gauging of a $\Z_N$ subgroup of the momentum $U(1)$ symmetry. In the case of Maxwell theory, a similar operation leads to a modification of the complexified coupling constant\cite{Hasan:2024aow}. In this setting, we can associate these codimension one defects with $(-1)$-form symmetries of the absolute theory.

\subsubsection*{Non-invertible $0$-form symmetries of SymTFT}
In addition to the defects with invertible fusion, we can also have non-invertible $0$-form symmetries of the SymTFT, which were used to construct non-invertible $(-1)$-form symmetries of the absolute theory\cite{Lin:2025oml}. The three dimensional SymTFT has action
\begin{equation}
    S_{\text{SymTFT}} = \frac{1}{2\pi} \int_{\mathcal{M}_3}  \Big( b_1 dA_1 + f_2dB_0 + f_0 dB_2 + \frac{k_A}{2\pi} f_0 A_1 f_2 \Big).
\end{equation}
where the uppercase letters are $U(1)$-valued fields and the lowercase letters are $\mathbb{R}$-valued fields. This TFT carries a non-invertible $0$-form symmetry generated by the operator 
\begin{equation}
\begin{aligned}
    V_p[\Sigma_2] &= \int \mathcal{D}\phi_0 \mathcal{D}\phi_1 \exp\left[ ip \int_{\Sigma_2} \left( B_2 + \frac{k_A}{2\pi} (\phi_0 f_2 - A_1 \phi_1 -\phi_1 d\phi_0) \right) \right], \\
    &= \exp\left( ip \int_{\Sigma_2} B_2 \right) \mathcal{A}[\Sigma_2, A_1, f_2] ,
\end{aligned}
\end{equation}
due to the presence of additional topological degrees of freedom ($\phi_1, \phi_0$) the operator obeys non-invertible fusion. This defect and how it modifies the physical theory was studied in detail in \cite{Lin:2025oml}.
It was observed that in addition to a shift of a parameter these could modify global symmetries of the theory. %\

\section{$(-2)$-form symmetry}\label{sec:-2form-general}

The defects we discussed in the last section were genuine defects in the SymTFT bulk. Beyond these, we can also have have gauge invariant non-genuine operators in the bulk of SymTFT. These ones require attaching a space-time filling bulk operator to the codimension one  defect, to make it well defined.  We would like to distinguish these from the genuine codimension one defects, which we associate with $(-1)$-form symmetries of the absolute theory. We observe that, in addition to shifting the parameters of the absolute theory, these non-genuine defects can modify the anomalies of global symmetries of the absolute theory, which can be verified from the bulk as well. Moreover, these defects can shift the SymTFT action by adding additional couplings, mimicking the notion of  $(-1)$-form symmetries in the absolute theory. But in this case, this is a $(-1)$-form symmetry of the SymTFT itself, which suggests the name -- \textit{$(-2)$-form symmetry} of the absolute theory. We emphasize that a general notion of lower-form ($(-m)$-form) symmetry was proposed in \cite{Heckman:2024oot}, defined as a $(-1)$-form symmetry of a higher-dimensional bulk system in which the theory of interest sits as a defect. The $(-2)$-form symmetries studied here are the special case in which that bulk system is the SymTFT of the absolute theory; our aim is to develop this SymTFT realization concretely and to exhibit its consequences in a range of examples.

One approach to studying these types of structures is in terms of functors valued in topological field theories (TFTs).  Here we would understand generalized symmetries as insertions of topological defects.  As it has been recognized in the literature (e.g. \cite[Section 2]{Heckman:2026beg}), in very general terms, such insertions may also be understood as couplings/stackings of the physical theory with some TFT.  For example, if the defect is supported at a fixed codimension, the corresponding TFT is a non-extended TFT with the same support. For a fixed physical $d$-dimensional theory $\mathsf{t}$, one may consider, at least abstractly, the $d$-category of TFT's $\mathcal{T}$ one can stack with $\mathsf{t}$.  Then codimension $1$ TFTs, i.e.~$0-$form symmetries of $\mathsf{t}$, correspond to objects in $\mathcal{T}$, codimension $2$ TFTs that sit at interfaces between codimension $1$ TFTs, i.e.~genuine and non-genuine $1-$form symmetries, are $1-$morphisms of $\mathcal{T}$, codimension $3$ TFTs are $2-$morphisms corresponding to $2-$form symmetries and so forth.  This categorical language easily incorporates both genuine and non-genuine defects, and so provides a good context to consider the bulk-filling defects and non-genuine codimension $1$ defects that are of interest in this work.

In Appendix~\ref{app:cat} we elaborate on some of the more technical aspects of this formulation.  For now, we summarize the dictionary between categorical structures and topological defects (for the $d=2$ case) in Table~\ref{tab:defect-dictionary}.

\begin{table}
\begin{center}
\begin{tabular}{|c|c|}\hline
    Defect in SymTFT's &  Category-theoretic structure \\ \hline
    Bulk-filling defect & Braided fusion category $\mathscr{B}$ 
     \\ Trivial bulk-filling defect & Trivial braided fusion category $\mathbbm{1}_{(3)}$ \\
     Surface defect & $(\mathscr{T}\boxtimes \mathscr{B},\mathscr{T}\boxtimes \mathscr{B}')-$bimodule fusion category $\mathcal{C}$
     \\
     Trivial surface defect & Regular bimodule fusion category $\mathscr{T}\boxtimes \mathscr{B}$
     \\
     Non-genuine surface defect & Bimodule fusion category $\mathcal{C}$ with $\mathscr{B}\neq \mathscr{B}'$. 
     \\
     & e.g. a $(\mathscr{T},\mathscr{T}\boxtimes \mathscr{B})$-bimodule
     \\
     Line defect & $(\mathcal{C},\mathcal{D})$-bimodule abelian category $\mathcal{A}$
     \\
     Non-genuine line defect & Bimodule with $\mathcal{C}\neq \mathcal{D}$ \\ & e.g. $(\mathscr{T},\mathcal{D})$-bimodule
     \\\hline
\end{tabular}
\captionof{table}{Dictionary between bulk defect operators in the $3d$ SymTFT and category theoretic structures in the Morita higher category $\mathbf{BrFus}$ of braided fusion categories.  Detailed definitions are given in appendix~\ref{app:cat}.}\label{tab:defect-dictionary}
\end{center}
\end{table}

In what follows next, we explore these non-genuine codimension one defects in the SymTFT in detail in various settings, and we make an attempt at understanding their implications from the perspective of the absolute theory.

\subsection{General prescription}\label{sec:gen_prescription}

We are trying to construct non-genuine operators that will modify the action of the SymTFT itself, hence we can firmly say that we are hunting for a $(-1)$-form symmetry of the SymTFT itself. We will interpret this as $(-2)$-symmetry of the absolute theory. The easiest way to get a $(-1)$-form symmetry within the SymTFT itself is to allow point operators in the $(d+1)$-dimensional SymTFT, which will generate a $((d+1)-1)$-form symmetry. Due to the presence of the point operators, the theory should decompose and then we can gauge this $d$-form symmetry -- undoing decomposition\cite{Sharpe:2019ddn} -- and engineer a quantum dual $(-1)$-form symmetry within the SymTFT itself\footnote{For the time being, we will only consider the SymTFT quiche \cite{Bhardwaj:2023bbf}, then we will `\textit{attach}' the physical boundary to it, completing it into the SymTFT sandwich.}.

The simplest example of such a decomposing theory is the following BF theory, 
\begin{equation}\label{eqn:decomposingBF}
    S_{\text{bulk}} = \frac{iN}{2\pi} \int_{\mathcal{M}_{d+1}} b_0 \wedge dc_d
\end{equation}
where both $b_0$ and $c_d$ are $U(1)$-valued fields. When viewed as a SymTFT of a $d$-dimensional quantum field theory, this theory could capture the data of $(d-1)$-form and $(-1)$-form symmetry. The equation of motion of $c_d$, tells us that this theory carries topological point operators $U_\alpha(x_0) = \exp\left(i\alpha b_0\right)$. Moreover, these topological point operators are charged under a $0$-form symmetry generated by the codimension one $c_d$ surfaces. 

We can gauge the $\Z_N^{(d)}$-form symmetry and engineer a quantum dual $(-1)$-form symmetry within the SymTFT. In order to gauge the $d$-form symmetry, we add the following coupling to our action, 
\begin{equation}
    S \subset \frac{iN}{2\pi} \int_{\mathcal{M}_{d+1}} b_0 \wedge
    A_{d+1}
\end{equation}
where $A_{d+1}$ is the background gauge field of the $d$-form symmetry. The total action becomes 
\begin{equation}
    S_{\text{gauged}} = \frac{iN}{2\pi} \int_{\mathcal{M}_{d+1}} b_0 \wedge dc_d +  \frac{iN}{2\pi} \int_{\mathcal{M}_{d+1}} b_0 \wedge
    A_{d+1} .
\end{equation}
Treating $A_{d+1}$ as a Lagrange multiplier allows us to set $b_0 = 0$. Hence, the action vanishes. In other words, we are projecting on a universe where the action vanishes or the theory is trivial. 

Next, we can add a theta angle, an analogue of discrete torsion, before gauging the $d$-form symmetry in the SymTFT, 
\begin{equation}
     S_{\text{gauged}} = \frac{iN}{2\pi} \int_{\mathcal{M}_{d+1}} b_0 \wedge dc_d +  \frac{iN}{2\pi} \int_{\mathcal{M}_{d+1}} b_0 \wedge
    A_{d+1} - \frac{iN}{2\pi} \int_{\mathcal{M}_{d+1}} \hat{b} \wedge A_{d+1}.
\end{equation}
Here, looking at the $A_{d+1}$ equation of motion, we can see that the value of $b_0$ has shifted by $\hat{b}$. This is a clear indication that we have projected onto a separate universe.  Integrating out the Lagrange multiplier $A_{d+1}$ reduces the Lagrangian to $\hat{b}\wedge dc_d$, which is a total derivative, so the action again vanishes.

The $(-1)$-form symmetry we have generated is trivial in the sense that shifting the expectation value of $b_0$ does not qualitatively change any of the physics. To make this non-trivial we must include further couplings to the SymTFT action in \eqref{eqn:decomposingBF}. 
\begin{equation}\label{eqn:generalsymtft}
    S_{\text{bulk}} := \frac{iN}{2\pi} \int_{\mathcal{M}_{d+1}} \left( a_p \wedge db_{d-p} + b_0 \wedge dc_d + \alpha~ b_0 \wedge \omega_{d+1} \right),
\end{equation}
where the last terms can be thought off as an anomalous coupling and $\omega_{d+1}$ is some $(d+1)$-form made out of $a_p$. 

We can play the same game of gauging the $d$-form symmetry generated by the point operators. Gauging the $d$-form symmetry without discrete torsion, projects us onto the universe where $b_0 =0$:
\begin{equation}\label{eqn:univ_1}
    \text{Universe 1: }~ S^1_{\text{bulk}} = \frac{iN}{2\pi} \int_{\mathcal{M}_{d+1}} a_p \wedge b_{d-p} 
\end{equation}
Adding discrete torsion shifts the value of the point operator and produces the action, 
\begin{equation}\label{eqn:univ_2}
    \text{Universe 2: }~ S^2_{\text{bulk}} = \frac{iN}{2\pi} \int_{\mathcal{M}_{d+1}}
    \left(a_p \wedge b_{d-p} + k~ \omega_{d+1} \right),
\end{equation}
where $k = \alpha \hat{b}$. To summarize, we have constructed a decomposing topological field theory, where the universes differ by coefficient of the twist term. Since we have obtained these universes by undoing decomposition, each universe carries a quantum dual $(-1)$-form symmetry that acts on the theory itself and modifies the action of the theory. Moreover, the actions in \eqref{eqn:univ_1} and \eqref{eqn:univ_2} can be treated as the SymTFT's of some $d$-dimensional QFT with a $(p-1)$-form symmetry. Hence, this $(-1)$-form symmetry of the SymTFT itself, should eventually boil down to a notion of $(-2)$-form symmetry for their respective absolute theories. 

One could argue that in the absence of the mixed anomaly, we can still gauge the $d$-form symmetry in the bulk and project down onto individual TFTs. In that case, every universe will have the same action, namely
\begin{equation}
    \frac{iN}{2\pi}\int_{\mathcal{M}_{d+1}} a_p db_{d-p}.
\end{equation}
In this setting as well, one can say there is a $(-1)$-form symmetry of the SymTFT that takes from one universe to the other but this is trivial. This only captures the shift of the value of the point operator, just like an ordinary $(-1)$-form symmetry. Naturally we do have genuine codimension one defects in this setting. Only when we have anomalies, the SymTFT actions in each universe differ completely. That is a non-trivial imprint of the anomaly and in this case we no longer have genuine operators, which is evident from the $b_0$ equation of motion, 
\begin{equation}
    dc_d + \alpha ~ \omega_{d+1} =0
\end{equation}
which tells us that the codimension one surface ($c_d$-surface) is no longer genuine and well defined on it's own, instead it must be attached to a space-time filling bulk.
\begin{eqnarray}
    \exp\Big( i \int_{\Sigma_d = \partial \Delta} c_d + \alpha \int_{\Delta_{d+1}} \omega_{d+1} \Big).
\end{eqnarray}
We would like to pause at this point and try to distinguish these non-genuine codimension one defects from genuine codimension one defects which we usually associate with $(-1)$-form symmetries. The conventional $(-1)$-form symmetries of absolute theory which we usually associate with the genuine co-dimension one defects of SymTFT, only takes care of the shift of the $\theta$-angle type parameter. Here, due to the BF pairing between $b_0$ and $c_d$, when we take the $b_0$ point operator through the $c_d$-surface, there is a shift to the value of $b_0$ (locally), in addition to that, there is a modification of the SymTFT action itself. We believe this is one of the key distinction between $(-1)$-form symmetries and the notion of $(-2)$-form symmetry. This total shift of the action is very similar to the notion of a $(-1)$-form symmetries, only this time it is a $(-1)$-form symmetry of the SymTFT itself and a $(-2)$-form symmetry of the absolute theory.

\subsection{Example: $2d$ toy model}

Next, let us specialize the previous discussion to the case $d=2$.
Starting from the action \eqref{eqn:generalsymtft}, 
in the special case $p=1$,
\begin{equation}\label{eqn:3d_Sym}
    S^{\text{Toy Model}}_{\text{SymTFT}} : = \frac{iN}{2\pi}\int_{\mathcal{M}_{3}} \left(  a_1 \wedge db_1 + b_0 \wedge dc_2 + \frac{1}{2} b_0 \wedge a_1 \wedge da_1 \right) .
\end{equation}
We have replaced the mixed anomaly term with 
\begin{equation}
    \frac{1}{2}\int b_0 a_1 da_1. 
\end{equation}
and we identified $\alpha = 1/2$. 
We simplify further by specializing to  $N=2$, so that our symmetries will eventually be $\Z_2$ instead of $\Z_N$. Now if we gauge the $\Z^{(d)}_2$-form symmetry in the bulk without discrete torsion we should project onto the universe where $b_0 =0$. In other words, the SymTFT action of that universe only consists of a single BF pair, 
\begin{equation}
    \text{Universe~ 1:} ~~ S^1_{3d} = \frac{i}{\pi}\int_{\mathcal{M}_{3}} a_1 \wedge db_1 ,
\end{equation} 
which captures a non-anomalous $\Z_2$ symmetry of a $2d$-absolute theory. Repeating the process with discrete torsion, we would project onto a universe where the $b_0$ shifts to $1$ (our symmetries our $\Z_2$ here), producing the following SymTFT action, 
\begin{equation}
    \text{Universe~ 2:} ~~ S^2_{3d} = \frac{i}{\pi}\int_{\mathcal{M}_{3}}  \left( a_1 \wedge db_1 + \frac{1}{2} a_1 \wedge da_1 \right),
\end{equation}
capturing the data of an anomalous $\Z_2$ symmetry. Now, we can add the physical boundary to our quiches. They differ only via the data of the anomaly. 

Just by looking at \eqref{eqn:3d_Sym}, one might argue that $3d$ SymTFT carries a codimension one defect in the bulk which would eventually generate a $(-1)$-form symmetry of the absolute theory, which should take care of the shift in the value of point operator. This naive conclusion is not fully correct because the codimension one operators in the bulk are not genuine operators of the SymTFT. To be more explicit, we postulate the following gauge transformations of the action \eqref{eqn:3d_Sym}, 
\begin{equation}
\begin{aligned}
    a_1 &\rightarrow a_1 + d\lambda_0 , \\
    b_1 &\rightarrow b_1 + d\mu_0 - na_1, \\ 
    b_0 &\rightarrow b_0 + 2 n, \\
    c_2 &\rightarrow c_2 + d\chi_1 - \frac{\lambda_0}{2}da_1.
\end{aligned}
\end{equation}
with this gauge transformation, the gauge invariant codimension one topological operator is given by the following:
\begin{equation}
    \exp \left(i \left[  \int_{\Sigma = \partial \Delta} c_2 + \frac{1}{2}\int_{\Delta} a_1 da_1 \right] \right),
\end{equation}
where the $c_2$ surface $\Sigma$  
must be attached to a space-time filling object -- making this a non-genuine operator. 

This example might seem a bit artificial but SymTFT's with similar structure can be obtained in real physical setting, in three dimensional ABJM theories or $4d$ generalized Yang-Mills theories \cite{Tanizaki:2019rbk}. The respective SymTFT's of these theories have already been discussed in the literature \cite{vanBeest:2022fss,Yu:2024jtk, Lin:2025oml}. As expected, the co-dimension one surfaces in the respective SymTFTs of these examples do become non-genuine.

\subsection{Example: $3d$ ABJM theories}

Our next example will be the three dimensional ABJM theories which arise on the world volume of M2 branes probing the $\mathbb{C}^4/\Z_k$ singularity \cite{Aharony:2008ug}. The presence of fractional M2 branes localized at the orbifold singularity leads to the $U(N+m)_k \times U(N)_{-k}$ theory \cite{Aharony:2008gk}. The global symmetries of the ABJM theories have been discussed before in the literature \cite{Bergman:2020ifi,vanBeest:2022fss}.

For our purposes, the relevant part of the SymTFT action is the sector involving the discrete one-form background and the fields $b_0,c_3$. A convenient way of formulating this sector is to first choose the global form of the ABJM theory. After this choice, the two-form field should not be treated as a completely off-shell continuum field. Instead, it is an on-shell discrete cocycle. We denote it by $B_2$ in order to emphasize this point. Thus
\begin{equation}
    B_2 \in Z^2(\mathcal{M}_4,\Gamma),
    \qquad
    B_2 \sim B_2+\delta \Lambda_1 ,
\end{equation}
where the coefficient group $\Gamma$ depends on the global form. For the usual $U(N)_k$ ABJM theory one has
\begin{equation}
    \Gamma=\Z_k ,
\end{equation}
while for the $SU(N)_k$ global form one instead has
\begin{equation}
    \Gamma=\Z_N .
\end{equation}
Equivalently, the BF term which imposes the flatness and quantization condition on the two-form field has already been solved. The SymTFT should therefore be viewed as a quantum theory obtained by summing over the on-shell sectors of $B_2$, while keeping $b_0$ and $c_3$ off-shell.

In this partially on-shell description, the relevant SymTFT action is
\begin{equation}\label{eqn:ABJMsymtft-onshell}
    S^{\text{ABJM}}_{\text{SymTFT}}
    =
    \int_{\mathcal{M}_4}
    \left(
        k\, b_0 \wedge dc_3
        -
        \Omega\, b_0 \wedge B_2 \wedge B_2
    \right).
\end{equation}
The corresponding path integral is schematically
\begin{equation}
    Z_{\Gamma}
    =
    \sum_{B_2\in Z^2(\mathcal{M}_4,\Gamma)}
    \int \mathcal{D}b_0\,\mathcal{D}c_3\,
    \exp\left[
        i\int_{\mathcal{M}_4}
        \left(
            k\, b_0 \wedge dc_3
            -
            \Omega\, b_0 \wedge B_2 \wedge B_2
        \right)
    \right].
\end{equation}
Here and below we use continuum notation for the product $B_2\wedge B_2$; in a fully discrete formulation it should be understood as the corresponding cup-product expression.

Although $B_2$ is not an off-shell continuum field, there is still a gauge redundancy associated with changing the representative of the same cocycle class. In continuum notation this takes the form
\begin{equation}\label{eqn:ABJM-gauge-transformation}
\begin{aligned}
    B_2 &\rightarrow B_2+d\Lambda_1,
    \qquad dB_2=0,\\
    c_3 &\rightarrow c_3+d\Lambda_2
    +\frac{2\Omega}{k}
    \left(
        \Lambda_1\wedge B_2
        +\frac{1}{2}\Lambda_1\wedge d\Lambda_1
    \right),\\
    b_0 &\rightarrow b_0 .
\end{aligned}
\end{equation}
The first line should be understood as a change of representative inside the space of on-shell $\Gamma$-valued cocycles, with $\Gamma=\Z_k$ for the $U(N)_k$ global form and $\Gamma=\Z_N$ for the $SU(N)_k$ global form. The transformation of $c_3$ is chosen so that the combination
\begin{equation}
    dc_3-\frac{\Omega}{k}B_2\wedge B_2
\end{equation}
is invariant on the locus $dB_2=0$. Indeed, under $B_2\rightarrow B_2+d\Lambda_1$ one has
\begin{equation}
    (B_2+d\Lambda_1)\wedge (B_2+d\Lambda_1)-B_2\wedge B_2
    =
    d\left(
        2\Lambda_1\wedge B_2+\Lambda_1\wedge d\Lambda_1
    \right),
\end{equation}
where we used $dB_2=0$.

We can now gauge the $\Z_k^{(3)}$ symmetry generated by the topological point operators in the SymTFT coming from the $b_0$ field. Gauging without discrete torsion projects onto the universe where $b_0$ is trivial, and hence there is no anomalous coupling involving the background field for the one-form symmetry. On the other hand, including discrete torsion selects a non-trivial value of the corresponding discrete parameter, which we denote by $\hat b$. In that universe the anomaly term becomes
\begin{equation}
    S_{\mathrm{anomaly}}
    =
    -
    \int_{\mathcal{M}_4}
    \Omega\, \hat b\, B_2\wedge B_2 .
\end{equation}
The only dependence on the global form is through the allowed on-shell values of $B_2$: for the $U(N)_k$ theory $B_2$ is a $\Z_k$ cocycle, while for the $SU(N)_k$ global form it is a $\Z_N$ cocycle.

The $b_0$ equation of motion in the theory \eqref{eqn:ABJMsymtft-onshell} is
\begin{equation}
    k\, dc_3
    =
    \Omega\, B_2\wedge B_2,
\end{equation}
or equivalently
\begin{equation}
    dc_3
    =
    \frac{\Omega}{k}\, B_2\wedge B_2 .
\end{equation}
Thus the operator obtained by integrating $c_3$ over a three-cycle is not genuine in a background with non-trivial $B_2$. The corresponding non-genuine operator must be dressed by a four-dimensional bulk term:
\begin{equation}
    \exp \left(
        i \left[
            \int_{\Sigma_3=\partial \Delta_4} c_3
            -
            \frac{\Omega}{k}
            \int_{\Delta_4} B_2\wedge B_2
        \right]
    \right).
\end{equation}
Because $B_2$ is restricted to be an on-shell cocycle, this expression is gauge invariant under the representative change \eqref{eqn:ABJM-gauge-transformation}. This is the same mechanism by which the $c_3$ operator becomes non-genuine in the presence of the mixed anomaly.

\subsection{Example: $4d$ generalized Yang-Mills theories }

Four dimensional generalized Yang-Mills theory \cite{Tanizaki:2019rbk} also has similar features. The five dimensional Symmetry TFT that probes the generalized symmetries of the generalized Yang Mills theory is given by the following\cite{Lin:2025oml}:
\begin{equation}
    S^{\text{gYM}}_{\mathrm{SymTFT}} = \frac{iN}{2\pi}\int_{\mathcal{M}_5}b_0 \wedge dc_4+ \frac{iN}{2\pi}\int_{\mathcal{M}_5} c_2 \wedge db_2+ \frac{\pi i (N-1)}{N}\int_{\mathcal{M}_5}\frac{db_0}{2\pi} \wedge \mathcal{P}(b_2) ,
\end{equation}
where the second term captures the $1$-form center symmetries of Yang-Mills theory. In addition we have the mixed anomaly involving the $(-1)$-form symmetry and $1$-form symmetry, which \textit{resembles} the mixed anomalies discussed in the last few examples. The Pontraygin square $\mathcal{P}(b_2) \sim 2 b_2^2$ is valued in $\mathbb{Z}_{2N}$. Hence, to be precise we should use integer cohomology instead of differential forms. Also the $b_0$ appearing in the lagrangian should be understood as the lift of $b_0$ to $\mathbb{Z}_{2N}$. 

Similar to our discussion in the $2d$ toy model, we can gauge the $\mathbb{Z}_N^{(4)}$ in the $5d$ bulk and project down to universes with $b_0 =0 $ or a shifted value of $b_0$, due to the presence of discrete torsion. Furthermore, just like the earlier example, the $c_4$ surfaces are no longer genuine. In order to establish that we write down the gauge transformations for the entire action, 
\begin{equation}
\begin{aligned}
     \delta b_2 &= d\Lambda_1, \\
     \delta c_2 &= d\tilde\Lambda_1, \\
     \delta c_4 &= d\Lambda_3 + \frac{2\pi (N-1)}{N^2} \big(b_2 \wedge  d\Lambda_1 + \frac{1}{2} d\Lambda_1 \wedge d\Lambda_1\big).
\end{aligned}
\end{equation}
With these gauge transformations, the gauge invariant codimension one topological operator is given by the following combination:
\begin{equation}
    \exp \left(i \left[  \int_{\mathcal{M}_4= \partial \mathcal{M}_5} c_4 - \frac{\pi (N-1)}{N^2}\int_{\mathcal{M}_5} \text{Bock}(\mathcal{P}(b_2)) \right] \right) ,
\end{equation}
where Bock denotes the Bockstein, $\text{Bock} : H^4(\mathcal{M}_5,\mathbb{Z}_{2N})\rightarrow H^5(\mathcal{M}_5,\mathbb{Z})$. One would naively think that the equation of motion for $b_0$ is $dc_4\sim d\mathcal{P}(b_2)$, but this is not precise. We have to use differential $\delta$ in integral cohomology, instead of ordinary deRham differential $d$. Moreover, $\mathcal{P}(b_2)$ is valued in $\mathbb{Z}_{2N}$, whereas, $c_4$ is in $U(1)$, so we have used the Bockstein. So, $\delta \mathcal{P}(b_2)$ should be understood as $\text{Bock}(\mathcal{P}(b_2))$ and the correct equation of motion for $b_0$ is thus, 
\begin{equation}
    \delta c_4\sim \text{Bock}(\mathcal{P}(b_2)) .
\end{equation}
The Bockstein measures the obstruction to lift a $\mathbb{Z}_{2N}$ valued cohomology element to U(1)-valued. 

This operator can be seen as the generator of `$(-2)$-form symmetry'. As emphasized earlier, the $c_4$ surface must be attached to a space-time filling object. The conventional $(-1)$-form symmetries are usually associated with the genuine co-dimension one defects of SymTFT, which is not the case here. This $(-1)$-form symmetry not only is shifting the $\theta$ angle for the $4d$ Yang-Mills, but it is also shifting the anomaly information.

\subsection{Absolute theory point of view on these examples}

Guided by these examples, we would to like make a few general comments from the perspective of the absolute theory. One key feature that is common in all these examples that there is a mixing between the background gauge fields of the $(d-1)$-form symmetry and some other global symmetry (either a 0-form symmetry or a 1-form symmetry in the examples we have discussed).  For convenience, we are going to list the relevant gauge transformations where such mixing takes place, 
\begin{equation}
    \begin{aligned}
        \text{$2d$ toy model}:~~  c_2 &\rightarrow c_2 + d\chi_1 - \frac{\lambda_0}{2}da_1, 
        \\
        \text{$3d$ ABJM}: ~~ c_3 &\rightarrow c_3 + d\Lambda_2+\frac{2\Omega}{k} (\Lambda_1 \wedge b_2+\tfrac12\Lambda_1 \wedge d\Lambda_1),
        \\
        \text{$4d$ gYM}: ~~ c_4 &\rightarrow c_4 + d\Lambda_3 + \frac{2\pi (N-1)}{N^2} \big(b_2 \wedge  d\Lambda_1 + \frac{1}{2} d\Lambda_1 \wedge d\Lambda_1\big).
    \end{aligned}
\end{equation}
All of these point towards the existence of a higher group structure or a mixed anomaly involving $(d-1)$-form symmetry. In \cite{Tanizaki:2019rbk}, it was shown that there  exists a 4-group involving the $(d-1)$-form symmetry in generalized Yang-Mills theory. The $1$-form symmetry of the Yang-Mills theory was extended by a $3$-form symmetry. Similarly, the existence of a higher group in ABJM theory was proposed in \cite{Yu:2024jtk}. In the same spirit, we can argue that, in the two dimensional example, the gauge transformations of the $1$-form symmetry involves the field strength of the $0$-form symmetry, once again this is a hallmark of 2-group symmetry \cite{Cordova:2018cvg, Benini:2018reh}.

From the SymTFT, in order to uncover this higher group structure we must impose Neumann boundary condition on topological point operators and Dirchlet boundary condition on $c_d ~ (d=2, 3, 4)$, on the symmetry boundary; this will produce the desired higher group. When the codimension one defect was genuine, imposing Dirichlet boundary condition on $c_d$ would have been equivalent to gauging a $(-1)$-form symmetry in the absolute theory. Quite naturally, this gauging produced a quantum dual $(d-1)$-form symmetry. But in this case, imposing Dirichlet boundary conditions\footnote{Whether we want to call this choice of boundary condition ``gauging a $(-2)$-form symmetry'' is pure semantics.} produces a complicated higher group structure that involves the $(d-1)$-form symmetry. This is a departure from the standard notion of $(-1)$-form symmetries. 

\subsubsection*{An attempt at generalization }
We can reverse this chain of thought and consider starting with a QFT carrying a $K^{(d-1)}$ symmetry and some other global symmetry $G^{(p)}$ $\left(0 \le p < (d-1)\right)$, with a mixed anomaly or a higher group involving them, gauging the $(d-1)$-form symmetry in these circumstances can lead us to a few interesting phenomena including the ones we have been observing. In the case of the higher group, $G^{(p)}$ is extended by $K^{(d-1)}$, 
\begin{equation}\label{eqn:highergroup_ses}
    0 \rightarrow K^{(d-1)} \rightarrow \mathfrak{G} \rightarrow G^{(p)} \rightarrow 0 . 
\end{equation}
This is what we have exemplified in our examples. The universes differ via the data of an anomaly of some other global symmetry. It will be interesting to find examples where the $(d-1)$-form symmetry has mixed anomaly with some other symmetries. In that case, gauging the $(d-1)$-form symmetry could explicitly break the quotient $G^{(p)}$ (or to a subgroup). If $ p= (d-2)$, undoing decomposition could potentially lead us to a non-invertible symmetry, very much like the Kaidi-Ohmori-Zheng construction of the Duality defects in $4d$ \cite{Kaidi:2021xfk}.

From the perspective of $d$-dimensional family of theories, we are taking a family of theories with some global symmetry $G^{(p)}$, indexed by $b_0$\footnote{In the $2d$ toy model example, this is an artificial label. For $4d$ generalized Yang-Mills theory, $b_0$ is basically $\theta$.}. There is a classical symmetry under the shift of $b_0$ (one can also say that we have taken a collection of identical theories), and we can either do the spurion analysis \cite{Brennan:2024tlw,Brennan:2026ira} or we can use the notion of $(-1)$-form symmetry to say
\begin{equation}
    \mathcal{U}^{(-1)}\mathcal{T}_{b_0} \cong \mathcal{T}_{b_0 + k},
\end{equation}
where $\mathcal{U}^{(-1)}$ is the generator of $(-1)$-form symmetry. The global symmetries of the individual theories can be probed via individual SymTFTs. The individual SymTFTs will be $(d+1)$ dimensional $G^{(0)}$-Dijkgraaf-Witten theories. We can define a new SymTFT that will essentially capture the symmetries of this whole family, 
\begin{equation}\label{eqn:Sym_family}
    S^{\text{family}}_{\text{SymTFT}} = \frac{iN}{2\pi} \int b_0\, dc_d + S^{G^{(0)}}_{sym} .
\end{equation}
Here we have assumed $b_0$ to be discrete, for a continuous family of theories, we will need to rewrite the first coupling using $\mathbb{R}$ and $U(1)$-valued gauge fields. The SymTFT captures the notion of the $(d-1)$-form symmetry (and the quantum dual $(-1)$-form symmetry), which is not involved in any sort of higher group structure. Now, we can play the game of undoing decomposition as in the previous section and obtain the individual SymTFTs. The genuine co-dimension one $c_2$ surface operator takes care of the shift of $b_0$. 

Moving on, if we have a scenario where the classical symmetry under the shift of $b_0$ is broken due to the presence of background gauge fields of $G^{(0)}$. 
\begin{equation}
    \mathcal{U}^{(-1)}\mathcal{T}_{b_0} \cong e^{\left( i \alpha \right)} \mathcal{T}_{b_0 + k},
\end{equation}
where $\exp {\left( i \alpha \right)}$ is the anomalous phase due to the anomaly in the space of coupling constants \cite{Cordova:2019jnf,Cordova:2019uob}. For example, in Yang-Mills theories, under the shift of $\theta \rightarrow \theta + 2\pi$, we pick up a phase $\exp\left(i\int B_2 \wedge B_2\right)$, where $B_2$ is the background field of the 1-form symmetry. We can just stick to two theories in this family, $b_0$ at $0$ and at $1$. We can model the shift of $b_0$ (or, introduction of $b_0$) as $\Z^{(-1)}_2$ symmetry. The SymTFT action that will probe the generalized symmetry structure of this family will have a structure to similar to \eqref{eqn:3d_Sym}, capturing the higher group structure involving the $(d-1)$-form symmetry. Once again, we will have a non-genuine surface operator, which will take care of the shift of $b_0$. Moreover, it will `\textit{add}' the anomaly. When we try to tune $b_0$ from 0 to 1, we are transitioning from a universe with a non-anomalous symmetry to a universe with an anomalous symmetry. 

We have observed the interesting interplay between decomposition and higher group structure. This leads us to the question of whether we can have mixed anomalies between $(d-1)$-form symmetries and some other global symmetries of our QFT. Allowing the global symmetry to be non-invertible could also be an interesting avenue, which we leave for future work.

\subsection{Example: shifting the associator of a fusion categorical symmetry}\label{subsec:assoc-shift}

In the examples discussed so far, the $(-2)$-form symmetry shifts the 't~Hooft anomaly coefficient of an \emph{invertible} symmetry --- a $\Z_2$ in the $2d$ toy model, the center one-form symmetry in $3d$ ABJM and $4d$ generalized Yang--Mills. It is natural to ask whether the same mechanism can capture more general modifications of the symmetry data, in particular the F-symbols (associators) of a non-invertible symmetry. We now construct a single parent SymTFT in which one and the same non-genuine codimension-one defect can be interpreted as either kind of $(-2)$-form symmetry, depending on the choice of topological boundary condition (Lagrangian algebra) on the symmetry boundary.

The two fusion categorical symmetries of interest are $\operatorname{Rep}(D_4)$ and $\operatorname{Rep}(Q_8)$. As Tambara--Yamagami categories \cite{TAMBARA1998692}, both take the form $\mathrm{TY}(\Z_2 \times \Z_2,\,\chi,\,\tau)$ with the same non-degenerate symmetric bicharacter $\chi$ on $A = \Z_2 \times \Z_2$, and hence the same fusion ring: four invertible objects $\{1\}\cup A$ together with a single non-invertible object $m$ obeying $m \otimes m = \bigoplus_{g \in A} g$. The two categories are distinguished only by the Frobenius--Schur sign $\tau = \pm 1$, which multiplies the non-trivial component of the associator with three non-invertible inputs,
\begin{equation}\label{eqn:Fmmm}
    [F^{m,m,m}_m]_{a,b} \;=\; \frac{\tau}{\sqrt{|A|}}\, \chi(a,b),
    \qquad a,b\in A.
\end{equation}
The choice $\tau = +1$ realizes $\operatorname{Rep}(D_4)$ --- the unique simple non-invertible representation $m$ of $D_4$ is of real type --- while $\tau = -1$ realizes $\operatorname{Rep}(Q_8)$, where the analogous representation is pseudo-real (quaternionic). The Grothendieck ring, the bicharacter $\chi$, and every other F-symbol of the category are identical between the two cases. Following \cite{Robbins:2025puq,Bergman:2026lnz}, the corresponding $3d$ SymTFTs can be written in terms of three $\Z_2$-valued 1-cocycles $a, b, c$ and their duals $\hat a, \hat b, \hat c$ as
\begin{align}
    S_{D_4} &= \pi \int_{X_3} \Big( \hat a\,\delta a + \hat b\,\delta b + \hat c\,\delta c + a\,b\,c \Big), \label{eqn:SymTFT_D4} \\
    S_{Q_8} &= \pi \int_{X_3} \Big( \hat a\,\delta a + \hat b\,\delta b + \hat c\,\delta c + a\,b\,c + \tfrac{1}{2}\,c\,\delta a + \tfrac{1}{2}\,c\,\delta b \Big), \label{eqn:SymTFT_Q8}
\end{align}
which differ only by the last two couplings of \eqref{eqn:SymTFT_Q8}. These encode the cocycle distinguishing the central $\Z_2$ extension $Q_8$ from $D_4$, and on the boundary side they are precisely what shifts $\tau \to -\tau$ in \eqref{eqn:Fmmm}.

Following the strategy of Section~\ref{sec:gen_prescription}, we engineer a $(-2)$-form symmetry interpolating between the two SymTFTs by promoting the relative twist to the value of an auxiliary $\Z_2$-valued $0$-form $b_0$ with BF partner $c_2$:
\begin{equation}\label{eqn:SymTFT_TY_parent}
    S_{\text{parent}}
    \;=\; \pi \int_{X_3} \Big( \hat a\,\delta a + \hat b\,\delta b + \hat c\,\delta c + a\,b\,c
        + b_0\,\delta c_2 + \frac{b_0}{2}\,\big(c\,\delta a + c\,\delta b\big) \Big),
\end{equation}
with $b_0 \in C^0(X_3,\Z_2)$ and $c_2 \in C^2(X_3,\Z_2)$. The BF coupling forces $b_0$ to be constant on connected components, so the bulk decomposes into two universes labelled by $b_0 \in \{0, 1\}$, which recover $S_{D_4}$ and $S_{Q_8}$, respectively; gauging the $\Z_2^{(2)}$ shift symmetry of $b_0$ with or without discrete torsion projects onto one of them.

Just as in the $2d$ toy model \eqref{eqn:3d_Sym}, the codimension-one $c_2$ surface that implements the shift $b_0 \to b_0 + 1$ is not a genuine operator. Gauge invariance of \eqref{eqn:SymTFT_TY_parent} forces the dressed combination
\begin{equation}\label{eqn:nongenuine_TY}
    \mathcal{U}^{(-1)}_{\text{bulk}}[\Sigma_2]
    \;=\;
    \exp\!\Big( i\pi \Big[ \int_{\Sigma_2} c_2 \;+\; \tfrac{1}{2} \int_{\Delta_3} \big(c\,\delta a + c\,\delta b\big) \Big] \Big),
    \qquad \Sigma_2 = \partial \Delta_3,
\end{equation}
to be attached to a three-dimensional filling region $\Delta_3$ carrying the relative twist. This is the hallmark of a $(-1)$-form symmetry of the bulk SymTFT, and hence of a $(-2)$-form symmetry of the $2d$ absolute theory.

\paragraph{Two readings of the same non-genuine defect.}
The key observation is that the parent SymTFT \eqref{eqn:SymTFT_TY_parent} admits several inequivalent Lagrangian algebras on the symmetry boundary, and the \emph{same} non-genuine operator $\mathcal{U}^{(-1)}_{\text{bulk}}$ in \eqref{eqn:nongenuine_TY} descends to physically distinct $(-2)$-form symmetries of the resulting $2d$ absolute theory, depending on the choice. We highlight two:

\begin{itemize}
\item \textbf{Dirichlet on $a, b, c$ --- anomaly shift of an invertible $\Z_2^3$ symmetry.}
With the purely electric Lagrangian algebra in \cite{Yu:2023nyn},
the boundary theory carries an invertible $\Z_2^{(a)} \times \Z_2^{(b)} \times \Z_2^{(c)}$ symmetry \cite{Robbins:2025puq}. In universe~1 the boundary inherits the mixed anomaly captured by $abc$ alone; in universe~2 the additional bulk couplings $\tfrac12 c\,\delta a + \tfrac12 c\,\delta b$ shift the anomaly cocycle by the class distinguishing $Q_8$ from $D_4$ in $H^3(\Z_2^3, U(1))$. In this reading $\mathcal{U}^{(-1)}_{\text{bulk}}$ realizes a $(-2)$-form symmetry of a $\Z_2^3$-symmetric theory that shifts its 't~Hooft anomaly --- a direct higher-rank analogue of the $\Z_2 / \Z_2^\omega$ story of the $2d$ toy model.

\item \textbf{Dirichlet on $\hat{a}, \hat{b}, c$ --- associator shift between $\operatorname{Rep}(D_4)$ and $\operatorname{Rep}(Q_8)$.}
With this symmetry boundary condition, the boundary realizes a non-invertible Tambara--Yamagami symmetry: $\operatorname{Rep}(D_4)$ in universe~1 and $\operatorname{Rep}(Q_8)$ in universe~2 \cite{Robbins:2025puq}. The two symmetry categories share the fusion ring $\mathrm{TY}(\Z_2 \times \Z_2,\chi)$ but carry inequivalent associators; concretely, $\mathcal{U}^{(-1)}_{\text{bulk}}$ flips the Frobenius--Schur sign $\tau \to -\tau$ in the F-symbol \eqref{eqn:Fmmm} while leaving every other piece of the categorical data, including the fusion rules and the bicharacter $\chi$, untouched. In this reading the same non-genuine operator acts as a $(-2)$-form symmetry that rewires the associator of a non-invertible symmetry while preserving its fusion ring.
\end{itemize}

The parent SymTFT supports further Lagrangian algebras, e.g., those realizing $D_4$ or $Q_8$ themselves (Dirichlet on two of $a, b, c$ and Neumann on the third), or their $\Z_4$ subgroups \cite{Yu:2023nyn, Robbins:2025puq, Bergman:2026lnz}. Each one yields another symmetry category on which $\mathcal{U}^{(-1)}_{\text{bulk}}$ acts as a $(-2)$-form symmetry mapping it to its $Q_8$-cousin counterpart.

\subsubsection*{Beyond Tambara--Yamagami: unitary vs.\ non-unitary categorifications.}
The construction above relates two \emph{unitary} fusion categories that share a Grothendieck ring. It is natural to ask whether $(-2)$-form symmetries can also relate a unitary categorification of a given fusion ring to a non-unitary one. A natural candidate is the Fibonacci fusion ring with simple objects $\{1, X\}$ and $X \otimes X = 1 \oplus X$, which admits exactly two solutions of the pentagon equation up to equivalence \cite{Kitaev:2005hzj, Rowell:2007dge}: the unitary Fibonacci category $\mathrm{Fib}$, with quantum dimension $d_X = \varphi = (1+\sqrt5)/2$, and its Galois conjugate, the non-unitary Yang--Lee category $\mathrm{YL}$, with $d_X = -1/\varphi$. The two are exchanged by the Galois action $\sqrt{5}\to -\sqrt{5}$, and their Drinfeld centers are inequivalent modular tensor categories \cite{Rowell:2007dge, davydov2011wittgroupnondegeneratebraided} realized, respectively, by the chiral algebras of the $(G_2)_1$ WZW model and the $(2,5)$ non-unitary minimal model; both categories also appear naturally as topological defect lines in two-dimensional CFTs \cite{Chang:2018iay,Thorngren:2021yso}. We do not have a $\Z_2$ toggle of the type \eqref{eqn:SymTFT_TY_parent} in this case, because the corresponding SymTFTs are not Dijkgraaf--Witten theories that differ by a discrete cocycle; in particular, the two centers have distinct chiral central charges modulo $8$, so the would-be parent topological theory cannot be a single bulk action with a $b_0$-controlled term. Nevertheless, the structural picture suggested by our analysis applies verbatim: there should exist some parent topological theory in which a non-genuine codimension-one defect interpolates between the boundary realizations $\mathrm{Fib}$ and $\mathrm{YL}$ of the same Grothendieck ring, and this defect would be a $(-2)$-form symmetry of the $2d$ absolute theory that exchanges unitarity for non-unitarity at the level of the symmetry category. We leave the explicit construction of such a parent SymTFT, and the corresponding (non-DW) bulk action, as a natural direction for future work.

\paragraph{Summary.}
The lesson of this subsection is that the bulk-side characterization of $(-2)$-form symmetries as $(-1)$-form symmetries of the SymTFT is genuinely \emph{prior}\footnote{In the category-theoretic language of Appendix~\ref{app:cat}, the non-genuine bulk defect $\mathcal{U}^{(-1)}_{\text{bulk}}$ corresponds to a single $1$-morphism in the higher category $\mathbf{T}$ of stackings of the SymTFT $\mathscr{T}$ with auxiliary $(d+1)$-dimensional TFTs. A choice of Lagrangian algebra on the symmetry boundary selects which object of $\mathbf{T}$ plays the role of the distinguished boundary theory $\mathsf{t}$. The two readings above thus correspond to two different choices of distinguished boundary theory sharing the \emph{same} bulk $1$-morphism, and the resulting $(-2)$-form symmetries differ only because they are obtained by composition, in the sense of the diagram \eqref{eq:t-boundary-theory}, with these two different distinguished objects. From this perspective, a $(-2)$-form symmetry is most invariantly defined as a bulk $1$-morphism, with its concrete realization as a transformation of an absolute $d$-dimensional theory recovered only after specifying a Lagrangian algebra.} to any specific symmetry-category description on the boundary. The same non-genuine codimension-one defect in the SymTFT can realize: an anomaly shift for an invertible symmetry, an associator shift between two unitary fusion categories with the same fusion ring, and --- one expects, more generally --- a transition between unitary and non-unitary categorifications of the same fusion ring. The universal feature is that $\mathcal{U}^{(-1)}_{\text{bulk}}$ shifts whatever boundary datum is encoded by the relative twist between the two SymTFTs.

\subsection{Dual picture: $d$-form symmetries}

A way in which one may gain additional insight into $(-2)$-form symmetries is by considering their quantum duals under gaugings, namely, $d$-form symmetries. In this case, the symmetry is locally parameterized by a flat $d$-form. The flatness condition in this case is automatic for dimensional reasons, but in general the form, being a connection, is defined only locally. Furthermore, gauging a symmetry generically requires making the symmetry parameter non-flat. On $d$-dimensional space $M$ on its own, this is not possible, but one may use the well-known approach of realizing $M$ as the boundary of a $(d+1)$-dimensional space $B$, so that the differential of the $d$-form can be nonzero on $B$.

Consider the concrete example of $d=2$. A $U(1)$ $2$-form symmetry should correspond to the action of a flat connection on a bundle gerbe over $M$. This is the next step in the higher-group categorization of $U(1)$ $1$-form symmetries as (shift) actions of flat connections on principal $U(1)$ bundles. 
Gauging thus sums over all such possible (isomorphism classes of) bundle gerbes. In a dual way, this matches the description of $(-2)$-form symmetries we gave above in terms of non-genuine defects, as here we are summing over two-dimensional defects which are boundaries of three-dimensional defects.

In very broad generality, $U(1)$ $d$-form symmetry actions may be read as shifting the Lagrangian by a locally-defined topological term (the flat connection on a bundle $d$-gerbe)
\begin{equation}
    B_{(d)}: \mathcal{L} \mapsto \mathcal{L}+{B}_{(d)}.
\end{equation}
We shall see a concrete realization of this perspective in Section~\ref{sec:romans-minus-two-topdown}.

\section{Club sandwich, RG flows and $(-2)$-form symmetry}\label{sec:rgflowand-2fs}
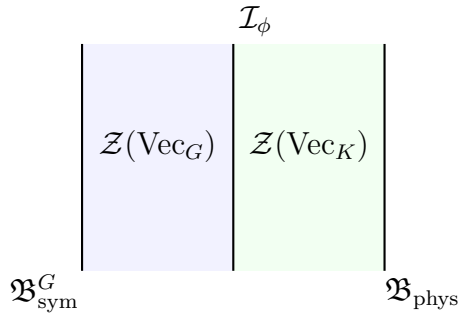
\begin{figure}[H]
    \centering
    \begin{tikzpicture}[scale=1.0]
    
    % Original diagram with two rectangles and interface
    % Draw lines and fill colors
    \fill[blue!5] (0,0) rectangle (2,3); % Blue region
    \fill[green!5] (2,0) rectangle (4,3); % Green region
    
    % Draw interface (thick black line)
    \draw[thick, black] (2,0) -- (2,3);
    
    % Draw boundary lines
    \draw[thick] (0,0) -- (0,3);
    \draw[thick] (4,0) -- (4,3);
    
    % Labels
    \node at (-0.5,-0.3) {{$\mathfrak{B}_\text{sym}^{G}$}}; % Label for left line
    \node at (4.5,-0.3) {{$\mathfrak{B}_\text{phys}$}}; % Label for right line
    \node at (2.3,3.3) {{$\mathcal{I}_\phi$}}; % Label for interface
    \node at (1.0, 1.65) {{${\mathcal{Z}}(\text{Vec}_G)$}};
    \node at (3.0, 1.65) {{${\mathcal{Z}}(\text{Vec}_K)$}};
    \end{tikzpicture}
    
    \caption{The club sandwich}
    \label{fig:club_sandwich}
\end{figure}

In the last few sections, we introduced the notion of $(-2)$-form symmetry, which introduces a twist term to the SymTFT action. In this section, we will use the club sandwich construction to emphasize the physical implications of such $(-2)$-form symmetries. The club sandwich -- an elegant extension of the SymTFT sandwich -- is designed to capture the trivial action of symmetry in the deep IR \cite{Bhardwaj:2023bbf,Antinucci:2025fjp,Lu:2025yru}. We note that the club sandwich can be viewed as a special case of the broader multi-sector / SymTree construction of \cite{Baume:2023kkf}, which appeared earlier; we adopt the club-sandwich language for concreteness, but the constructions below fit naturally into that more general framework. For example, consider a quantum field theory with symmetry $G$ acting effectively in the UV. In the IR, if the operators charged under $H\subset G$ are absent, then that particular subgroup acts trivially. In a situation like this, we can construct the following club sandwich Fig.~\ref{fig:club_sandwich}. 
We can summarize the situation with the following short exact sequence of groups (more generally, an exact sequence of fusion categories), 
\begin{equation}
    0 \longrightarrow H \longrightarrow G \longrightarrow K \longrightarrow 0,
\end{equation}
where $H$ is the symmetry that acts trivially in the IR. For the particular example of UV symmetry group $\Z_4$, we can construct two different such club sandwiches that effectively capture two different renormalization group flows. In the IR theory, these differ by the presence of an anomaly of the effectively acting $\Z_2$ subgroup. They essentially encode the two different categorical extensions for $\Z_4$, 
\begin{equation}
\begin{aligned}
    0 \longrightarrow \text{Vec}(\Z_2) &\longrightarrow \text{Vec}(\Z_4) \longrightarrow \text{Vec}(\Z_2, \omega) \longrightarrow 0 ,\\
    0 \longrightarrow \text{Vec}(\Z_2) &\longrightarrow \text{Vec}(\Z_4) \longrightarrow \text{Vec}(\Z_2) \longrightarrow 0 .
\end{aligned}    
\end{equation}
Here the arrows represent monoidal functors between fusion categories.  Restricting to objects, both sequences reduce to the short exact sequence of groups $1\rightarrow\Z_2\rightarrow\Z_4\rightarrow\Z_2\rightarrow 1$.
The club sandwich is constructed by condensing a subalgebra in the right slab of the SymTFT bulk.  For the case with an anomalous $\Z_2^\omega$ phase we use the condensable algebra\footnote{We are using the standard nomenclature for the anyons in the Drinfeld center of $\Z_4$, 
\begin{align*}
    \mathcal{Z}(\text{Vec}_{\Z_4}) = \{ e^km^l ~ ~ | ~ ~ k,l \in (0,1,2,3)\} \, . 
\end{align*}} $1\oplus e^2m^2$, while for the non-anomalous $\Z_2$ phase we condense $1\oplus m^2$.  
In other words, we can construct the club sandwich via half gauging. For example, consider the club sandwich with $\Z_4$ and $\Z^\omega_2$. To construct the club sandwich via half gauging, let's start with the $\Z_4$ SymTFT given by, 
\begin{equation}
    S_{\Z_4} = \pi\int_{\mathcal{M}_3} \left( a_1 \cup \delta b_1 + \hat{a}_1 \cup \delta \hat{b}_1 - \frac{1}{2} a_1 \cup \delta \hat{b}_1 \right).
\end{equation}
By gauging the 1-form symmetry generated by, $\exp\left[i \pi \oint \left( a_1 + \hat{b}_1 \right) \right]$ in the half space we can recover the club sandwich. The half gauging enforces the condition, $(a_1 + \hat{b}_1) =0$ in the half space. Substituting this condition back into the action above produces the SymTFT for $\Z^\omega_2$ in the half space. In the same spirit, we can also construct the club sandwich with $\Z_4$ and $\Z_2$ SymTFT's by half gauging the $1$-form symmetry generated by $\exp\left(i\pi \oint \hat{b}_1 \right)$, setting $\hat{b}_1 =0$ in the half space.

\subsection{Quarter gauging in SymTFT}
\begin{figure}[H]
    \centering
    \begin{tikzpicture}[scale=1]
    
    % Original diagram with two rectangles and interface
    % Draw lines and fill colors
   \fill[blue!15] (0,0) rectangle (4,3); % Blue region
    \fill[blue!15] (2,0) rectangle (4,1.5); % Green region
    \fill[red!15] (2,1.5) rectangle (4,3);
    
    % Draw interface (thick black line)
    \draw[thick, black] (2,1.5) -- (2,3);
    \draw[thick, black] (2,1.5) -- (4,1.5);
    
    % Draw boundary lines
    \draw[thick] (0,0) -- (0,3);
    \draw[thick] (4,0) -- (4,3);

    % Labels
    \node at (-0.5,-0.3) {{$\mathfrak{B}_\text{sym}$}}; % Label for left line
    \node at (4.5,-0.3) {{$\mathfrak{B}_\text{phys}$}}; % Label for right line
%    \node at (2.3,3.3) {\footnotesize{$\mathcal{I}_\phi$}}; % Label for interface
    \node at (1.0, 1.5) {{${\mathcal{Z}}(\text{Vec}_G)$}};
    \node at (3.0, 2.3) {{${\mathcal{Z}}(\text{Vec}_H)$}};
    
    % Equal sign
    \node at (5.5,1.5) {$\rightarrow$}; % Equal sign

    % Vertical line T
    \draw[thick, black] (7,0) -- (7,3); % Vertical line T
    \node at (7.5,2.2) {$\text{T}_{IR}$}; % Label for line T
    \node at (7.5,0.8) {$\text{T}_{UV}$};
    \fill[orange] (7,1.5) circle (2pt); 
    
    \end{tikzpicture}
    
    \caption{The orange blob on the right hand side is a RG interface. The physical boundaries attached with $\text{Vec}_G$ and $\text{Vec}_H$ are different. To reduce clutter with notations we have chosen not to add another label. }
    \label{fig:club_sandwich_3}
\end{figure}
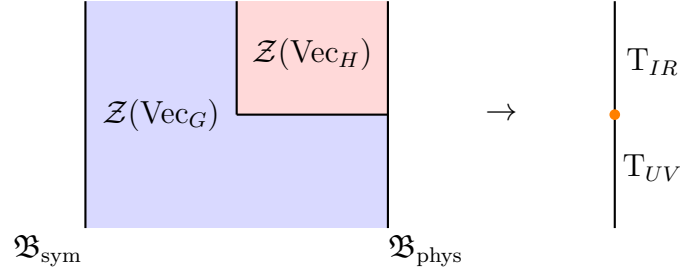

\begin{figure}[H]
    \centering
    \begin{tikzpicture}[scale=1]
    
    % Original diagram with two rectangles and interface
    % Draw lines and fill colors
    \fill[blue!15] (0,0) rectangle (4,3); % Blue region
    \fill[green!15] (2,0) rectangle (4,1.2); % Green region
    \fill[red!15] (2,1.8) rectangle (4,3);
    
    % Draw interface (thick black line)
    \draw[thick, gray] (2,0) -- (2,1.2);
    \draw[thick, black] (2,1.8) -- (2,3);

    % Horizontal line in right (green) region
    \draw[thick, gray] (2,1.2) -- (4,1.2);
    \draw[thick, black] (2,1.8) -- (4,1.8);
    
    % Draw boundary lines
    \draw[thick] (0,0) -- (0,3);
%    \draw[thick, gray] (4,0) -- (4,1.5);
    \draw[thick, black] (4,0) -- (4,3);

%    \fill[black] (2,1.5) circle (2pt);
%    \fill[orange] (4,1.5) circle (2pt);
    
    % Labels
    \node at (-0.5,-0.3) {{$\mathfrak{B}_\text{sym}$}}; % Label for left line
    \node at (4.5,-0.3) {{$\mathfrak{B}_\text{phys}$}}; % Label for right line
    \node at (2.1,3.3) {{$\mathcal{I}_{\phi''}$}}; % Label for interface
    \node at (2.1,- 0.5) {{$\mathcal{I}_{\phi'}$}}; % Label for interface
    \node at (1.0, 1.5) {{${\mathcal{Z}}(\text{Vec}_{G})$}};
    \node at (3.0, 2.3) {{${\mathcal{Z}}(\text{Vec}_{H})$}};
    \node at (3.0, 0.7) {{${\mathcal{Z}}(\text{Vec}_{K})$}};

    \node at (5.5,1.5) {$\rightarrow$}; % Equal sign

    % Vertical line T
    \draw[thick, black] (7,0) -- (7,3); % Vertical line T
    \node at (7.5,2.6) {$\text{T}_{IR_2}$}; % Label for line T
    \node at (7.5,0.3) {$\text{T}_{IR_1}$};
    \draw[thick, black] (7,0) -- (7,3); % Vertical line T
    \node at (7.5, 1.5) {$\text{T}_{UV}$};
    \fill[orange] (7,1) circle (2pt); 
    \fill[orange] (7,2) circle (2pt);
    \end{tikzpicture}
    
    \caption{We have performed different gaugings in different regions in this setting.}
    \label{fig:club_sandwich_5}
\end{figure}
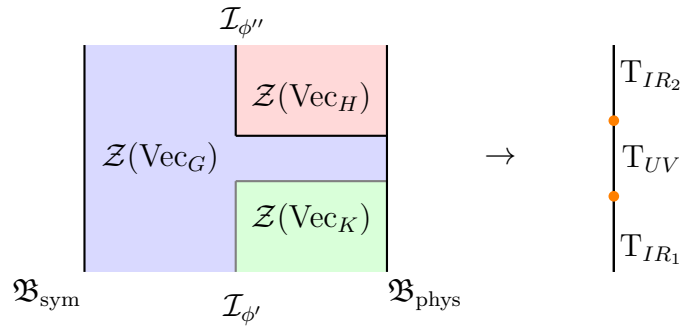

\begin{figure}[H]
    \centering
    \begin{tikzpicture}[scale=1.5]
    
    % Original diagram with two rectangles and interface
    % Draw lines and fill colors
    \fill[blue!15] (0,0) rectangle (2,3); % Blue region
    \fill[green!15] (2,0) rectangle (4,1.5); % Green region
    \fill[red!15] (2,1.5) rectangle (4,3);
    
    % Draw interface (thick black line)
    \draw[thick, gray] (2,0) -- (2,1.5);
    \draw[thick, black] (2,1.5) -- (2,3);

    % Horizontal line in right (green) region
    \draw[dashed, blue] (2,1.5) -- (4,1.5);
    
    % Draw boundary lines
    \draw[thick] (0,0) -- (0,3);
%    \draw[thick, gray] (4,0) -- (4,1.5);
    \draw[thick, black] (4,0) -- (4,3);

    \fill[black] (2,1.5) circle (2pt);
    \fill[orange] (4,1.5) circle (2pt);
    
    % Labels
    \node at (-0.5,-0.3) {{$\mathfrak{B}_\text{sym}$}}; % Label for left line
    \node at (4.5,-0.3) {{$\mathfrak{B}_\text{phys}$}}; % Label for right line
    \node at (3.0,1.73) {{$\mathcal{I}$}}; % Label for interface
    \node at (1.0, 1.5) {{${\mathcal{Z}}(\text{Vec}_{G})$}};
    \node at (3.0, 2.5) {{${\mathcal{Z}}(\text{Vec}_{K})$}};
    \node at (3.0, 0.7) {{${\mathcal{Z}}(\text{Vec}_{H})$}};
    \end{tikzpicture}
    \caption{After fusing the interfaces $\mathcal{I}_{\phi''}$ and $\mathcal{I}_{\phi'}$.}
    \label{fig:club_sandwich_6}
\end{figure}
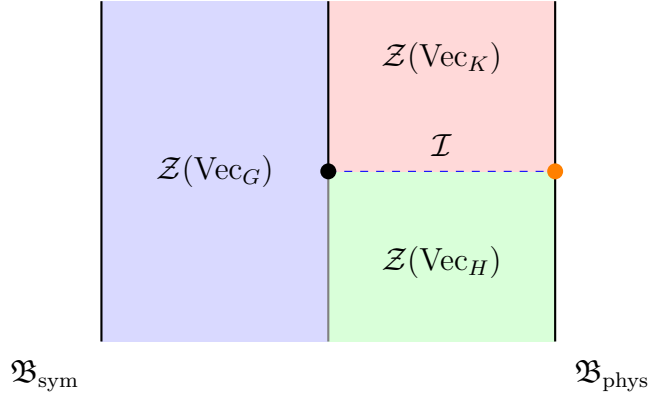

To build the club sandwich, we essentially do a half gauging in the bulk of SymTFT. Moving on, instead of gauging the 1-form symmetry in half space, we can do a \textit{quarter gauging} of the 1-form symmetry. In other words, starting from the SymTFT for $\text{Vec}(G)$ defined on $\mathcal{M}_3 := M_2 \times [0,1]$, we can gauge a non-anomalous 1-form symmetry of the bulk (in other words, condensing a particular condensable algebra of the Drinfeld center) in the upper right quarter of $\mathcal{M}_3$. Collapsing the sandwich\footnote{We have attached different physical boundaries to each half of the club quiche to make it a club sandwich.} at this stage will produce a scenario where the $G$ symmetric UV theory and $H$ symmetric IR theory are separated by an RG interface introduced in \cite{Gaiotto:2012np}, see Fig.~\ref{fig:club_sandwich_3}. The resulting configuration, in which two topological boundaries and an interface meet along a corner, is an instance of the SymTh/SymTFT-with-corners picture developed in \cite{Cvetic:2024dzu}; in particular, the data localized at the corner where the interface meets the symmetry boundary can be analyzed using the cornering construction of that work.
We can play the same game by gauging a different $1$-form symmetry (or by choosing a different condensable algebra) in the lower right half of $\mathcal{M}_3$, as in the figure below.
\begin{figure}[H]
    \centering
    \begin{tikzpicture}[scale=1]
    
    % Original diagram with two rectangles and interface
    % Draw lines and fill colors
   \fill[blue!15] (0,0) rectangle (4,3); % Blue region
    \fill[green!15] (2,0) rectangle (4,1.5); % Green region
    \fill[blue!15] (2,1.5) rectangle (4,3);
    
    % Draw interface (thick black line)
    \draw[thick, gray] (2,0) -- (2,1.5);
%    \draw[thick, black] (2,1.5) -- (2,3);

    % Horizontal line in right (green) region
    \draw[thick, gray] (2,1.5) -- (4,1.5);
    
    % Draw boundary lines
    \draw[thick] (0,0) -- (0,3);
%    \draw[thick, gray] (4,0) -- (4,1.5);
    \draw[thick, black] (4,0) -- (4,3);

%    \fill[black] (2,1.5) circle (2pt);
%    \fill[orange] (4,1.5) circle (2pt);
    
    % Labels
    \node at (-0.5,-0.3) {{$\mathfrak{B}_\text{sym}$}}; % Label for left line
    \node at (4.5,-0.3) {{$\mathfrak{B}_\text{phys}$}}; % Label for right line
    \node at (2.1,- 0.5) {{$\mathcal{I}_{\phi'}$}}; % Label for interface
    \node at (1.0, 1.5) {{${\mathcal{Z}}(\text{Vec}_G)$}};
%    \node at (3.0, 2.3) {{${\mathcal{Z}}(\text{Vec}_{\Z^\omega_2})$}};
    \node at (3.0, 0.7) {{${\mathcal{Z}}(\text{Vec}_K)$}};
    \node at (5.5,1.5) {$\rightarrow$}; % Equal sign

    % Vertical line T
    \draw[thick, black] (7,0) -- (7,3); % Vertical line T
    \node at (7.5,2.2) {$\text{T}_{UV}$}; % Label for line T
    \node at (7.5,0.8) {$\text{T}_{IR}$};
    \fill[orange] (7,1.5) circle (2pt); 
    
    \end{tikzpicture}
    
    \caption{The orange blob is a RG interface. The IR theory in this setting is different from the previous case in Fig. \ref{fig:club_sandwich_3}.}
    \label{fig:club_sandwich_4}
\end{figure}
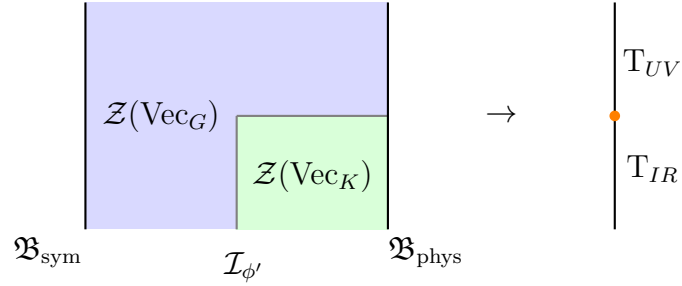
Taking this a step further, gauging different 1-form symmetries simultaneously in different regions of $\mathcal{M}_3$ allows us to have a situation where both of the cases discussed above co-exist, Fig.~\ref{fig:club_sandwich_5}.  
Since the procedure carried out is topological, we can fuse the interfaces $\mathcal{I}_{\phi''}$ and $\mathcal{I}_{\phi'}$ to get the Fig~\ref{fig:club_sandwich_6},
The interface $\mathcal{I}$ in Fig.~\ref{fig:club_sandwich_6}, represented by the blue dashed line is the product of two topological interfaces. 
\begin{equation}
    \mathcal{I} := \mathcal{I}_{\phi''} \otimes  \mathcal{I}_{\phi'} 
\end{equation}
We believe this setting can capture symmetry data of two different RG flows triggered from the UV theory. Specifically, we demand $H$ to carry some anomaly and $K$ to be anomaly free for our purposes. We would like to give some examples for our quarter gauging construction where we end up with $H$ being an anomalous group.
\subsection{Examples}
Our first example will be $\text{Vec}_{\Z_4}$. The condensable algebras of $\mathcal{Z}(\text{Vec}_{\Z_4})$ are well known \cite{Wen:2022tkg,Bhardwaj:2024qrf,Li:2023knf,Li:2022jbf}, using them we can construct a situation similar to Fig. \ref{fig:club_sandwich_5}.
\subsubsection*{$\mathbb{Z}_4 \rightarrow {\mathbb Z}_2 / {\mathbb Z}_2^{\omega}$ example}

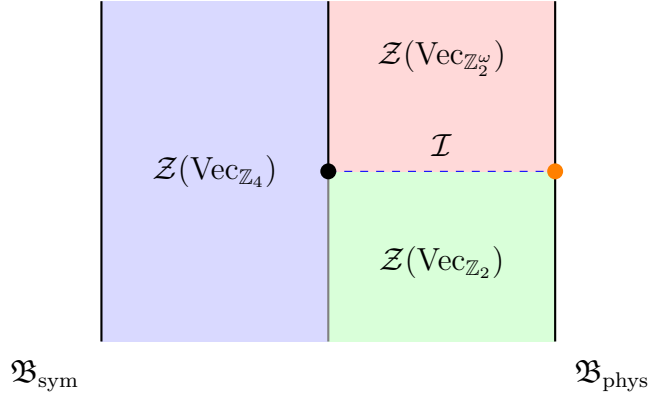
\begin{figure}[H]
    \centering
    \begin{tikzpicture}[scale=1.5]
    
    % Original diagram with two rectangles and interface
    % Draw lines and fill colors
    \fill[blue!15] (0,0) rectangle (2,3); % Blue region
    \fill[green!15] (2,0) rectangle (4,1.5); % Green region
    \fill[red!15] (2,1.5) rectangle (4,3);
    
    % Draw interface (thick black line)
    \draw[thick, gray] (2,0) -- (2,1.5);
    \draw[thick, black] (2,1.5) -- (2,3);

    % Horizontal line in right (green) region
    \draw[dashed, blue] (2,1.5) -- (4,1.5);
    
    % Draw boundary lines
    \draw[thick] (0,0) -- (0,3);
%    \draw[thick, gray] (4,0) -- (4,1.5);
    \draw[thick, black] (4,0) -- (4,3);

    \fill[black] (2,1.5) circle (2pt);
    \fill[orange] (4,1.5) circle (2pt);
    
    % Labels
    \node at (-0.5,-0.3) {{$\mathfrak{B}_\text{sym}$}}; % Label for left line
    \node at (4.5,-0.3) {{$\mathfrak{B}_\text{phys}$}}; % Label for right line
    \node at (3.0,1.73) {{$\mathcal{I}$}}; % Label for interface
    \node at (1.0, 1.5) {{${\mathcal{Z}}(\text{Vec}_{\Z_4})$}};
    \node at (3.0, 2.5) {{${\mathcal{Z}}(\text{Vec}_{\Z^\omega_2})$}};
    \node at (3.0, 0.7) {{${\mathcal{Z}}(\text{Vec}_{\Z_2})$}};
    \end{tikzpicture}
    \caption{Starting from the $3d$ SymTFT of a theory with $\Z_4$ symmetry, we have gauged different 1-form symmetries. The physical boundary attached with $\text{Vec}(\Z_2)$ and $\text{Vec}(\Z^\omega_2)$ are different.}
    \label{fig:club_sandwich_7}
\end{figure}

Starting from the $\Z_4$ SymTFT defined on $\mathcal{M}_3 := M_2 \times [0,1]$, we can gauge 1-form symmetry generated by, $\exp\left[i \pi \oint \left( a_1 + \hat{b}_1 \right) \right]$ in the upper right half of $\mathcal{M}_3$. By gauging the $1$-form symmetry generated by $\exp\left(i\pi \oint \hat{b}_1 \right)$ in the lower right half of $\mathcal{M}_3$, and then fusing the interfaces $\mathcal{I}_{e^2m^2}$ and $\mathcal{I}_{m^2}$ we end up with Fig.~\ref{fig:club_sandwich_7}, 
The interface $\mathcal{I}$, represented by the blue dashed line is the product of two topological interfaces (in the ambient Drinfeld center). 
\begin{equation}
    \mathcal{I} := \mathcal{I}_{e^2m^2} \otimes  \mathcal{I}_{m^2} .
\end{equation}
There is a boundary term ${i\pi} \int a \cup \hat{a} $ term on $\mathcal{I}_{e^2m^2}$, but we suspect upon fusion with $\mathcal{I}_{m^2}$, the $2d$ boundary term will not survive. Individually (before fusion), the interfaces had two separate boundary conditions $(a_1 + \hat{b}_1) = 0$ and $\hat{b}_1=0$. As they collide, we should naively have $a_1 =0$ and $\hat{b}_1=0$, which would trivialize the $a\cup a$ coupling. Moreover, the topological theories on either side of the interface are not related via discrete gauging. All of these, point towards the fact that the interface $\mathcal{I}$ is a totally reflective interface.

If we now collapse the UV side of the club sandwich, we end up recovering a situation where we have the SymTFTs of $\Z_2$ and $\Z^\omega_2$ separated by a topological interface. Adding the relevant physical boundaries to these will produce two IR phases separated by a conformal interface. From the absolute theory perspective, we have turned on two different relevant deformations in half spaces. As a result, we end up with a situation where we have a $\Z_2$ symmetric theory and a $\Z^\omega_2$ theory separated by an interface, which is the product of two RG interfaces, hence a conformal interface. 

From the discussion in the previous section, we have observed that there is a notion of $(-1)$-form symmetry that can connect $\text{Vec}(\Z_2)$ and $\text{Vec}(\Z^\omega_2)$, that topological operation will boil down as a $(-2)$-form symmetry of the absolute theories. Now in this setting, we do observe a  physical significance of $(-2)$-form symmetry. It is connecting two different IR phases, that can not be connected via a symmetry preserving path. 

\subsubsection*{$D_4 \rightarrow (\mathbb{Z}_2 \times \mathbb{Z}_2 / \mathbb{Z}_2 \times^\omega \mathbb{Z}_2)$ example}
\begin{figure}[H]
    \centering
    \begin{tikzpicture}[scale=1.2]
    
    % Original diagram with two rectangles and interface
    % Draw lines and fill colors
    \fill[blue!15] (0,0) rectangle (2,3); % Blue region
    \fill[green!15] (2,0) rectangle (4,1.5); % Green region
    \fill[red!15] (2,1.5) rectangle (4,3);
    
    % Draw interface (thick black line)
    \draw[thick, gray] (2,0) -- (2,1.5);
    \draw[thick, black] (2,1.5) -- (2,3);

    % Horizontal line in right (green) region
    \draw[dashed, blue] (2,1.5) -- (4,1.5);
    
    % Draw boundary lines
    \draw[thick] (0,0) -- (0,3);
%    \draw[thick, gray] (4,0) -- (4,1.5);
    \draw[thick, black] (4,0) -- (4,3);

    \fill[black] (2,1.5) circle (2pt);
    \fill[orange] (4,1.5) circle (2pt);
    
    % Labels
    \node at (-0.5,-0.3) {{$\mathfrak{B}_\text{sym}$}}; % Label for left line
    \node at (4.5,-0.3) {{$\mathfrak{B}_\text{phys}$}}; % Label for right line
    \node at (3.0,1.73) {{$\mathcal{I}$}}; % Label for interface
    \node at (1.0, 1.5) {{${\mathcal{Z}}(\text{Vec}_{D_4})$}};
    \node at (3.0, 2.5) {{${\mathcal{Z}}(\text{Vec}_{\Z_4 })$}};
    \node at (3.0, 0.7) {{${\mathcal{Z}}(\text{Vec}_{\Z_2\times \Z_2})$}};
    \end{tikzpicture}
    \caption{ By repeating the quarter gauging procedure, we have obtained the $\text{Vec}(\Z_2 \times \Z_2)$ and $\text{Vec}(\Z_4)$ theory separated by an interface.}
   \label{fig:club_sandwich_8}
\end{figure}
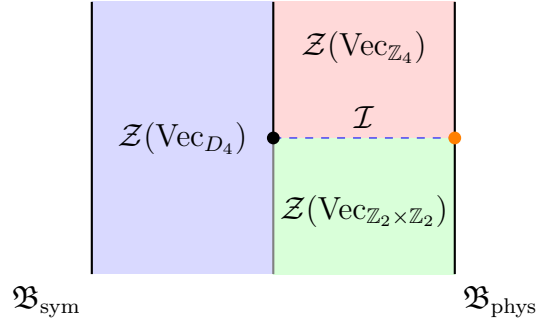

As our second example, we would like to consider the $D_4$ symmetry. The SymTFT action for this symmetry has recently been discussed in \cite{He:2016xpi, Yu:2023nyn, Franco:2024mxa,Kaidi:2023maf,deWildPropitius:1995cf, Robbins:2025puq,Bergman:2026lnz,Xue:2026hji}. 
\begin{equation}
    S_{\text{SymTFT}} = {\pi} \int_{\mathcal{M}_3} \left( \hat{a} \cup \delta a + \hat{b} \cup \delta b + \hat{c} \cup \delta c + a \cup b \cup c \right) .
\end{equation}
By condensing different condensable algebras of Drinfeld center in two different quarters of the SymTFT, one can obtain a scenario depicted in Fig.~\ref{fig:club_sandwich_8},
where we have assumed that the chosen symmetry Lagrangian algebra is given by, 
\begin{equation}
    \mathcal{L}_{\text{Sym}} = 1 \oplus e_G \oplus e_R \oplus e_{RG} \oplus 2m_B .
\end{equation}
Then the relevant condensable algebras in this case are given by the following, 
\begin{equation}
    \mathcal{A}_1 = 1 \oplus e_R, \qquad \mathcal{A}_2 = 1 \oplus e_{GB} .
\end{equation}
The condensable algebras of the Drinfeld center for $D_4$ can be found in \cite{Bhardwaj:2024qrf}. Using the dictionary between $\mathcal{Z}(D_4)$ notation and the RGB notation of the anyons presented in \cite{Robbins:2025urk}, we can equivalently write down, $\mathcal{A}_1 = 1 \oplus U_b $ and $\mathcal{A}_2 = 1 \oplus U_{ac}$, where $U_b = \exp\left(i\pi\oint b \right) $. Hence, condensing $\mathcal{A}_1$ is equivalent to imposing $b =0$ in that region and condensing $\mathcal{A}_2$ is equivalent to imposing $(a+c) =0$. Imposing these in the SymTFT action leads us to Figure~\ref{fig:club_sandwich_7}, as expected.

We can reduce the UV side of the club sandwich, in doing so we recover a situation where we have the SymTFTs of $\text{Vec}_{\Z_2 \times \Z_2}$ and $\text{Vec}_{\Z_4}$ separated by an interface. The symmetry boundary condition on the $\text{Vec}_{\Z_4}$, dictated by the fusion of $\mathfrak{B}_{\text{sym}} \otimes \mathcal{I}_{\mathcal{A}_2}$, gives us the boundary condition that realizes a $\Z_2 \times \Z_2$ symmetry with a mixed anomaly in the IR theory \cite{Robbins:2025urk}.  

Unlike in the $\mathbb{Z}_4\rightarrow\mathbb{Z}_2/\mathbb{Z}_2^\omega$ example above, the interface constructed here is not totally reflective, but only partially reflective.  This is because there is a common $\mathbb{Z}_2$ (generated by $U_a$ or equivalently $U_c$) that survives both condensations.  A $U_a$ line operator can exist in any of the three bulk regions and can be pulled through the dashed blue interface $\mathcal{I}$ of Figure~\ref{fig:club_sandwich_8}.

This could be framed as another example of $(-2)$-form symmetry, where we are relating a non-anomalous phase with a $\Z_2 \times \Z_2$ symmetry with an anomalous phase of the same symmetry. One might argue that this process is like `\textit{going up the RG flow}' due to the fact the UV symmetry group cannot be uniquely predicted. Same IR phases can be obtained obtained from different UV theories, for example, if we have a $\Z_2$ symmetric phase, that can be obtained from a UV theory with a $\Z_4$ symmetry or a $D_4$ symmetry.

\subsection{Alternate construction via nested discrete gauging}

In the last few sections, we started with a three dimensional TFT with topological point operator. We engineered a non-trivial $(-1)$-form symmetry of SymTFT by undoing decomposition within SymTFT. In this section, we propose another construction of $(-1)$-form symmetry for a $3d$ topological field theories without using the ideas of decomposition. We will primarily work with $3d$ TFTs. The construction we propose is by no means a general construction.  

The $(-1)$-form symmetry we have been discussing relates the $\Z_2$ Dijkgraaf-Witten (DW) theory without any twist to the $\Z_2$ DW theory with a twist term. A similar feat can be achieved by a two-step process. The two-step process involves gauging global symmetries (both invertible and non-invertible) of the $3d$ theory. As we emphasized earlier, instead of being completely general, we would like to work with examples. 

To begin with, once again we consider the $\Z_2$ DW theory. This theory admits six condensation surfaces (including the trivial one) constructed by 1-gauging 1-form symmetries of the theory \cite{Roumpedakis:2022aik}. The non-trivial ones include the invertible electro-magnetic duality defect, while the rest have non-invertible fusion. We can construct the $\Z_4$ DW theory by gauging one of these non-invertible condensation surfaces over the three dimensional space-time \cite{KNBalasubramanian:2025vum}. 
\begin{equation}\label{eqn:-2fsZ4}
    \mathbb{X}_1: ~ \text{Vec}(\Z_2) \longrightarrow \text{Vec}(\Z_4).
\end{equation}
where we have labeled the operation of gauging the non-invertible condensation surface as $\mathbb{X}_1$. At this point, once we have the $\Z_4$ theory, we can gauge a non-anomalous 1-form symmetry of this theory to obtain the $\Z_2$ theory with a twist.  
\begin{equation}
    \mathbb{X}_2: ~ \text{Vec}(\Z_4) \longrightarrow \text{Vec}(\Z_2, \omega).
\end{equation}
Now if we nest these two operations, 
\begin{equation}\label{eq:x1x2}
     \mathbb{X}_2 \cdot \mathbb{X}_1: ~ \text{Vec}(\Z_2) \longrightarrow \text{Vec}(\Z_2, \omega).
\end{equation}
This nested discrete gauging operation indeed replicates the action of adding a twist term to $\Z_2$ DW theory. This purely topological manipulation replicates the action of $(-1)$-form symmetry of the SymTFT we discussed in the last section. We remark the notation in (\ref{eq:x1x2}) indicates topological manipulations, not strict tensor functors.

In order to be more explicit, we will work with the $\Z_2 \times \Z_2$ DW theory. The discrete gauging procedures in this example involve only invertible symmetries. To be more specific, the $\Z_2 \times \Z_2$ theory carries a $\Z^{(0)}_2$ swap symmetry, which can be constructed via higher gauging. Starting from the $\Z_2 \times \Z_2$ theory with the following action, 
\begin{equation}
    S = \pi \int_{\mathcal{M}} \left( a \cup \delta\tilde{a} +  b \cup \delta\tilde{b} \right).
\end{equation}
We 1-gauge the subgroup of 1-form symmetry on $\Sigma$ to construct the swap symmetry\cite{Cordova:2024mqg,Barkeshli:2014cna,Tantivasadakarn:2022ceu},
\begin{equation}
    \mathcal{D}(\Sigma) \sim \sum_{\gamma, \delta \in H_1(\Sigma, \Z_2)}  (-1)^{<\gamma, \delta>} \eta_a(\gamma) \eta_b(\delta)
\end{equation}
where we have denoted the $a$-lines ($b$-lines) as $\eta_a(\gamma)$ ($\eta_b(\delta)$) and $(-1)^{<\gamma, \delta>}$ denotes discrete torsion. We can use Poincare duality to rewrite the expression above, 
\begin{equation}
    \mathcal{D}(\Sigma) \sim \sum_{A, B \in H^1(\Sigma, \Z_2)} \exp\left( i\pi \int_\Sigma a\cup A + b\cup B + A\cup B \right),
\end{equation}
where $A$ and $B$ are the Poincare duals of $\gamma$ and $\delta$. Moreover, the last term encodes the discrete torsion. Integrating out $B$ sets $A = b$. 
\begin{equation}
    \mathcal{D}(\Sigma) \sim \exp\left( i\pi\int_\Sigma a\cup b \right) .
\end{equation}
Now we gauge this invertible 0-form symmetry over $\mathcal{M}$. In order to that, we add the following term to the action,  
\begin{equation}
    S' \subset \pi\int_\mathcal{M} a\cup b \cup c ,
\end{equation}
where $c$ is the background gauge field of this 0-form symmetry. The total action becomes, 
\begin{equation}
    S + S' = \pi \int_\mathcal{M} \left( a \cup \delta\tilde{a} +  b \cup \delta\tilde{b} + a \cup b \cup c \right).
\end{equation}
Next we add the term $\pi \int_{\mathcal{M}} c\cup \delta \tilde{c}$ to our action, where $\tilde{c}$ is like a Lagrange multiplier that will enforce the flatness condition on $c$. The total action takes the following form, 
\begin{equation}
    \bar{S} = \pi \int_{\mathcal{M}} \left( a \cup \delta\tilde{a} +  b \cup \delta\tilde{b} + c \cup \delta \tilde{c} + a \cup b \cup c \right),
\end{equation}
which is the action of a $D_4$ gauge theory. Hence, we can identify the operation of gauging the swap symmetry as $\mathbb{X}_1$ in this case:
\begin{equation}
    \mathbb{X}_1: ~ \text{Vec}(\Z_2 \times \Z_2) \longrightarrow \text{Vec}(D_4).
\end{equation}
Once we have the $D_4$ theory, we can gauge the invertible 1-form symmetry of this theory that will lead us to the $\Z_2 \times \Z_2$ theory with mixed anomaly. The 1-form symmetry that we need to gauge is the following:
\begin{equation}
    \exp\left( i\pi \oint b + c \right). 
\end{equation}
To gauge this symmetry, we add the following term to our action, $\pi \int z_2 \cup (b+c)$, where $z_2$ is the background gauge field of the 1-form symmetry. Treating $z_2$ as Lagrange multiplier, we can set $(b+c) = 0$, 
\begin{equation}
    \bar{S} = \pi \int_{\mathcal{M}} \left( a \cup \delta \tilde{a} + b \cup \delta \tilde{b} - b \cup \delta \tilde{c} - a \cup b^2 \right),
\end{equation}
and then doing a field redefinition, $(\tilde{b} - \tilde{c}) = \hat{b}$, we get
\begin{equation}
\begin{aligned}
    \bar{S} &= \pi \int_{\mathcal{M}} \left(  a \cup \delta \tilde{a} + b \cup \delta \hat{b} - a \cup b^2 \right),
    \\
    &= \pi \int_\mathcal{M} \left( a \cup \delta \tilde{a} + b \cup \delta \hat{b} - \frac{1}{2} a \cup d\hat{b} \right).
\end{aligned}    
\end{equation}
which is the action of $\Z_2 \times \Z_2$ theory with a mixed anomaly. Hence, we can identify this discrete gauging with $\mathbb{X}_2$,
\begin{equation}
    \mathbb{X}_2:~ \text{Vec}(D_4) \longrightarrow \text{Vec}(\Z_2 \times \Z_2, \omega).
\end{equation}
Next, we stack the two operations, 
\begin{equation}
    \mathbb{X}_2 \cdot \mathbb{X}_1 : ~ \text{Vec}(\Z_2 \times \Z_2)  \longrightarrow \text{Vec}(\Z_2 \times \Z_2, \omega) .
\end{equation}
We have provided another example where we can introduce an anomalous twist term to our SymTFT action. We would like to claim that this nested discrete gauging operation is a manifestation of a $(-1)$-form symmetry of the SymTFT itself.

This alternate construction solidifies our statement in the previous section about the ambiguity in identifying the UV embedding of the phases we have been relating via a $(-2)$-form symmetry. As we can see, in \eqref{eqn:-2fsZ4}, the topological operation, relating $\text{Vec}(\Z_2)$ and $\text{Vec}(\Z_2, \omega)$ was obtained by going through $\text{Vec}(\Z_4)$. One could instead have embedded these inside a $\text{Vec}(D_4)$ symmetry as well, giving a different UV origin.

\section{A holographic SymTFT for
$(-2)$-form symmetry}
\label{sec:romans-minus-two-topdown}

Recall that for an ordinary $\mathbb{Z}_k$ or $U(1)$ $p$-form symmetry, the SymTFT is simply a one-dimensional higher BF-theory, with a $(p+1)$-form gauge field and its canonical dual. From this perspective, the constructions discussed in Section~\ref{sec:-2form-general}, though realizing $(-2)$-form symmetries for associated absolute QFTs, are not really  ``SymTFTs for $(-2)$-form symmetries" conventionally. In this section, we describe a logically different realization, which can be regarded as a genuine SymTFT for $(-2)$-form symmetries. 

The basic idea, first suggested in the discussion of \cite{Yu:2024jtk}, is that the Romans mass of massive type IIA, naturally read as the field strength of a $(-1)$-form gauge potential of the bulk gravity theory, descends to a $(-2)$-form symmetry of the dual three-dimensional Chern--Simons-matter theory. More generally, the holographic dictionary in which a $(-1)$-form symmetry of the bulk gravity theory descends to a $(-2)$-form symmetry of the boundary CFT was proposed on general AdS/CFT grounds in \cite[Section~5.1]{Heckman:2024oot}. The construction below makes the Romans-mass proposal concrete: we derive the explicit four-dimensional SymTFT for this $(-2)$-form symmetry from a string-theoretic reduction on the internal $S^6$, study its two natural polarizations, identify the dual quantum $3$-form symmetry and its brane origin, and show that shifts of the Romans-mass background induce shifts of the 't~Hooft anomaly coefficient of the boundary one-form symmetry.

\subsubsection*{Romans mass as a $(-2)$-form symmetry background in AdS$_4$/CFT$_3$}
Intuitively, the background field for $(-2)$-form symmetry, should be a ``$(-1)$-form gauge field," with a 0-form field-strength. One celebrated example of this type is the Romans mass in the
massive type IIA string theory \cite{Romans:1985tz}. It is a quantized RR
zero-form field strength, sourced by D8-branes.  Formally one may imagine
\begin{equation}
        F_0 = d C_{-1},
        \label{eq:F0formal}
\end{equation}
where \(C_{-1}\) is (formally!) a ``\((-1)\)-form gauge potential.''  Of course
\(C_{-1}\) is only a formal notation: the actual ten-dimensional datum is
the quantized zero-form flux \(F_0\).  Nevertheless, from the viewpoint
of a three-dimensional holographic QFT, this is precisely the correct
degree for the background field strength of a \((-2)\)-form symmetry.
Thus the Romans mass is not merely an emergent \((-1)\)-form symmetry of
an abstract SymTFT; it is the ten-dimensional string-theoretic
background field strength for a negative-degree symmetry.

A particularly clean example is given by the massive IIA
\(AdS_4\times S^6\) backgrounds\footnote{This background is obtained from the dyonic
\(\mathrm{ISO}(7)\) gauging of supergravity \cite{Guarino:2015jca}.}.  In this branch, the proposed boundary
duals are three-dimensional super-Chern--Simons-matter theories with
simple gauge group \(SU(N)\), adjoint matter, and Chern--Simons level
identified with the quantized Romans mass \cite{Guarino:2015jca}.  In
the conventions of \cite{Guarino:2015jca}, the mass parameter $m$ deforming IIA supergravity is given by 
\begin{equation}
        m = F_0
        =
        \frac{\nu}{2\pi \ell_s},
\end{equation}
where the dimensionless integer
\begin{equation}
        \nu \equiv 2\pi \ell_s F_0 \in {\mathbb Z}
        \label{eq:RomansLevelS6}
\end{equation}
will descend to the Chern-Simons level holographically.

At the \(\mathcal N=2\) point, one may take the candidate dual to be the
IR fixed point of the \(SU(N)_k\) \(\mathcal N=2\)
Chern--Simons-matter theory with three adjoint chiral multiplets
\(X,Y,Z\) and superpotential
\begin{equation}
        W = \operatorname{Tr} X[Y,Z].
\end{equation}
The precise matter content will not play an essential role below.  What was already in the literature~\cite{Gaiotto:2009mv} is the dictionary that changing
the Romans mass changes the Chern--Simons level of the boundary theory.
This \(S^6\) example is especially useful because the internal space has
no independent two- or four-cycles, so the Romans-mass/CS-level relation
is not obscured by the Page-charge mixing familiar from the
\(AdS_4\times \mathbb{CP}^3\) ABJM/GT branch.

Let \(\mathcal M_4\) denote the four-dimensional \(AdS_4\)
bulk.  The magnetic dual of \(F_0\) in ten dimensions is the ten-form RR
flux \(F_{10} \sim \star F_0 \).  The kinetic term for $F_0$ schematically reads
\begin{equation}
	F_0 \wedge F_{10}.
\end{equation}
Following the discussion in \cite{Yu:2024jtk}, reducing the above generalized Maxwell-type action on internal $S^6$ at asymptotic boundary of \(AdS_4\) amounts to taking the topological limit for the resulting 4D Maxwell-type action, resulting in the following BF-term 
\begin{equation}
        S_{\mathrm{Rom}}^{\mathrm{top}}
        =
        \int_{\mathcal M_4}
        F_0 \wedge f_4 ,
      \label{eq:RomansTopForm}
\end{equation}
where $f_4$ is a $\mathbb{R}$-valued gauge field coming from the half-magnetic formulation of the generalized Maxwell theory \cite{Yu:2024jtk}, which can be understood as inherited from the reduction of $F_{10}$ on $S^6$, meaning $F_{10} = f_4 \wedge \omega_6$ for $\omega_6$ a volume form on $S^6$. 

To be more explicit, the quantum theory of 10D kinetic term of $F_0$ reduced on $S^6$ can be schematically as 
\begin{equation}\label{eq: 4d maxwell for f0}
    \sum_{F_0 \in \mathbb{Z}} \exp \left[ \frac{i}{2e^2}\int_{\mathcal{M}_4}F_0\wedge \star_4 F_0 \right]
\end{equation}
where we have substituted the conventional path integral measure $\int \mathcal{D}A_{p+1}$ for a $(p+1)$-form $U(1)$ gauge field into a discrete sum over $F_0$, illustrating it is a field-strength of a ``$(-1)$-form gauge field''. The half-magnectic formulation \cite{Witten:1995gf,Yu:2024jtk} for this theory reads
\begin{equation}
    \sum_{F_0 \in \mathbb{Z}} \int \mathcal{D}f_4 \exp \left[ \int_{\mathcal{M}_4}\frac{ie^2}{2}f_4\wedge \star_4 f_4+ F_0\wedge f_4 \right],
\end{equation}
which is straightforward to check that integrating out $f_4$ reproduces \eqref{eq: 4d maxwell for f0}. The topological limit of this theory is due to the infinite volume of the $S^6$ at the asymptotic boundary, i.e. conformal boundary of the AdS$_4$:
\begin{equation}
    e^2 \sim \frac{1}{\mathrm{Vol}(S^6)}\to 0.
\end{equation}
The resulting theory thus reduces to the action \eqref{eq:RomansTopForm}, with $F_0$ an integer and $f_4$ $\mathbb{R}$-valued field. 

Imagining $F_0$ as the field-strength of a $(-1)$-form gauge connection, this BF theory is exactly the $U(1)$ SymTFT \cite{Antinucci:2024zjp,Brennan:2024fgj}, but for a $(-2)$-form symmetry. Equivalently, after using the dimensionless variable
\(\nu=2\pi\ell_s F_0\), we may write the same topological sector
schematically as
\begin{equation}
        S_{\mathrm{Rom}}^{\mathrm{top}}
        =
    \frac{1}{2\pi}     \int_{\mathcal M_4}
        \nu\, \wedge  \widehat{f}_4,
        \qquad
        \nu\in\mathbb Z,
        \label{eq:RomansTopInteger}
\end{equation}
where  \( \widehat{f}_4 \) is normalized so that the spacetime-filling operator $\exp( i\int_{\mathcal{M}_4} \widehat f_4)$ comes from a unit flux 9-brane with the Wess-Zumino coupling 
\(\exp( i\int_{\mathcal{M}_4\times S^6} F_{10})\).

A fixed boundary CFT is obtained by imposing a Dirichlet condition for
this negative-degree background field strength,
\begin{equation}
        \nu\big|_{\partial \mathcal M_4}=k.
        \label{eq:RomansDirichlet}
\end{equation}
The boundary theory selected by this condition is
\begin{equation}
        \mathcal T_{N,k}
        =
        SU(N)_k \ \text{Chern--Simons-matter theory}.
\end{equation}
However, unlike the $(-1)$-form symmetry case where SymTFT provides a 3D topological operator which shifts this parameter for the 3D QFT, in this case there is only a 4D operator $\exp(i\int_{\mathcal{M}_4} \widehat f_4)$, which is spacetime filling for the SymTFT itself. The Romans mass is thus a holographic background
for a \((-2)\)-form symmetry of the boundary 3D theory, with the 4D symmetry operator $\exp(i\int_{\mathcal{M}_4} \widehat f_4)$.

\subsubsection*{A quantum 3-form symmetry}
There is another natural boundary condition for the same topological
sector.  Instead of fixing the electric variable \(\nu\), one may impose a
Dirichlet boundary condition on the magnetic $\mathbb{R}$-valued gauge field \(\widehat{f}_4\).  Since
\(\widehat{f}_4\) is a top-form in the 4D bulk, its boundary value can not be understood as usual, i.e. fixing a 4-form profile on the 3D boundary, but instead
should be understood in the inflow sense: it specifies the periods
of \(\widehat{f}_4\) on four-chains in \(\mathcal M_4\) whose boundaries lie on the
asymptotic boundary \(M_3=\partial\mathcal M_4\).  We denote this
background by
\begin{equation}
        \widehat f_4 \;=\; B_4 ,
        \label{eq:f4Dirichlet}
\end{equation}
where \(B_4\) is the background field for the quantum 3-form symmetry
obtained by gauging the \((-2)\)-form symmetry.  In this magnetic
boundary condition, the Romans-mass variable is no longer fixed at the
boundary.  Instead, one sums over the allowed flux 9-brane sectors,
\begin{equation}
        Z[B_4]
        =
        \sum_{\nu\in\Gamma}
        \exp\!\left(
            i \int_{\mathcal M_4}
            \nu\, B_4
        \right)
        Z_{\mathcal T_{N,\nu}},
        \qquad
        \Gamma\subset \mathbb Z ,
        \label{eq:sumOverRomansLevels}
\end{equation}
where \(\Gamma\) denotes the subset of Romans-mass fluxes included in the
gauging.  Therefore, the magnetic boundary condition does not produce decomposed 3D QFTs with multiple universes as in the $(D-1)$-form symmetry case. Instead, it produces a sum over a 4D-3D coupled system, namely 3D CFTs with
different Chern--Simons levels, each of which is coupled to an invertible 4D TFT. The partition function in this form indeed looks subtle, which we plan to investigate further in the future.

Recalling that $\widehat{f}_4$ couples to the flux 9-brane, we can thus deduce a direct
ten-dimensional origin for the charged objects of this quantum 3-form symmetry.  A flux 9-brane wrapping \(S^6\) and a four-chain
\(\Delta_4\subset\mathcal M_4\) gives the 
insertion
\begin{equation}
        \mathcal V_\alpha(\Delta_4)
        =
        \exp\!\left(
            i\,\alpha
            \int_{\Delta_4} \widehat f_4
        \right)
        \qquad
        \alpha\in\mathbb R/\mathbb Z .
        \label{eq:fluxNineBraneBulk}
\end{equation}
If the four-chain ends on the asymptotic boundary,
\begin{equation}
        \partial\Delta_4=\Sigma_3\subset M_3 ,
        \label{eq:fluxbraneBoundary}
\end{equation}
then the same flux 9-brane reduces to a three-dimensional object
supported on \(\Sigma_3\).  This is precisely the charged object of the
dual 3-form symmetry.  In particular, when \(\Sigma_3=M_3\), the
object is spacetime-filling from the viewpoint of the boundary QFT, as
expected for a 3-form symmetry in three dimensions.

This should be contrasted with the ordinary D8-brane.  A D8-brane is the
electric source for the Romans mass and can produce jumps of \(F_0\) in
ten dimensions.  However, the object naturally associated with the
Dirichlet boundary condition for \(f_4\), and hence with the quantum
3-form symmetry after gauging the \((-2)\)-form symmetry, is the flux
9-brane described above.  In other words, the brane ending on the
asymptotic boundary in the magnetic polarization is not the D8 wall but
the fluxbrane whose Wess--Zumino coupling is built from \(F_{10}\).

\subsubsection*{$(-2)$-form symmetry background shifting 't Hooft anomaly}
As discussed in previous sections, one possible realization of $(-2)$-form symmetry for a $D$-dimensional QFT is to consider its $(D+1)$-dimensional topological operator identified with the $(D+1)$-dimensional invertible TFT capturing its 't Hooft anomaly. We now show that for this Chern--Simons-matter theory, this interpretation also works. 

We return to the electric polarization, in which the boundary value
of \(\nu\) is fixed:
\begin{equation}
        \nu\big|_{M_3}=k .
\end{equation}
In this polarization the boundary theory is a single member
\(\mathcal T_{N,k}\) of the family.  The important point is that \(k\) is
not merely a label of the Lagrangian.  It also appears as an anomaly
coefficient in the SymTFT of the boundary theory.  Since the matter in
the \(SU(N)_k\) theory is in the adjoint representation, the theory has a
\(\mathbb Z_N^{(1)}\) center one-form symmetry.  Coupling this symmetry
to a background two-form field \(B_2\), the corresponding four-dimensional
inflow term is schematically \cite{Hsin:2018vcg},
\begin{equation}
        S_{\mathrm{center}}[B_2;k]
        =
        \, k\,
        \mathfrak A_N[B_2],
        \qquad
        \mathfrak A_N[B_2]
        =
        \frac{N-1}{2N}
        \int_{\mathcal M_4}
        \mathcal P(B_2),
        \label{eq:centerAnomalyRomans}
\end{equation}
up to the standard spin-structure and normalization refinements.  Thus
the Chern--Simons level \(k\) is the coefficient of the 't Hooft anomaly
for the one-form symmetry, valued modulo the appropriate periodicity.

Using the holographic dictionary where \(k \) is the boundary condition value for \( \nu=2\pi\ell_s F_0\), this anomaly
term should be combined with the Romans-mass SymTFT as
\begin{equation}
        S_{\mathrm{top}}
        =
       \frac{1}{2\pi}
        \int_{\mathcal M_4}
        \nu
        \left(
            \widehat f_4
            +
            \frac{N-1}{2N}\mathcal P(B_2)
        \right)
        +\cdots ,
        \label{eq:RomansHigherGroupForm}
\end{equation}
where the dots denote other topological couplings not relevant for the
present discussion.

This formula makes the \((-2)\)-form interpretation manifest.  The
Romans mass \(\nu\) is the field strength of the formal
\((-1)\)-form background \(C_{-1}\), hence it is the background field
strength for a \((-2)\)-form symmetry of the three-dimensional QFT.
But the same \(\nu\) also multiplies the anomaly inflow term for the
ordinary one-form symmetry.  Therefore changing the \((-2)\)-form
background shifts the 't Hooft anomaly data of the boundary theory:
\begin{equation}
        \nu\to \nu+\ell
        \quad\Longrightarrow\quad
        S_{\mathrm{inflow}}[B_2;\nu]
        \to
        S_{\mathrm{inflow}}[B_2;\nu]
        +
        \,\ell\,\mathfrak A_N[B_2].
        \label{eq:minusTwoShiftsAnomaly}
\end{equation}
This is precisely the phenomenon encountered in the
examples in Section~\ref{sec:-2form-general}: a
\((-2)\)-form symmetry is detected by the fact that its background does
not only label different theories, but changes the SymTFT action itself,
or equivalently changes the anomaly coefficients of ordinary generalized
symmetries.

\section{Related examples: Coupled bulk-boundary systems}\label{sec:bulk/boundary_system}
In this section, we study the action of $(-1)$-form symmetries on bulk-boundary coupled systems. Although this topic is not part of the main thread of the paper, it is closely motivated and we believe it is a natural extension of topics discussed in the rest of the paper. Our goal here is not to give a complete treatment, but rather to record some preliminary observations that we expect to explore in future work\footnote{Coupled bulk--boundary systems with boundary (edge) modes, and the corner structures they give rise to, have been studied in \cite{ Cvetic:2024dzu, Acharya:2023bth, Heckman:2017uxe}.}.

\subsection{Imposing boundary conditions using edge mode TFTs}

To define quantum field theories on non-closed manifold we must impose Dirichlet boundary condition to ensure the boundary variation cancels in order to have a well posed variational problem. The boundary condition can be imposed by hand or by adding a boundary TFT \cite{Arbalestrier:2025jsg, Kapustin:2009av, Choi:2023xjw, Cordova:2023ent}. As discussed in \cite{Arbalestrier:2025jsg}, for $4d$ Maxwell theory (with the $\theta$-term) defined on 4-manifold $\mathcal{M}_4$ with boundary, to have well defined variational problem we must impose either Dirichlet or Neumman boundary condition.

Consider the Maxwell action (including the $\theta$-term) on a non-closed manifold $\mathcal{M}_4$,
\begin{equation}
\begin{aligned}
    S_{\text{Maxwell}} &= \int_{\mathcal{M}_4} \left( \frac{1}{4\pi^2} F \wedge * F + \frac{i\theta}{8\pi^2} F \wedge F \right), \\
     \delta S_{\text{Maxwell}} &\sim \text{bulk terms} + 2\int_{\partial \mathcal{M}_4} \delta A \wedge (*F + \theta F). 
\end{aligned}
\end{equation}
As emphasized, imposing Dirichlet boundary conditions on $A$ kills the boundary variation. The same boundary conditions can be imposed by adding an edge mode topological field theory on the boundary. The three dimensional TFT we need to add to the boundary is the following, 
\begin{equation}
    S_{\text{Edge Mode}} = \int_{\partial \mathcal{M}_4} \Phi \wedge dA ,
\end{equation}
where is $\Phi$ is a $U(1)$ valued field localized on the boundary. Taking the variation of the edge mode into account, 
\begin{equation}
    \delta  S_{\text{Edge Mode}} = \int_{\partial \mathcal{M}_4} \Big( \delta \Phi \wedge dA + \Phi \wedge d \delta A \Big) =   \int_{\partial \mathcal{M}_4} \Big( \delta \Phi \wedge dA - d\Phi \wedge \delta A \Big) . 
\end{equation}
and combining the boundary variations from the Maxwell action and the edge mode theory, 
\begin{equation}
    \delta (S_{\text{Maxwell}} + S_{\text{Edge Mode}} ) \sim \int_{\partial \mathcal{M}_4} \Big( \delta A \wedge (\Pi - d\Phi) + \delta \Phi \wedge dA \Big).
\end{equation}
The $\Phi$ equation of motion, i.e.~the coefficient of $\delta\Phi$, enforces $dA =0$ on the boundary. This is equivalent to the Dirichlet boundary condition. Using this trick and using much more general version of the boundary TFT \cite{Kapustin:2009av}, we can come up with a plethora of boundary conditions that can be imposed.  

A similar analysis also holds for two dimensional free boson theories defined on a manifold with a boundary. Consider the free boson theory on a non-closed manifold,
\begin{equation}
\begin{aligned}
      S_{\text{free boson}} &= \frac{1}{4\pi} \int_{\mathcal{M}_2} dX \wedge * dX,
      \\
      \delta S &= \frac{1}{2\pi} \int_{\mathcal{M}_2} d(\delta X) \wedge *dX = - \frac{1}{2\pi} \int_{\mathcal{M}_2} \delta X \wedge d * dX + \frac{1}{2\pi} \int_{\partial \mathcal{M}_2} \delta X \wedge *dX .
\end{aligned}
\end{equation}
The boundary variation vanishes under Dirichlet and Neumann boundary conditions. 
\begin{equation}
    \text{Dirichlet}: ~\delta X = 0 , \qquad \text{Neumann}:~ *d X =0
\end{equation}
We can try to replicate the trick of imposing the boundary conditions using a boundary TFT (in this case, a one dimensional TFT.). The relevant TFT in this scenario is given by\footnote{ We can also add a more general version of this TFT,
\begin{equation*}
    S_{\text{edge mode}} = \frac{ir}{2\pi}\int_{\partial \mathcal{M}_2} \Psi \wedge dX,
\end{equation*} 
but then one can do a field redefinition $\bar{\Psi} = r \Psi$, which will alter the quantization of the field.
}, 
\begin{equation}
    S_{\text{edge mode}} = \frac{1}{2\pi}\int_{\partial \mathcal{M}_2} \Psi \wedge dX,
\end{equation}
where $\Psi$ is a $U(1)$-valued scalar localized on the boundary. The combined boundary variation is then 
\begin{equation}
\begin{aligned}
    \delta ( S_{\text{free boson}} + S_{\text{edge mode}} )  &= \text{bulk} +   \frac{1}{2\pi} \int_{\partial \mathcal{M}_2} \delta X \wedge *dX + \frac{1}{2\pi} \int_{\partial \mathcal{M}_2} \delta \Psi \wedge dX - d\Psi \wedge \delta X ,
    \\
    &= \text{bulk} +  \frac{1}{2\pi} \int_{\partial \mathcal{M}_2} \Big( \delta X \wedge (*dX - d\Psi) +  \delta \Psi \wedge dX \Big).
\end{aligned}    
\end{equation}
The boundary equations of motion enforce
\begin{equation}
    \Psi: ~ dX =0 , \qquad X: ~ * dX = d\Psi.
\end{equation}
This is equivalent to enforcing Dirichlet boundary condition on $X$. We can impose Neumann boundary conditions by adding further coupling to our edge mode action and making them dynamical.

\subsubsection*{Gauging $\Z_2 \subset U(1)$}
We know that gauging a $\Z_2$ subgroup of the momentum $U(1)$ symmetry of the free boson theory rescales the radius of the theory. We would like to repeat this procedure in the presence of the boundary and observe how the boundary gets modified under the gauging. In order to gauge $\Z_2$, we would like to couple to the background gauge field in the half space and add a coupling $\frac{1}{2\pi} \int \phi_1 \wedge *dX$ to our action. In addition we add a $\frac{i}{\pi} \int \phi_1 \wedge d\phi_0$, where $\phi_0$ acts like a lagrange multiplier to enforce the flatness condition on $\phi_1$ and restricts the holonomy to $\Z_2$. The combined action is
\begin{equation}
    \frac{1}{4\pi} \int_{\mathcal{M}_2} dX \wedge * dX + \frac{i}{2\pi} \int_{\partial \mathcal{M}_2} \phi_1 \wedge *dX  + \frac{i}{\pi} \int_{\partial \mathcal{M}_2} \phi_1 \wedge d\phi_0.
\end{equation}
Integrating out $\phi_1$ will impose a delta function $\delta\left( *dX + 2 d\phi_0 \right)$, which in turn will give us the free boson theory written in terms of $\phi_0$. 

Hence, on the left we have $\int d\phi_0 \wedge *d\phi_0$, and on the right of the interface we have $\int dX \wedge *dX$.  These are coupled via the interaction $\int_{\Sigma} \phi_0 dX$. When we fuse this interface with our boundary TFT, it is modified as 
\begin{equation}\label{eqn:actiononedge}
    \int \left( \Psi \wedge dX  + \phi_0 \wedge dX \right).
\end{equation}
This is the action of the $\Z_2$-gauging on the bulk boundary coupled system, which rescales the radius of the free boson theory in the bulk and modifies the edge mode on the boundary.

This $\Z_2$ gauging of the free boson theory can be thought off as an action of $(-1)$-form symmetry, since this is a rescaling of a parameter of the theory by topological manipulations. Moreover, we can associate this procedure with a codimension one defect in the SymTFT for the free boson theory \cite{Robbins:2025urk}. In presence of boundary, the global symmetries of the bulk-boundary coupled system can be probed using Boundary SymTFT \cite{Copetti:2024onh,Copetti:2024dcz,Bhardwaj:2024igy}. Naturally, one can also analyze the action of $(-1)$-form symmetries on the bulk-boundary coupled system using this technology.

\subsection{Boundary SymTFT}
\begin{figure}[H]
    \centering
    \begin{tikzpicture}[scale=1.35]
    
    % Original diagram with two rectangles and interface
    % Draw lines and fill colors
   \fill[gray!15] (0,0) rectangle (4,3); % Blue region
%    \fill[blue!10] (2,0) rectangle (4,1.5); % Green region
%    \fill[red!15] (2,1.5) rectangle (4,3);
    
    % Draw interface (thick black line)
%    \draw[thick, black] (2,1.5) -- (2,3);
%    \draw[thick, black] (2,1.5) -- (4,1.5);
    
    % Draw boundary lines
    \draw[thick] (0,0) -- (0,3);
    \draw[thick] (4,0) -- (4,3);
    \draw[dashed] (0,3) -- (4,3);
    \fill[red] (0,3) circle (3pt);
    \fill[blue] (4,3) circle (3pt);
    
    % Labels
    \node at (-0.5,-0.3) {{$\mathfrak{B}^{\text{gapped}}_\text{sym}$}}; % Label for left line
    \node at (4.5,-0.3) {{$\mathfrak{B}_\text{phys}$}}; % Label for right line
    \node at (4.4, 3) {$\text{T}^1_C$};
    \node at (-0.4, 3) {$\text{T}^2_C$};
    \node at (2, 1.5) {$S^{3d}_{\text{SymTFT}}$};

    \node at (2.1,3.3) {\footnotesize{$\mathfrak{B}'_{\text{gapped}}$}}; % Label for interface

    % Equal sign
    \node at (5.5,1.5) {$\rightarrow$}; % Equal sign

    % Vertical line T
    \draw[thick, black] (7,0) -- (7,3); % Vertical line T
    \node at (7.5,3.0) {$\text{T}_{\text{Edge}}$}; % Label for line T
    \node at (7.6,0.8) {$\text{T}_{\text{FB}}(R)$};
    \fill[orange] (7,3) circle (3pt); 
    
    \end{tikzpicture}
    
    \caption{Boundary SymTFT}
    \label{fig:boundary_symtft_1}
\end{figure}
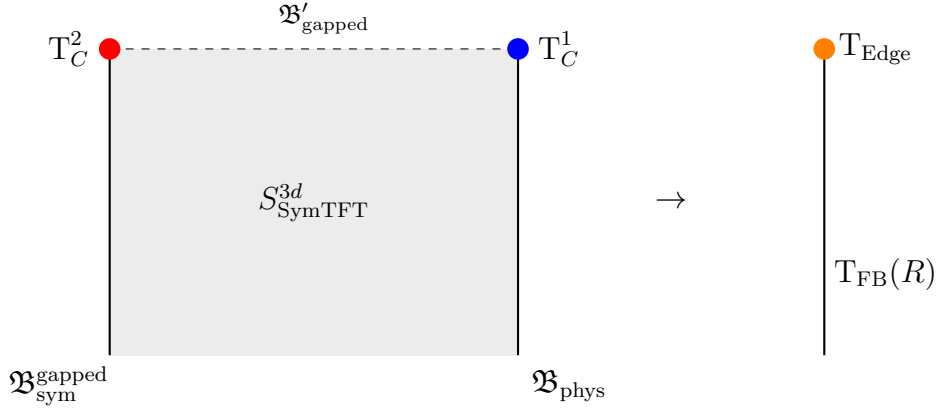

The bulk SymTFT in the figure below (Fig. \ref{fig:boundary_symtft_1}) can be written in a couple of ways using $\R$-valued gauge fields. Following \cite{Antinucci:2024zjp,Brennan:2024fgj}, 
\begin{equation}
    S^{3d}_{\text{SymTFT}} = 
    \frac{1}{2\pi} \int \Big( b_1 \wedge dB_1 + a_1 \wedge dA_1 - B_1 \wedge dA_1 \Big),
\end{equation}
where $a_1, b_1$ are $\R$ valued and $A_1,B_1$ are $U(1)$ valued gauge fields\footnote{The SymTFT for free boson can also be written using $\mathbb{R}$ gauge fields.:
\begin{equation}
    S = \frac{1}{2\pi} \int b_1 \wedge da_1.
\end{equation}
Here, both $b_1$ and $a_1$ are real-valued gauge fields.}. 
\subsubsection*{Codimension one defects of SymTFT}
In free boson theory (or Maxwell theory), we have a notion of $(-1)$-form symmetry, which acts on the theory by rescaling the radius (modifying the complexified coupling constant).
In the SymTFT description, these correspond to codimension one condensation defects constructed via higher gauging. As emphasized in section \ref{sec:codim-1-defects-symtft}, these do not act on point operators by construction but they have the ability to act on line operators. In doing so, they can modify the chosen Lagrangian algebra on the symmetry boundary. 

The defect that is relevant for our purposes was constructed in \cite{Robbins:2025urk}, by 1-gauging a $\Z$ subgroup of the $\Z\times U(1)$ 1-form symmetry in the bulk \cite{Robbins:2025urk,Kaidi:2022cpf}. 
\begin{equation}
    \mathcal{D}(\Sigma) \sim \sum_{\gamma_1 \in H_1(\Sigma,\Z)} W^1_1(\gamma_1) \otimes U^1_{1/2}(\gamma_1),
\end{equation}
where, $W^1_m = \exp \left(im \oint B_1\right)$ and $U^1_\alpha = \exp \left( i\alpha \oint b_1\right)$. In other words, this defect will only act on $W^1_1$ and modify it to a $U^1_{\frac{1}{2}}$ line. 

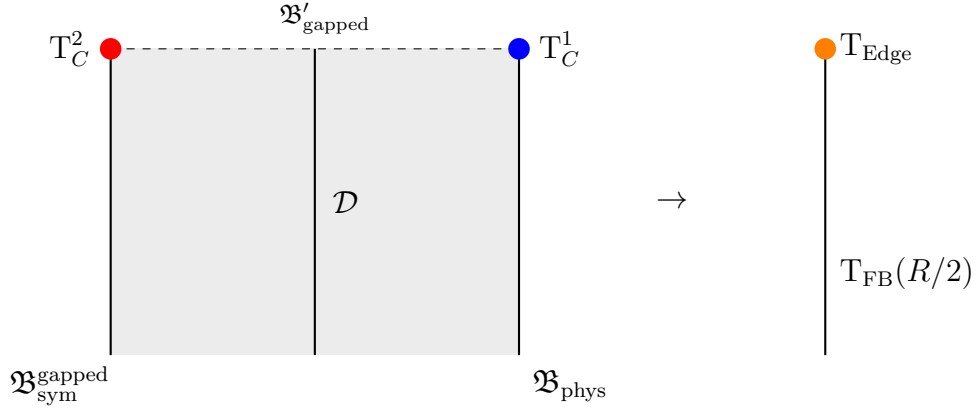
\begin{figure}[H]
    \centering
    \begin{tikzpicture}[scale=1.35]
    
    % Original diagram with two rectangles and interface
    % Draw lines and fill colors
   \fill[gray!15] (0,0) rectangle (4,3); % Blue region
%    \fill[blue!10] (2,0) rectangle (4,1.5); % Green region
%    \fill[red!15] (2,1.5) rectangle (4,3);
    
    % Draw interface (thick black line)
%    \draw[thick, black] (2,1.5) -- (2,3);
%    \draw[thick, black] (2,1.5) -- (4,1.5);
    
    % Draw boundary lines
    \draw[thick] (0,0) -- (0,3);
    \draw[thick] (4,0) -- (4,3);
    \draw[thick] (2,0) -- (2,3);
    
    \draw[dashed] (0,3) -- (4,3);
    \fill[red] (0,3) circle (3pt);
    \fill[blue] (4,3) circle (3pt);
    
    % Labels
    \node at (-0.5,-0.3) {{$\mathfrak{B}^{\text{gapped}}_\text{sym}$}}; % Label for left line
    \node at (4.5,-0.3) {{$\mathfrak{B}_\text{phys}$}}; % Label for right line
    \node at (4.4, 3) {$\text{T}^1_C$};
    \node at (-0.4, 3) {$\text{T}^2_C$};
    \node at (2.3, 1.5) {$\mathcal{D}$};

    \node at (2.1,3.3) {\footnotesize{$\mathfrak{B}'_{\text{gapped}}$}}; % Label for interface

    % Equal sign
    \node at (5.5,1.5) {$\rightarrow$}; % Equal sign

    % Vertical line T
    \draw[thick, black] (7,0) -- (7,3); % Vertical line T
    \node at (7.5,3.0) {$\text{T}_{\text{Edge}}$}; % Label for line T
    \node at (7.8,0.8) {$\text{T}_{\text{FB}}(R/2)$};
    \fill[orange] (7,3) circle (3pt); 
    
    \end{tikzpicture}
    
    \caption{Placing $\mathcal{D}$ vertically, produces a $(-1)$-form symmetry of the absolute theory (the bulk boundary coupled system). }
    \label{fig:boundary_symtft_2}
\end{figure}

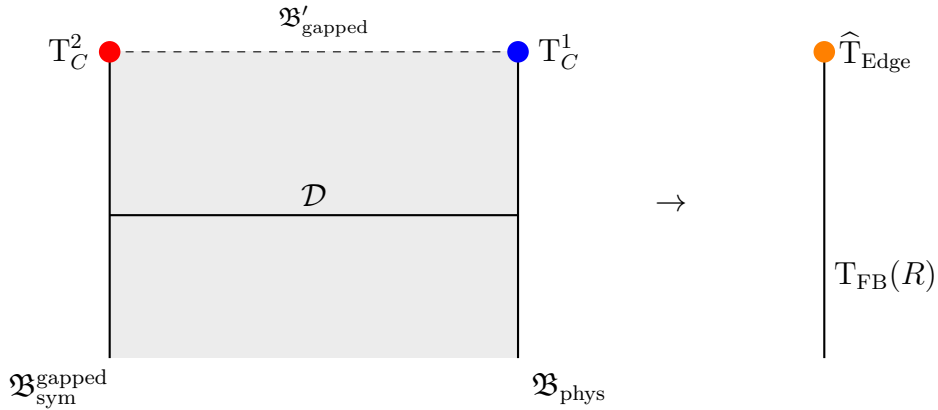
\begin{figure}[H]
    \centering
    \begin{tikzpicture}[scale=1.35]
    
    % Original diagram with two rectangles and interface
    % Draw lines and fill colors
   \fill[gray!15] (0,0) rectangle (4,3); % Blue region
%    \fill[blue!10] (2,0) rectangle (4,1.5); % Green region
%    \fill[red!15] (2,1.5) rectangle (4,3);
    
    % Draw interface (thick black line)
%    \draw[thick, black] (2,1.5) -- (2,3);
%    \draw[thick, black] (2,1.5) -- (4,1.5);
    
    % Draw boundary lines
    \draw[thick] (0,0) -- (0,3);
    \draw[thick] (4,0) -- (4,3);
    \draw[thick] (0,1.4) -- (4,1.4);
    
    \draw[dashed] (0,3) -- (4,3);
    \fill[red] (0,3) circle (3pt);
    \fill[blue] (4,3) circle (3pt);
    
    % Labels
    \node at (-0.5,-0.3) {{$\mathfrak{B}^{\text{gapped}}_\text{sym}$}}; % Label for left line
    \node at (4.5,-0.3) {{$\mathfrak{B}_\text{phys}$}}; % Label for right line
    \node at (4.4, 3) {$\text{T}^1_C$};
    \node at (-0.4, 3) {$\text{T}^2_C$};
    \node at (2.0, 1.6) {$\mathcal{D}$};

    \node at (2.1,3.3) {\footnotesize{$\mathfrak{B}'_{\text{gapped}}$}}; % Label for interface

    % Equal sign
    \node at (5.5,1.5) {$\rightarrow$}; % Equal sign

    % Vertical line T
    \draw[thick, black] (7,0) -- (7,3); % Vertical line T
    \node at (7.5,3.0) {$\widehat{\text{T}}_{\text{Edge}}$}; % Label for line T
    \node at (7.6,0.8) {$\text{T}_{\text{FB}}(R)$};
    \fill[orange] (7,3) circle (3pt); 
    
    \end{tikzpicture}
    
    \caption{Placing $\mathcal{D}$ horizontally, produces a $(-1)$-form symmetry for $\mathfrak{B}'_{\text{gapped}}$. Inducing an action only on the topological corner. Fusing, $\mathfrak{B}'_{\text{gapped}} \otimes \mathcal{D} = \mathfrak{B}''_{\text{gapped}}$.  }
    \label{fig:boundary_symtft_3}
\end{figure}

When placed vertically to $\mathfrak{B}^{\text{gapped}}_{\text{sym}}$ (as in Fig. \ref{fig:boundary_symtft_2}), the codimension one defect acts like a $(-1)$-form symmetry of the absolute theory (bulk-boundary coupled system). In this case, rescaling the radius of the free boson theory. Furthermore, it also acts on the edge mode theory and modifies it, in the same spirit as \eqref{eqn:actiononedge}.
Instead, we can place the codimension one defect horizontally in the SymTFT bulk (as in Fig. \ref{fig:boundary_symtft_3}), in this way, it will not act on the symmetry $\mathfrak{B}^{\text{gapped}}_{\text{sym}}$. However, it can still be fused with $\mathfrak{B}'_{\text{gapped}}$. 
Fusing the defect $\mathcal{D}$ with $\mathfrak{B}'_{\text{gapped}}$ will modify the lagrangian algebra on that boundary, producing a new boundary condition (or boundary theory). This action can be thought of as a $(-1)$-form symmetry for this boundary slab. Collapsing the whole SymTFT, we obtain the free boson theory (at the same radius) with a different edge mode description. To summarize, we have 0-form symmetry of the SymTFT bulk, when placed vertically generates the $(-1)$-form symmetry that acts on the bulk-boundary coupled system. Moreover, when the same defect is placed horizontally in the SymTFT bulk, does not modify any parameters of the theory, instead only acts on the boundary condition.

\section{Conclusion and Outlook}\label{sec:conclusion}

In this work, we have made an attempt to extend the paradigm of lower form symmetries \cite{Vandermeulen:2022edk} by introducing the notion of $(-2)$-form symmetries. We have analyzed the non-genuine codimension one defects of SymTFT and their implications on the physical theory through various examples. While the genuine counterparts of these defects, which we associate with $(-1)$-form symmetries, are responsible for shifting a defining parameter of the theory, we observed that these non-genuine operators, in addition to shifting the value of the parameter, are capable of relating two different theories that differ via anomaly data. We have exemplified this through various bottom-up examples in different dimensions. 

We have further motivated this notion of $(-2)$-form symmetries by leveraging the bulk-boundary correspondence of the SymTFT. In this framework, topological defects filling the SymTFT bulk, which encode $(-1)$-form symmetries of the SymTFT itself, give rise to non-trivial actions on the absolute theory. Crucially, this action is realized through the lens of non-genuine topological operators with $d$-dimensional support. This physical description highlights a universal mechanism for realizing global symmetries of negative degree by reinterpreting the base theory as a defect or interface mediating between families of higher-dimensional topological field theories, as anticipated in \cite{Heckman:2024oot}. In category-theoretic terms, this mechanism corresponds to explicitly lifting the degree of the physical theory. By delooping the target category of defects and theories to realize the physical theory as a higher morphism (specifically, a boundary condition within a pointed $(d+2)$-category of stacked higher-dimensional theories), we establish a general mathematical blueprint. Ultimately, this demonstrates how we can define global symmetries of arbitrary negative degree by incorporating the original theory into a broader, generalized descent mechanism.

Building on these insights, and motivated by the construction in \cite{Cvetic:2024dzu}, it would be interesting to take this discussion a step further: treating the $(d+1)$-dimensional SymTFT ($\text{Sym}_{d+1}$) as an absolute theory, can we write down a `\textit{SymTFT of a SymTFT}'---a $(d+2)$-dimensional Symmetry TFT ($\text{Sym}_{d+2}$) that captures the global symmetries of this $\text{Sym}_{d+1}$? If so, then following the usual logic, $0$-form symmetries of $\text{Sym}_{d+2}$ will give rise to $(-1)$-form symmetries of $\text{Sym}_{d+1}$. In other words, the data of the non-genuine operators that we are interested in studying would be captured by defects living in $(d+2)$. Moreover, this would suggest a path towards realizing $(-3)$-form symmetries. Realizing this category-theoretic picture more systematically, and formalizing this higher-dimensional defect data properly, remain compelling directions for future work.

In order to uncover the physical significance of $(-2)$-form symmetries we have proposed a quarter gauging construction within SymTFT. Using the fact that the club sandwich can be used to model renormalization group flows \cite{Bhardwaj:2023bbf,Antinucci:2025fjp,Lu:2025yru}, the quarter gauging operation we have proposed, can capture two distinct IR phases arising from different renormalization group flows of a common theory in the UV. We have presented different examples of this quarter gauging construction. This construction enable us to realize that $(-2)$-from symmetries relate two distinct IR phases that can not connected via a symmetry preserving path. We have also proposed an alternate construction of this using nested discrete gauging.  

We have observed that symmetry categories with the same Grothendieck ring but different associators are related by $(-2)$-form symmetries, for e.g. in our $\Z_2/ \Z_2^{\omega}$ example in section \ref{sec:-2form-general}. One of the immediate generalization is to find examples of $(-2)$-form symmetries that modify the data of the associator of a non-invertible symmetry. In other words, we are looking at generalizations of the short exact sequence in \eqref{eqn:highergroup_ses}, where some anomalous fusion category would be extending by the $(d-1)$-form symmetry forming a non-anomalous multi fusion catgeory. Beyond that, we further believe that $(-2)$-form symmetries might relate theories with unitary and non-unitary symmetry. Consider the Fibonacci symmetry, $\{1, X\}$ with
$X^2 = 1=X$.  This has two F-symbol solutions, the Fibonacci category and the Yang-Lee category. The former is unitary, the latter is not. We suggest that $(-2)$-form symmetries can relate unitary and non-unitary symmetries. It will be interesting to explore this further.

We have also presented a holographic, top-down realization of $(-2)$-form symmetry in Section~\ref{sec:romans-minus-two-topdown}. There the quantized Romans mass of massive type IIA string theory \cite{Romans:1985tz}, formally the field strength of a $(-1)$-form gauge potential, plays the role of a background for a $(-2)$-form symmetry of the three-dimensional boundary CFT. Concretely, in the $AdS_4 \times S^6$ background dual to $SU(N)_k$ super-Chern--Simons-matter theory \cite{Guarino:2015jca}, reducing the ten-dimensional Romans-mass kinetic term on the internal $S^6$ produces a genuine four-dimensional SymTFT for a $(-2)$-form symmetry, whose spacetime-filling symmetry operator descends from a flux $9$-brane. The two natural polarizations of this SymTFT realize the expected dual picture: the electric boundary condition fixes the Chern--Simons level $k$, while the magnetic one gauges the $(-2)$-form symmetry, producing a quantum $3$-form symmetry as a sum over Romans-mass sectors coupled to an invertible four-dimensional TFT. Most importantly, the Romans mass simultaneously controls the inflow term for the $\Z_N^{(1)}$ center one-form symmetry, so shifting the $(-2)$-form background shifts the 't~Hooft anomaly of an ordinary one-form symmetry, which is the same hallmark of $(-2)$-form symmetries that we encountered in the bottom-up examples.

As a final remark, from Hypothesis H \cite[Section 2.5]{Sati:2013rxa}, it is believed that M theory brane charges are quantized in a non-abelian cohomology theory, called unstable 4-cohomotopy \cite[Theorem 3.1]{Giotopoulos:2024xcg}. M5 brane worldvolume is relatively flux quantized in the quaternionic Hopf fibration $h_{\mathbb{H}}:S^7\rightarrow S^4$ with the super Bianchi identity given by following \cite[Proposition 3.18]{Giotopoulos:2024sit}, 
\begin{equation}
    dH_3^s=\phi^\ast G_4^s
\end{equation}
where $\phi$ is the super-embedding map from the M5 super-worldvolume to the super-spacetime $(D=11)$ . The super Bianchi identity also implies the self dual nature of the $H_3$ flux\cite[Remark 3.19]{Giotopoulos:2024sit}. However, when we `\textit{put}' a Chern-Simons $U(1)$ gauge field on the worldvolume (also known as the `\textit{magnetized M5}' brane), then we get the relative quantization in the twistorial fibration $t_{\mathbb{H}}:CP^3 \rightarrow S^4$. The on-shell field content remains the same since the super Bianchi identity,
\begin{equation}
    dH_3^s=\phi^\ast G_4^s-F_2^s\wedge F_2^s
\end{equation}
implies the Chern-Simons equation of motion $F_2=0$\cite[Section 2]{Sati:2025ucv}. 

As a consequence of this, the global picture \footnote{Flux quantization of a theory implies transition data for consistent gluing of gauge potentials on (higher) intersections of submersions via gauge transformations and higher gauge transformations, thus making the theory globally well defined. Thus once we change the flux quantization rule, the global picture of the theory changes.} of the theory changes, so it could be possible that the global anomaly structures might also change. The flux quantization conditions encode when the local Bianchi identities can be promoted to globally quantized cocycles in relative non-abelian cohomology. We speculate it might be possible to identify this as a $(-2)$-form symmetry. The globally different structures should lie in the space of QFTs as distinct entities. If this is true, then we can have these negative degree symmetry structures between various quantization conditions, which control the global picture via flux quantization.

\subsection*{Acknowledgments}

We are grateful to Jonathan Heckman and Ling Lin for reading the draft carefully. X.Y.~ thanks Jonathan Heckman, Vinicius Nevoa, Sanjay Raman, and Jake McNamara for valuable discussion. S.R.~ thanks Ibrahima Bah and Mohammad Akhond for remarks on a previous version of the draft. X.Y.~thanks the Simons Collaboration on Global Categorical Symmetries Annual Meeting
2025 for the hospitality during part of this work. S.R.~thanks the Dept of Physics at Virginia Tech for the hospitality during part of this work. E.S.~and X.Y.~were partially supported by NSF grant PHY-2310588.

\appendix

\section{Gauge invariance of SymTFT action for ABJM}\label{sec:app1}

In this appendix we would like the gauge transformations for $4d$ SymTFT action of ABJM theories. The SymTFT action takes the following form:
\begin{equation}
    S^{\text{ABJM}}_{\text{SymTFT}} = \int \left(k b_2 dc_1 + kb_0 dc_3 -\Omega b_0 b_2 b_2\right).
\end{equation}
The postulated gauge transformations are the following,
\begin{equation}
\begin{aligned}
    b_2 &\rightarrow b_2 + d\lambda_1, \\
    c_1 &\rightarrow c_1 + d\chi_0 + 2\frac{\Omega}{k} b_0 \lambda_1 \\
    c_3 &\rightarrow c_3 + d\chi_2 + \frac{2\Omega}{k} b_2 \lambda_1 + \frac{\Omega}{k} \lambda_1 d\lambda_1.
\end{aligned}    
\end{equation}
We replace these in the action,
\begin{equation}
    \begin{aligned}
        S' = \int \Big( k (b_2 + d\lambda_1) (dc_1 + 2\frac{\Omega}{k} db_0 \lambda_1 + 2\frac{\Omega}{k} b_0 d\lambda_1) + kb_0(dc_3 + \frac{2\Omega}{k} db_2 \lambda_1 + \frac{2\Omega}{k} b_2 d\lambda_1 + \frac{\Omega}{k} d\lambda_1 d\lambda_1  ) \\
        - \Omega b_0 (b_2 + d\lambda_1)(b_2 + d\lambda_1) \Big),
    \end{aligned}
\end{equation}
\begin{equation}
    \begin{aligned}
        \delta S' = \int \Big( 2\Omega  b_2 db_0 \lambda_1 + 2\Omega  b_2 b_0 d\lambda_1 +  d\lambda_1 dc_1 + 2\Omega  d\lambda_1 db_0 \lambda_1 + 2\Omega  d\lambda_1 b_0 d\lambda_1  \\
        2\Omega b_0 db_2 \lambda_1 + 2\Omega b_0 b_2 d\lambda_1 +  \Omega b_0 d\lambda_1 d\lambda_1 \\   
        -\Omega b_0 b_2 d\lambda_1 - \Omega b_0 d\lambda_1 b_2 - \Omega b_0 d\lambda_1 d\lambda_1 \Big).
    \end{aligned}
\end{equation}
The term $k d\lambda_1 dc_1$ can be written as a total derivative $d(k d\lambda_1 c_1)$, hence we ignore it.
\begin{equation}
    \begin{aligned}
        \delta S' = \int \Big( 2\Omega  b_2 db_0 \lambda_1 + 2\Omega  b_2 b_0 d\lambda_1 + 2\Omega  d\lambda_1 db_0 \lambda_1 + 2\Omega  d\lambda_1 b_0 d\lambda_1 + 2\Omega b_0 db_2 \lambda_1  \\
        +  2\Omega b_0 b_2 d\lambda_1    
        -\Omega b_0 b_2 d\lambda_1 - \Omega b_0 d\lambda_1 b_2  \Big).
    \end{aligned}
\end{equation}
Using $b_0 d\lambda_1 b_2 = b_0 b_2 d\lambda_1$, we cancel the last three terms, leaving 
\begin{equation}
    \begin{aligned}
        \delta S' &= \int \Big(2\Omega  b_2 db_0 \lambda_1 + 2\Omega  b_2 b_0 d\lambda_1 + 2\Omega  d\lambda_1 db_0 \lambda_1 + 2\Omega  d\lambda_1 b_0 d\lambda_1 + 2\Omega b_0 db_2 \lambda_1 \Big), 
        \\
        &= 2\Omega \int \Big( b_2 db_0 \lambda_1 + b_2 b_0 d\lambda_1 + b_0 db_2 \lambda_1 +  d\lambda_1 db_0 \lambda_1 +  d\lambda_1 b_0 d\lambda_1  \Big).
    \end{aligned}
\end{equation}
The first three terms can be packaged as $d(b_0 b_2 \lambda_1)$, hence
\begin{equation}
\begin{aligned}
     \delta S' &= 2\Omega \int \Big( d\lambda_1 db_0 \lambda_1 +  d\lambda_1 b_0 d\lambda_1 \Big), \\
     &= 2\Omega \int d\lambda_1 (db_0 \lambda_1 + b_0 d\lambda_1) ,\\
     &= 2\Omega \int d\lambda_1 d(b_0 \lambda_1), \\
     &= 2\Omega \int d(d\lambda_1 (b_0 \lambda_1)) .
\end{aligned}     
\end{equation}
Once again, this is a total derivative.

\section{Categorical details and definitions}  \label{app:cat}

In this appendix we will briefly outline a more categorical approach to $(-2)$-form symmetries.  The higher categories involved below technically require further conditions such as enrichments and geometric structures, but for the present purposes the following schematic picture will suffice.  We also restrict to finite symmetries.

In order to get our point across without introducing unnecessary complications, 
we will describe a defect supported at fixed codimension by a non-extended topological field theory with the same support.
As in section~\ref{sec:-2form-general}, we will start with a $d$-dimensional physical theory $\mathsf{t}$.  We then consider, at least abstractly, the $d$-category of TFT's $\mathcal{T}$ one can stack $\mathsf{t}$ with. Intuitively, an object $x$ in $\mathcal{T}$ is a codimension $1$ TFT, a $1$-morphism $f: x \to y$ is a codimension $2$ TFT acting as an interface between the TFT's $x$ and $y$, and so forth. Respectively, these describe the $0-$, $1-$, and higher-form symmetries of $\mathsf{t}$. This includes both genuine and non-genuine operators. For example, a $1$-morphism $f: 1 \to x$, where $1$ is the trivial codimension $1$ operator, is a codimension $2$ topological operator at the end (boundary) of $x$. If, furthermore, $x=1$, then $f$ is a free codimension $2$ operator and therefore a genuine $1$-form symmetry operator.

Since codimension $1$ TFT's may also be tensored, $\mathcal{T}$ is a monoidal $d$-category. As was done in the higher geometric side in \cite{Perez-Lona:2025xxm,PerezLona:2026tba}, to define a category-theoretic symmetry parameterized by, for example, some monoidal $d$-category $\mathcal{C}$, it suffices to define an action functor, meaning a monoidal functor of monoidal $d$-categories
\begin{equation}
    F: \mathcal{C} \to \mathcal{T}.
\end{equation}
An equivalent way to phrase this monoidal functor is as a functor of $(d+1)$-categories, simply by taking the delooping on both sides as
\begin{equation}
    BF: B\mathcal{C} \to B\mathcal{T}.
\end{equation}
By $B\mathcal{C}$, we mean the $(d+1)$-category with a single object $\bullet\in\mathrm{ob} (B\mathcal{C})$ such that the endomorphism $d$-category $B\mathcal{C}(\bullet, \bullet)=\mathcal{C}$ is our monoidal $d$-category. Likewise, $B\mathcal{T}$ is the $(d+1)$-category with a single object $\mathsf{t}\in \mathrm{ob}(B\mathcal{T})$ such that $B\mathcal{T}(\mathsf{t},\mathsf{t}) = \mathcal{T}$.

Now, to make room for $(-1)$-form symmetries, one can consider including $B\mathcal{T}$ into a wider $(d+1)$ category $\mathsf{T}$. One possibility is the $(d+1)$-category of stackings of $\mathsf{t}$ with some $d$-dimensional TFT $\mathbbm{t}$. Equivalently, these are theories obtained from $\mathsf{t}$ by inserting a spacetime filling ($d$-dimensional) topological defect. Notice that $\mathsf{t}$ is a distinguished object of $\mathsf{T}$ since it is the result of stacking $\mathsf{t}$ with the trivial TFT $\mathbbm{1}_{(d)}$, denoted as $\mathsf{t} \otimes \mathbbm{1}_{(d)}$. This makes $\mathsf{T}$ a pointed $(d+1)$-category, which simply means it is equipped with a functor from the terminal $(d+1)$-category $*$
\begin{eqnarray}
    \mathsf{t}&:& * \to \mathsf{T},\\
    && \bullet \mapsto \mathsf{t} \otimes \mathbbm{1}_{(d)}.
\end{eqnarray}
which has a single object $\bullet$ and only trivial morphisms.

In this way, the objects of $\mathsf{T}$ form a family of $d$-dimensional theories, each of which is obtained from $\mathsf{t}$ by stacking to a $d$-dimensional TFT, the $1$-morphisms are topological interfaces between theories, and so forth. This is a closely related to the picture presented in \cite{Stockall:2025ngz}, where more general families of theories, not necessarily related by stackings with $d$-dimensional TFT's, are considered. Stackings of TFT's were also explored in \cite{Bhardwaj:2024xcx}. Also, notice we are not requiring the stacked TFT to be invertible. This conceptual point has already been advocated for in \cite{Teixeira:2025qsg}.

Now, let us consider some $(d+1)$-category $\mathcal{D}$, which is furthermore pointed $d_0: * \to \mathcal{D}$. Physically, a pointed functor of pointed $(n+1)$-categories,
\begin{equation}
    G_* : (\mathcal{D},d_0) \to (\mathsf{T},\mathsf{t}\otimes \mathbbm{1}_{(d)}),
\end{equation}
therefore describes a system of $(-1)-$, $0-$, $1-$, and higher-form symmetries of $\mathsf{t}$. We use the condition of pointed functors to unambiguously interpret this as symmetries \textit{of} $\mathsf{t}$. In particular, if $\mathcal{D}$ is discrete, meaning it only has identity morphisms, then the functor $G$ simply specifies a family of physical theories parameterized by the objects of $\mathcal{D}$, among which $G_*(d_0)=\mathsf{t}$ is distinguished. As such, this describes a pure $(-1)$-form symmetry of $\mathsf{t}$.

To make room for $(-2)$-form symmetries, we want to repeat the same process, namely, to deloop the target higher category by reinterpreting it as the endomorphisms of some object in a category of one degree higher. The SymTFT $\mathscr{T}$ of $\mathsf{t}$ allows us to do exactly this. As described in Section~\ref{sec:codim-1-defects-symtft}, the $0$-form symmetries of the SymTFT are directly related to the $(-1)$-form symmetries of the absolute theory, and as such we may assume there is a tensoring of TFT's $\mathbbm{t}\otimes \mathbbm{t}'$ one can stack $\mathsf{t}$ with. Schematically, the physical theory $\mathsf{t}$ is equivalently a boundary condition of its SymTFT $\mathscr{T}$, that is, a (non-topological) interface between the trivial $(d+1)$-dimensional TFT $\mathbbm{1}_{(d+1)}$, and $\mathscr{T}$
\begin{equation}\label{eq:t-boundary-theory}
    \begin{tikzcd}
                    & \mathsf{t} \arrow[ld] \arrow[rd] &             \\
\mathbbm{1}_{(d+1)} &                                  & \mathscr{T}
\end{tikzcd}.
\end{equation}
If we consider a topological interface $\mathbbm{t}:\mathscr{T}\to \mathscr{T}$ with $d$-dimensional support within $\mathscr{T}$, denoted as
\begin{equation}
    \begin{tikzcd}
            & \mathbbm{t} \arrow[ld] \arrow[rd] &             \\
\mathscr{T} &                                   & \mathscr{T}
\end{tikzcd},
\end{equation}
we can compose it with $\mathsf{t}$ so as to obtain another boundary theory, namely $\mathsf{t}\otimes \mathbbm{t}$
\begin{equation}
    \begin{tikzcd}
                    &                                  & \mathsf{t}\otimes \mathbbm{t} \arrow[ld] \arrow[rd] \arrow[lldd, bend right] \arrow[rrdd, bend left] &                                   &             \\
                    & \mathsf{t} \arrow[ld] \arrow[rd] &                                                                                                      & \mathbbm{t} \arrow[ld] \arrow[rd] &             \\
\mathbbm{1}_{(d+1)} &                                  & \mathscr{T}                                                                                          &                                   & \mathscr{T}
\end{tikzcd}.
\end{equation}
This description trivially  allows us to endow the objects of $\mathsf{T}$ with a product structure, and therefore we can lift the category-theoretic degree once more. More specifically, we may now look at the $(d+2)$-category $\mathbf{T}$ of TFT's obtained by stacking the $(d+1)$-dimensional SymTFT $\mathscr{T}$ of $\mathsf{t}$ with some other $(d+1)$-dimensional TFT. Again, this category is pointed
\begin{eqnarray}
    \mathscr{T} &:& * \to \mathbf{T},
    \\
    && \bullet \mapsto \mathscr{T},
\end{eqnarray}
so that a pointed functor of pointed $(d+2)$-categories
\begin{equation}\label{eq:d2functor}
    H_*: (\mathcal{E},e) \to (\mathbf{T},\mathscr{T})
\end{equation}
is physically understood as encoding $(-1)-$, $0-$, and higher-form symmetries of $\mathscr{T}$. 

By comparing category-theoretic degrees, we would want to interpret this as describing $(-2)-$, $(-1)-$, $0-$, and higher-form symmetries of $\mathsf{t}$. But one must be more careful in this case, since not every defect encoded by a functor such as (\ref{eq:d2functor}) acts explicitly on $\mathsf{t}$ as a boundary theory. This is most clearly the case for pure $(-1)$-form symmetries of $\mathscr{T}$ which, as mentioned above, are encoded by a pointed functor from a discrete $(d+2)$-category $X$
\begin{equation}
    H'_*: (X,x_0) \to (\mathbf{T}, \mathscr{T}).
\end{equation}
This describes a family of $(d+1)$-dimensional topological field theories obtained from stacking $\mathscr{T}\otimes \mathbbm{T}$ the SymTFT $\mathscr{T}$ with some other $(d+1)$-dimensional TFT $\mathbbm{T}$. In this case, the SymTFT $\mathscr{T}$ is the distinguished object as $H'_*(x_0)= \mathscr{T}$. However, while $\mathscr{T}$ has $\mathsf{t}$ as a boundary theory, this is not necessarily the case for some other $\mathscr{T}\otimes \mathbbm{T}$ --- all we can say is that $\mathsf{t}$ is now an interface as
\begin{equation}
    \begin{tikzcd}
            & \mathsf{t} \arrow[ld] \arrow[rd] &                                 \\
\mathbbm{T} &                                  & \mathbbm{T} \otimes \mathscr{T}
\end{tikzcd},
\end{equation}
as a result of tensoring the diagram (\ref{eq:t-boundary-theory}) with $\mathbbm{T}$ and the identity defect $\mathbbm{T}\leftarrow \mathbbm{1}_{d} \to \mathbbm{T}$. This complicates obtaining an absolute $d$-dimensional theory, since so far we had been describing $\mathsf{t}$ as a boundary theory as in the quiche picture.

This issue is ameliorated if we look at the image of the connected component of $e_0 \in \mathrm{ob}(\mathcal{E})$ under the functor (\ref{eq:d2functor}). Indeed, consider a $1$-morphism $f: e_0 \to e_1$ in $\mathcal{E}$. Its image in $\mathbf{T}$ is
\begin{equation}
 H_*:   \left( e_0 \xrightarrow{f} e_1 \right) \quad \mapsto \quad \left(\begin{tikzcd}
            & \mathbbm{t} \arrow[ld] \arrow[rd] &                                 \\
\mathscr{T} &                                   & \mathscr{T} \otimes \mathbbm{T}
\end{tikzcd}\right).
\end{equation}
The theory $\mathscr{T}$ itself, as was the case for $\mathsf{t}$, is associated with the trivial $(d+1)$-dimensional TFT $\mathscr{T}\otimes \mathbbm{1}_{(d+1)}$. It follows that, physically, the image of $f:e_0 \to e_1$ describes a non-genuine $d$-dimensional defect $\mathbbm{t}$ attached to the boundary of a defect $\mathbbm{T}$ with $(d+1)$-dimensional support.

But then this may be composed with (\ref{eq:t-boundary-theory}) as
\begin{equation}
    \begin{tikzcd}
                    &                                  & \mathsf{t}\circ \mathbbm{t} \arrow[ld] \arrow[rd] \arrow[lldd, bend right] \arrow[rrdd, bend left] &                                   &                                 \\
                    & \mathsf{t} \arrow[ld] \arrow[rd] &                                                                                                    & \mathbbm{t} \arrow[ld] \arrow[rd] &                                 \\
\mathbbm{1}_{(d+1)} &                                  & \mathscr{T}                                                                                        &                                   & \mathscr{T} \otimes \mathbbm{T}
\end{tikzcd}.
\end{equation}
This is again a boundary condition for a $(d+1)$-dimensional TFT, which is notably not necessarily the SymTFT $\mathscr{T}$ we started with, and thus we may collapse complete the sandwhich to obtain an absolute $d$-dimensional physical theory. All in all, this describes the action of certain $(-1)$-form symmetries of the SymTFT $\mathscr{T}$ on $\mathsf{t}$ through the lenses of non-genuine topological operators supported as dimension $d$. This is what we refer to as $(-2)$-form symmetries. 

Specializing to two-dimensional physical theories, we can leverage the correspondence between braided fusion categories and three-dimensional SymTFT's to construct a dictionary of operators and category-theoretic structures, in the spirit of \cite[Table 6]{Bergman:2026lnz}. Briefly, this amounts to a symmetries interpretation of the image of the endofunctor
\begin{equation}
    \mathscr{T} \boxtimes - : \mathbf{BrFus}\to \mathbf{BrFus},
\end{equation}
of the Morita $4$-category of braided fusion categories $\mathbf{BrFus}$ \cite{Johnson-Freyd:2015sza,Brochier:2018kxc}, for which the distinguished object is $\mathscr{T}\boxtimes \mathbbm{1}_{(3)}\in \mathbf{BrFus}$. This is based on the fact that braided fusion categories define $3d$ TQFT's \cite{Reshetikhin:1991tc,Turaev+2016}. We summarize this in Table~\ref{tab:defect-dictionary}. Notice in particular that since $\mathscr{T}$ has a dual, and since a $\mathcal{Z(C)}$-module is equivalently a $(\mathcal{C},\mathcal{C})$-bimodule for $\mathcal{C}$ a fusion category, a $(\mathscr{T}\boxtimes \mathbbm{1},\mathscr{T}\boxtimes \mathcal{Z}(\mathcal{C}))$-bimodule category $\mathcal{D}$ has a corresponding $(\mathcal{C},\mathcal{C})$-bimodule category $\mathcal{M}$, which is how more general bulk surface operators were defined in \cite[Section 5]{Bergman:2026lnz}. It follows that our dictionary is an extension of said characterization by allowing more general families of stacked $3d$ TFT's.

To conclude this section, we highlight that this process of lifting category-theoretic degrees of defects/TFT's may be iterated so as to obtain similar notions of symmetries of negative degrees. For example, it was observed in \cite{Antinucci:2024zjp}, in the context of continuous symmetries, that one may in fact have several different possible SymTFT's. These are all however related to a $(d+2)$-dimensional topological field theory, which the authors called the ``Anomaly Polynomial TFT.'' This would suggest a path towards defining some notion of $(-3)$-form symmetries. Another approach is to consider the SymTFT of the SymTFT itself; we touch on this route in Section~\ref{sec:conclusion}. More generally, it appears plausible to define lower-degree symmetries by properly including our starting physical theory in a descent mechanism, of which the latter case is a special case. The relation between descent and conserved charges was highlighted in \cite{Borsten:2025pbx}.

\providecommand{\href}[2]{#2}\begingroup\raggedright\endgroup

%\bibliographystyle{JHEP}
%\bibliography{refs}

\begin{thebibliography}{100}

\bibitem{Gaiotto:2014kfa}
D.~Gaiotto, A.~Kapustin, N.~Seiberg and B.~Willett, \emph{{Generalized global symmetries}}, \href{https://doi.org/10.1007/JHEP02(2015)172}{\emph{JHEP} {\bfseries 02} (2015) 172}, [\href{https://arxiv.org/abs/1412.5148}{{\ttfamily 1412.5148}}].

\bibitem{Sharpe:2015mja}
E.~Sharpe, \emph{{Notes on generalized global symmetries in QFT}}, \href{https://doi.org/10.1002/prop.201500048}{\emph{Fortsch. Phys.} {\bfseries 63} (2015) 659--682}, [\href{https://arxiv.org/abs/1508.04770}{{\ttfamily 1508.04770}}].

\bibitem{Cordova:2022ruw}
C.~Cordova, T.~T. Dumitrescu, K.~Intriligator and S.-H. Shao, \emph{{Snowmass white paper: Generalized symmetries in quantum field theory and beyond}},  in \emph{{Snowmass 2021}}, 5, 2022, \href{https://arxiv.org/abs/2205.09545}{{\ttfamily 2205.09545}}.

\bibitem{Schafer-Nameki:2023jdn}
S.~Schafer-Nameki, \emph{{ICTP lectures on (non-)invertible generalized symmetries}}, \href{https://doi.org/10.1016/j.physrep.2024.01.007}{\emph{Phys. Rept.} {\bfseries 1063} (2024) 1--55}, [\href{https://arxiv.org/abs/2305.18296}{{\ttfamily 2305.18296}}].

\bibitem{Brennan:2023mmt}
T.~D. Brennan and S.~Hong, \emph{{Introduction to generalized global symmetries in QFT and particle physics}},  \href{https://arxiv.org/abs/2306.00912}{{\ttfamily 2306.00912}}.

\bibitem{Bhardwaj:2023ayw}
L.~Bhardwaj and S.~Schafer-Nameki, \emph{{Generalized charges, part II: Non-invertible symmetries and the symmetry TFT}}, \href{https://doi.org/10.21468/SciPostPhys.19.4.098}{\emph{SciPost Phys.} {\bfseries 19} (2025) 098}, [\href{https://arxiv.org/abs/2305.17159}{{\ttfamily 2305.17159}}].

\bibitem{Freed:2022iao}
D.~S. Freed, \emph{{Introduction to topological symmetry in QFT}}, \href{https://doi.org/10.1090/pspum/107/01946}{\emph{Proc. Symp. Pure Math.} {\bfseries 107} (2024) 93--106}, [\href{https://arxiv.org/abs/2212.00195}{{\ttfamily 2212.00195}}].

\bibitem{Bah:2022xfv}
I.~Bah, D.~Freed, G.~W. Moore, N.~Nekrasov, S.~S. Razamat and S.~Sch{\"a}fer-Nameki, \emph{{Snowmass whitepaper: Physical mathematics 2021}},  \href{https://arxiv.org/abs/2203.05078}{{\ttfamily 2203.05078}}.

\bibitem{Bhardwaj:2023kri}
L.~Bhardwaj, L.~E. Bottini, L.~Fraser-Taliente, L.~Gladden, D.~S.~W. Gould, A.~Platschorre et~al., \emph{{Lectures on generalized symmetries}}, \href{https://doi.org/10.1016/j.physrep.2023.11.002}{\emph{Phys. Rept.} {\bfseries 1051} (2024) 1--87}, [\href{https://arxiv.org/abs/2307.07547}{{\ttfamily 2307.07547}}].

\bibitem{Shao:2023gho}
S.-H. Shao, \emph{{What's done cannot be undone: TASI lectures on non-invertible symmetries}},  \href{https://arxiv.org/abs/2308.00747}{{\ttfamily 2308.00747}}.

\bibitem{Gomes:2023ahz}
P.~R.~S. Gomes, \emph{{An introduction to higher-form symmetries}}, \href{https://doi.org/10.21468/SciPostPhysLectNotes.74}{\emph{SciPost Phys. Lect. Notes} {\bfseries 74} (2023) 1}, [\href{https://arxiv.org/abs/2303.01817}{{\ttfamily 2303.01817}}].

\bibitem{McGreevy:2022oyu}
J.~McGreevy, \emph{{Generalized symmetries in condensed matter}}, \href{https://doi.org/10.1146/annurev-conmatphys-040721-021029}{\emph{Ann. Rev. Condensed Matter Phys.} {\bfseries 14} (2023) 57--82}, [\href{https://arxiv.org/abs/2204.03045}{{\ttfamily 2204.03045}}].

\bibitem{Chang:2018iay}
C.-M. Chang, Y.-H. Lin, S.-H. Shao, Y.~Wang and X.~Yin, \emph{{Topological defect lines and renormalization group flows in two dimensions}}, \href{https://doi.org/10.1007/JHEP01(2019)026}{\emph{JHEP} {\bfseries 01} (2019) 026}, [\href{https://arxiv.org/abs/1802.04445}{{\ttfamily 1802.04445}}].

\bibitem{Bhardwaj:2022yxj}
L.~Bhardwaj, L.~E. Bottini, S.~Schafer-Nameki and A.~Tiwari, \emph{{Non-invertible higher-categorical symmetries}}, \href{https://doi.org/10.21468/SciPostPhys.14.1.007}{\emph{SciPost Phys.} {\bfseries 14} (2023) 007}, [\href{https://arxiv.org/abs/2204.06564}{{\ttfamily 2204.06564}}].

\bibitem{Iqbal:2024pee}
N.~Iqbal, \emph{{Jena lectures on generalized global symmetries: principles and applications}},  7, 2024, \href{https://arxiv.org/abs/2407.20815}{{\ttfamily 2407.20815}}.

\bibitem{Costa:2024wks}
D.~Costa et~al., \emph{{Simons lectures on categorical symmetries}},  11, 2024, \href{https://arxiv.org/abs/2411.09082}{{\ttfamily 2411.09082}}.

\bibitem{Cordova:2019jnf}
C.~C{\' o}rdova, D.~S. Freed, H.~T. Lam and N.~Seiberg, \emph{{Anomalies in the space of coupling constants and their dynamical applications I}}, \href{https://doi.org/10.21468/SciPostPhys.8.1.001}{\emph{SciPost Phys.} {\bfseries 8} (2020) 001}, [\href{https://arxiv.org/abs/1905.09315}{{\ttfamily 1905.09315}}].

\bibitem{Cordova:2019uob}
C.~C{\' o}rdova, D.~S. Freed, H.~T. Lam and N.~Seiberg, \emph{{Anomalies in the space of coupling constants and their dynamical applications II}}, \href{https://doi.org/10.21468/SciPostPhys.8.1.002}{\emph{SciPost Phys.} {\bfseries 8} (2020) 002}, [\href{https://arxiv.org/abs/1905.13361}{{\ttfamily 1905.13361}}].

\bibitem{Hellerman:2006zs}
S.~Hellerman, A.~Henriques, T.~Pantev, E.~Sharpe and M.~Ando, \emph{{Cluster decomposition, T-duality, and gerby CFT's}}, \href{https://doi.org/10.4310/ATMP.2007.v11.n5.a2}{\emph{Adv. Theor. Math. Phys.} {\bfseries 11} (2007) 751--818}, [\href{https://arxiv.org/abs/hep-th/0606034}{{\ttfamily hep-th/0606034}}].

\bibitem{Caldararu:2010ljp}
A.~Caldararu, J.~Distler, S.~Hellerman, T.~Pantev and E.~Sharpe, \emph{{Non-birational twisted derived equivalences in abelian GLSMs}}, \href{https://doi.org/10.1007/s00220-009-0974-2}{\emph{Commun. Math. Phys.} {\bfseries 294} (2010) 605--645}, [\href{https://arxiv.org/abs/0709.3855}{{\ttfamily 0709.3855}}].

\bibitem{Sharpe:2014tca}
E.~Sharpe, \emph{{Decomposition in diverse dimensions}}, \href{https://doi.org/10.1103/PhysRevD.90.025030}{\emph{Phys. Rev. D} {\bfseries 90} (2014) 025030}, [\href{https://arxiv.org/abs/1404.3986}{{\ttfamily 1404.3986}}].

\bibitem{Sharpe:2019ddn}
E.~Sharpe, \emph{{Undoing decomposition}}, \href{https://doi.org/10.1142/S0217751X19502336}{\emph{Int. J. Mod. Phys. A} {\bfseries 34} (2020) 1950233}, [\href{https://arxiv.org/abs/1911.05080}{{\ttfamily 1911.05080}}].

\bibitem{Tanizaki:2019rbk}
Y.~Tanizaki and M.~\"Unsal, \emph{{Modified instanton sum in QCD and higher-groups}}, \href{https://doi.org/10.1007/JHEP03(2020)123}{\emph{JHEP} {\bfseries 03} (2020) 123}, [\href{https://arxiv.org/abs/1912.01033}{{\ttfamily 1912.01033}}].

\bibitem{Nguyen:2021yld}
M.~Nguyen, Y.~Tanizaki and M.~{\"U}nsal, \emph{{Semi-Abelian gauge theories, non-invertible symmetries, and string tensions beyond $N$-ality}}, \href{https://doi.org/10.1007/JHEP03(2021)238}{\emph{JHEP} {\bfseries 03} (2021) 238}, [\href{https://arxiv.org/abs/2101.02227}{{\ttfamily 2101.02227}}].

\bibitem{Nguyen:2021naa}
M.~Nguyen, Y.~Tanizaki and M.~\"Unsal, \emph{{Noninvertible 1-form symmetry and Casimir scaling in 2D Yang-Mills theory}}, \href{https://doi.org/10.1103/PhysRevD.104.065003}{\emph{Phys. Rev. D} {\bfseries 104} (2021) 065003}, [\href{https://arxiv.org/abs/2104.01824}{{\ttfamily 2104.01824}}].

\bibitem{Komargodski:2020mxz}
Z.~Komargodski, K.~Ohmori, K.~Roumpedakis and S.~Seifnashri, \emph{{Symmetries and strings of adjoint QCD$_{2}$}}, \href{https://doi.org/10.1007/JHEP03(2021)103}{\emph{JHEP} {\bfseries 03} (2021) 103}, [\href{https://arxiv.org/abs/2008.07567}{{\ttfamily 2008.07567}}].

\bibitem{Cherman:2020cvw}
A.~Cherman and T.~Jacobson, \emph{{Lifetimes of near eternal false vacua}}, \href{https://doi.org/10.1103/PhysRevD.103.105012}{\emph{Phys. Rev. D} {\bfseries 103} (2021) 105012}, [\href{https://arxiv.org/abs/2012.10555}{{\ttfamily 2012.10555}}].

\bibitem{Robbins:2020msp}
D.~Robbins, E.~Sharpe and T.~Vandermeulen, \emph{{A generalization of decomposition in orbifolds}}, \href{https://doi.org/10.1007/JHEP10(2021)134}{\emph{JHEP} {\bfseries 21} (2020) 134}, [\href{https://arxiv.org/abs/2101.11619}{{\ttfamily 2101.11619}}].

\bibitem{Robbins:2021xce}
D.~G. Robbins, E.~Sharpe and T.~Vandermeulen, \emph{{Anomaly resolution via decomposition}}, \href{https://doi.org/10.1142/S0217751X21502201}{\emph{Int. J. Mod. Phys. A} {\bfseries 36} (2021) 2150220}, [\href{https://arxiv.org/abs/2107.13552}{{\ttfamily 2107.13552}}].

\bibitem{Sharpe:2021srf}
E.~Sharpe, \emph{{Topological operators, noninvertible symmetries and decomposition}}, \href{https://doi.org/10.4310/ATMP.2023.v27.n8.a2}{\emph{Adv. Theor. Math. Phys.} {\bfseries 27} (2023) 2319--2407}, [\href{https://arxiv.org/abs/2108.13423}{{\ttfamily 2108.13423}}].

\bibitem{Pantev:2022kpl}
T.~Pantev, D.~G. Robbins, E.~Sharpe and T.~Vandermeulen, \emph{{Orbifolds by 2-groups and decomposition}}, \href{https://doi.org/10.1007/JHEP09(2022)036}{\emph{JHEP} {\bfseries 09} (2022) 036}, [\href{https://arxiv.org/abs/2204.13708}{{\ttfamily 2204.13708}}].

\bibitem{Pantev:2022pbf}
T.~Pantev and E.~Sharpe, \emph{{Decomposition in Chern-Simons theories in three dimensions}}, \href{https://doi.org/10.1142/S0217751X2250227X}{\emph{Int. J. Mod. Phys. A} {\bfseries 37} (2022) 2250227}, [\href{https://arxiv.org/abs/2206.14824}{{\ttfamily 2206.14824}}].

\bibitem{Lin:2022xod}
L.~Lin, D.~G. Robbins and E.~Sharpe, \emph{{Decomposition, condensation defects, and fusion}}, \href{https://doi.org/10.1002/prop.202200130}{\emph{Fortsch. Phys.} {\bfseries 70} (2022) 2200130}, [\href{https://arxiv.org/abs/2208.05982}{{\ttfamily 2208.05982}}].

\bibitem{Perez-Lona:2023llv}
A.~Perez-Lona and E.~Sharpe, \emph{{Three-dimensional orbifolds by 2-groups}}, \href{https://doi.org/10.1007/JHEP08(2023)138}{\emph{JHEP} {\bfseries 08} (2023) 138}, [\href{https://arxiv.org/abs/2303.16220}{{\ttfamily 2303.16220}}].

\bibitem{Pantev:2024kva}
T.~Pantev, E.~Sharpe and X.~Yu, \emph{{Chern-Simons theory, decomposition, and the A model}}, \href{https://doi.org/10.1007/JHEP10(2024)112}{\emph{JHEP} {\bfseries 10} (2024) 112}, [\href{https://arxiv.org/abs/2406.18633}{{\ttfamily 2406.18633}}].

\bibitem{Pantev:2023dim}
T.~Pantev and E.~Sharpe, \emph{{Decomposition and the Gross{\textendash}Taylor string theory}}, \href{https://doi.org/10.1142/S0217751X23501567}{\emph{Int. J. Mod. Phys. A} {\bfseries 38} (2023) 2350156}, [\href{https://arxiv.org/abs/2307.08729}{{\ttfamily 2307.08729}}].

\bibitem{Santilli:2024dyz}
L.~Santilli and R.~J. Szabo, \emph{{Higher form symmetries and orbifolds of two-dimensional Yang\textendash{}Mills theory}}, \href{https://doi.org/10.1007/s11005-025-01905-4}{\emph{Lett. Math. Phys.} {\bfseries 115} (2025) 15}, [\href{https://arxiv.org/abs/2403.03119}{{\ttfamily 2403.03119}}].

\bibitem{Yu:2024jtk}
X.~Yu, \emph{{Gauging in parameter space: A top-down perspective}}, \href{https://doi.org/10.1103/638n-qwnm}{\emph{Phys. Rev. D} {\bfseries 112} (2025) 025020}, [\href{https://arxiv.org/abs/2411.14997}{{\ttfamily 2411.14997}}].

\bibitem{Najjar:2024vmm}
M.~Najjar, L.~Santilli and Y.-N. Wang, \emph{{(\ensuremath{-}1)-form symmetries from M-theory and SymTFTs}}, \href{https://doi.org/10.1007/JHEP03(2025)134}{\emph{JHEP} {\bfseries 03} (2025) 134}, [\href{https://arxiv.org/abs/2411.19683}{{\ttfamily 2411.19683}}].

\bibitem{Robbins:2025urk}
D.~Robbins and S.~Roy, \emph{{({\ensuremath{-}}1)-form symmetries and anomaly shifting from symmetry topological field theory}}, \href{https://doi.org/10.1103/2zk8-jd5t}{\emph{Phys. Rev. D} {\bfseries 112} (2025) 105020}, [\href{https://arxiv.org/abs/2505.14807}{{\ttfamily 2505.14807}}].

\bibitem{Lin:2025oml}
L.~Lin, D.~Robbins and S.~Roy, \emph{{Decomposition and (non-invertible) ({\ensuremath{-}}1)-form symmetries from the symmetry topological field theory}}, \href{https://doi.org/10.1007/JHEP09(2025)131}{\emph{JHEP} {\bfseries 09} (2025) 131}, [\href{https://arxiv.org/abs/2503.21862}{{\ttfamily 2503.21862}}].

\bibitem{Perez-Lona:2025add}
A.~Perez-Lona, E.~Sharpe, X.~Yu and H.~Zhang, \emph{{Total instanton restriction via multiverse interference: Noncompact gauge theories and ({\ensuremath{-}}1)-form symmetries}}, \href{https://doi.org/10.1007/JHEP05(2026)214}{\emph{JHEP} {\bfseries 05} (2026) 214}, [\href{https://arxiv.org/abs/2508.00050}{{\ttfamily 2508.00050}}].

\bibitem{Sharpe:2022ene}
E.~Sharpe, \emph{{An introduction to decomposition}},  \href{https://arxiv.org/abs/2204.09117}{{\ttfamily 2204.09117}}.

\bibitem{Aloni:2024jpb}
D.~Aloni, E.~Garc\'\i{}a-Valdecasas, M.~Reece and M.~Suzuki, \emph{{Spontaneously broken (-1)-form U(1) symmetries}}, \href{https://doi.org/10.21468/SciPostPhys.17.2.031}{\emph{SciPost Phys.} {\bfseries 17} (2024) 031}, [\href{https://arxiv.org/abs/2402.00117}{{\ttfamily 2402.00117}}].

\bibitem{Kapustin:2014gua}
A.~Kapustin and N.~Seiberg, \emph{{Coupling a QFT to a TQFT and duality}}, \href{https://doi.org/10.1007/JHEP04(2014)001}{\emph{JHEP} {\bfseries 04} (2014) 001}, [\href{https://arxiv.org/abs/1401.0740}{{\ttfamily 1401.0740}}].

\bibitem{Heckman:2024oot}
J.~J. Heckman, M.~H{\"u}bner and C.~Murdia, \emph{{On the holographic dual of a topological symmetry operator}}, \href{https://doi.org/10.1103/PhysRevD.110.046007}{\emph{Phys. Rev. D} {\bfseries 110} (2024) 046007}, [\href{https://arxiv.org/abs/2401.09538}{{\ttfamily 2401.09538}}].

\bibitem{Antinucci:2024zjp}
A.~Antinucci and F.~Benini, \emph{{Anomalies and gauging of U(1) symmetries}}, \href{https://doi.org/10.1103/PhysRevB.111.024110}{\emph{Phys. Rev. B} {\bfseries 111} (2025) 024110}, [\href{https://arxiv.org/abs/2401.10165}{{\ttfamily 2401.10165}}].

\bibitem{Cvetic:2024dzu}
M.~Cveti{\v{c}}, R.~Donagi, J.~J. Heckman, M.~H{\"u}bner and E.~Torres, \emph{{Cornering relative symmetry theories}}, \href{https://doi.org/10.1103/PhysRevD.111.085026}{\emph{Phys. Rev. D} {\bfseries 111} (2025) 085026}, [\href{https://arxiv.org/abs/2408.12600}{{\ttfamily 2408.12600}}].

\bibitem{Borsten:2025pbx}
L.~Borsten, D.~Kanakaris and H.~Kim, \emph{{Symmetries beget symmetries: ghostly higher-form symmetries and the descent equation}}, \href{https://doi.org/10.1007/JHEP05(2026)200}{\emph{JHEP} {\bfseries 05} (2026) 200}, [\href{https://arxiv.org/abs/2509.15978}{{\ttfamily 2509.15978}}].

\bibitem{Stockall:2025ngz}
D.~Stockall and M.~Yu, \emph{{Geometric categories for continuous gauging}},  \href{https://arxiv.org/abs/2511.08254}{{\ttfamily 2511.08254}}.

\bibitem{Heckman:2026beg}
J.~J. Heckman, R.~J. Hicks and C.~Murdia, \emph{{Generalized complexity distances and non-invertible symmetries}},  \href{https://arxiv.org/abs/2604.14275}{{\ttfamily 2604.14275}}.

\bibitem{Bergman:2026lnz}
O.~Bergman, J.~J. Heckman, M.~H{\"u}bner, D.~Migliorati, X.~Yu and H.~Y. Zhang, \emph{{On the SymTFTs of finite non-abelian symmetries}},  \href{https://arxiv.org/abs/2603.12323}{{\ttfamily 2603.12323}}.

\bibitem{Perez-Lona:2025xxm}
A.~Perez-Lona, \emph{{Higher-form symmetries as higher automorphism bundles}},  \href{https://arxiv.org/abs/2509.15301}{{\ttfamily 2509.15301}}.

\bibitem{Teixeira:2025qsg}
D.~Teixeira and M.~Yu, \emph{{Mutual influence of symmetries and topological field theories}},  \href{https://arxiv.org/abs/2507.06304}{{\ttfamily 2507.06304}}.

\bibitem{PerezLona:2026tba}
A.~Perez-Lona, \emph{{{Magnetic Higher-Form Symmetries as Dual Fundamental $\infty$-Groupoids \textit{(to appear)}}}},  2026.

\bibitem{Baume:2023kkf}
F.~Baume, J.~J. Heckman, M.~H\"ubner, E.~Torres, A.~P. Turner and X.~Yu, \emph{{SymTrees and Multi-Sector QFTs}}, \href{https://doi.org/10.1103/PhysRevD.109.106013}{\emph{Phys. Rev. D} {\bfseries 109} (2024) 106013}, [\href{https://arxiv.org/abs/2310.12980}{{\ttfamily 2310.12980}}].

\bibitem{Bhardwaj:2023bbf}
L.~Bhardwaj, L.~E. Bottini, D.~Pajer and S.~Schafer-Nameki, \emph{{The club sandwich: Gapless phases and phase transitions with non-invertible symmetries}}, \href{https://doi.org/10.21468/SciPostPhys.18.5.156}{\emph{SciPost Phys.} {\bfseries 18} (2025) 156}, [\href{https://arxiv.org/abs/2312.17322}{{\ttfamily 2312.17322}}].

\bibitem{KNBalasubramanian:2025vum}
M.~K.~N.~Balasubramanian, M.~Buican, C.~Delcamp and R.~Radhakrishnan, \emph{{Gauging non-invertible symmetries in (2+1)d topological orders}},  \href{https://arxiv.org/abs/2507.01142}{{\ttfamily 2507.01142}}.

\bibitem{Romans:1985tz}
L.~J. Romans, \emph{{Massive N=2a supergravity in ten-dimensions}}, \href{https://doi.org/10.1016/0370-2693(86)90375-8}{\emph{Phys. Lett. B} {\bfseries 169} (1986) 374}.

\bibitem{Bhardwaj:2023wzd}
L.~Bhardwaj and S.~Schafer-Nameki, \emph{{Generalized charges, part I: Invertible symmetries and higher representations}}, \href{https://doi.org/10.21468/SciPostPhys.16.4.093}{\emph{SciPost Phys.} {\bfseries 16} (2024) 093}, [\href{https://arxiv.org/abs/2304.02660}{{\ttfamily 2304.02660}}].

\bibitem{Cherman:2021nox}
A.~Cherman, T.~Jacobson and M.~Neuzil, \emph{{Universal deformations}}, \href{https://doi.org/10.21468/SciPostPhys.12.4.116}{\emph{SciPost Phys.} {\bfseries 12} (2022) 116}, [\href{https://arxiv.org/abs/2111.00078}{{\ttfamily 2111.00078}}].

\bibitem{Roumpedakis:2022aik}
K.~Roumpedakis, S.~Seifnashri and S.-H. Shao, \emph{{Higher gauging and non-invertible condensation defects}}, \href{https://doi.org/10.1007/s00220-023-04706-9}{\emph{Commun. Math. Phys.} {\bfseries 401} (2023) 3043--3107}, [\href{https://arxiv.org/abs/2204.02407}{{\ttfamily 2204.02407}}].

\bibitem{Kong:2013aya}
L.~Kong, \emph{{Anyon condensation and tensor categories}}, \href{https://doi.org/10.1016/j.nuclphysb.2014.07.003}{\emph{Nucl. Phys. B} {\bfseries 886} (2014) 436--482}, [\href{https://arxiv.org/abs/1307.8244}{{\ttfamily 1307.8244}}].

\bibitem{Kong:2014qka}
L.~Kong and X.-G. Wen, \emph{{Braided fusion categories, gravitational anomalies, and the mathematical framework for topological orders in any dimensions}},  \href{https://arxiv.org/abs/1405.5858}{{\ttfamily 1405.5858}}.

\bibitem{Else:2017yqj}
D.~V. Else and C.~Nayak, \emph{{Cheshire charge in (3+1)-dimensional topological phases}}, \href{https://doi.org/10.1103/PhysRevB.96.045136}{\emph{Phys. Rev. B} {\bfseries 96} (2017) 045136}, [\href{https://arxiv.org/abs/1702.02148}{{\ttfamily 1702.02148}}].

\bibitem{Gaiotto:2019xmp}
D.~Gaiotto and T.~Johnson-Freyd, \emph{{Condensations in higher categories}},  \href{https://arxiv.org/abs/1905.09566}{{\ttfamily 1905.09566}}.

\bibitem{Gaiotto:2020iye}
D.~Gaiotto and J.~Kulp, \emph{{Orbifold groupoids}}, \href{https://doi.org/10.1007/JHEP02(2021)132}{\emph{JHEP} {\bfseries 02} (2021) 132}, [\href{https://arxiv.org/abs/2008.05960}{{\ttfamily 2008.05960}}].

\bibitem{Kaidi:2022cpf}
J.~Kaidi, K.~Ohmori and Y.~Zheng, \emph{{Symmetry TFTs for non-invertible defects}}, \href{https://doi.org/10.1007/s00220-023-04859-7}{\emph{Commun. Math. Phys.} {\bfseries 404} (2023) 1021--1124}, [\href{https://arxiv.org/abs/2209.11062}{{\ttfamily 2209.11062}}].

\bibitem{Hasan:2024aow}
A.~Hasan, S.~Meynet and D.~Migliorati, \emph{{SL$_{2}$({\ensuremath{\mathbb{R}}}) symmetries of SymTFT and non-invertible U(1) symmetries of Maxwell theory}}, \href{https://doi.org/10.1007/JHEP12(2024)131}{\emph{JHEP} {\bfseries 12} (2024) 131}, [\href{https://arxiv.org/abs/2405.19218}{{\ttfamily 2405.19218}}].

\bibitem{Arbalestrier:2025jsg}
A.~Arbalestrier, R.~Argurio, G.~Galati and E.~Paznokas, \emph{{4d Maxwell on the edge: global aspects of boundary conditions and duality}}, \href{https://doi.org/10.1007/JHEP03(2026)010}{\emph{JHEP} {\bfseries 03} (2026) 010}, [\href{https://arxiv.org/abs/2510.19551}{{\ttfamily 2510.19551}}].

\bibitem{Niro:2022ctq}
P.~Niro, K.~Roumpedakis and O.~Sela, \emph{{Exploring non-invertible symmetries in free theories}}, \href{https://doi.org/10.1007/JHEP03(2023)005}{\emph{JHEP} {\bfseries 03} (2023) 005}, [\href{https://arxiv.org/abs/2209.11166}{{\ttfamily 2209.11166}}].

\bibitem{vanBeest:2022fss}
M.~van Beest, D.~S.~W. Gould, S.~Schafer-Nameki and Y.-N. Wang, \emph{{Symmetry TFTs for 3d QFTs from M-theory}}, \href{https://doi.org/10.1007/JHEP02(2023)226}{\emph{JHEP} {\bfseries 02} (2023) 226}, [\href{https://arxiv.org/abs/2210.03703}{{\ttfamily 2210.03703}}].

\bibitem{Aharony:2008ug}
O.~Aharony, O.~Bergman, D.~L. Jafferis and J.~Maldacena, \emph{{N=6 superconformal Chern-Simons-matter theories, M2-branes and their gravity duals}}, \href{https://doi.org/10.1088/1126-6708/2008/10/091}{\emph{JHEP} {\bfseries 10} (2008) 091}, [\href{https://arxiv.org/abs/0806.1218}{{\ttfamily 0806.1218}}].

\bibitem{Aharony:2008gk}
O.~Aharony, O.~Bergman and D.~L. Jafferis, \emph{{Fractional M2-branes}}, \href{https://doi.org/10.1088/1126-6708/2008/11/043}{\emph{JHEP} {\bfseries 11} (2008) 043}, [\href{https://arxiv.org/abs/0807.4924}{{\ttfamily 0807.4924}}].

\bibitem{Bergman:2020ifi}
O.~Bergman, Y.~Tachikawa and G.~Zafrir, \emph{{Generalized symmetries and holography in ABJM-type theories}}, \href{https://doi.org/10.1007/JHEP07(2020)077}{\emph{JHEP} {\bfseries 07} (2020) 077}, [\href{https://arxiv.org/abs/2004.05350}{{\ttfamily 2004.05350}}].

\bibitem{Cordova:2018cvg}
C.~C{\' o}rdova, T.~T. Dumitrescu and K.~Intriligator, \emph{{Exploring 2-Group Global Symmetries}}, \href{https://doi.org/10.1007/JHEP02(2019)184}{\emph{JHEP} {\bfseries 02} (2019) 184}, [\href{https://arxiv.org/abs/1802.04790}{{\ttfamily 1802.04790}}].

\bibitem{Benini:2018reh}
F.~Benini, C.~C\'ordova and P.-S. Hsin, \emph{{On 2-Group Global Symmetries and their Anomalies}}, \href{https://doi.org/10.1007/JHEP03(2019)118}{\emph{JHEP} {\bfseries 03} (2019) 118}, [\href{https://arxiv.org/abs/1803.09336}{{\ttfamily 1803.09336}}].

\bibitem{Kaidi:2021xfk}
J.~Kaidi, K.~Ohmori and Y.~Zheng, \emph{{Kramers-Wannier-like duality defects in (3+1)D gauge theories}}, \href{https://doi.org/10.1103/PhysRevLett.128.111601}{\emph{Phys. Rev. Lett.} {\bfseries 128} (2022) 111601}, [\href{https://arxiv.org/abs/2111.01141}{{\ttfamily 2111.01141}}].

\bibitem{Brennan:2024tlw}
T.~D. Brennan, \emph{{Constraints on symmetry-preserving gapped phases from coupling constant anomalies}}, \href{https://doi.org/10.1103/PhysRevD.110.L041701}{\emph{Phys. Rev. D} {\bfseries 110} (2024) L041701}, [\href{https://arxiv.org/abs/2404.11660}{{\ttfamily 2404.11660}}].

\bibitem{Brennan:2026ira}
T.~D. Brennan and K.~Intriligator, \emph{{Generalized families of QFTs}},  \href{https://arxiv.org/abs/2602.09105}{{\ttfamily 2602.09105}}.

\bibitem{TAMBARA1998692}
D.~Tambara and S.~Yamagami, \emph{Tensor categories with fusion rules of self-duality for finite abelian groups}, \href{https://doi.org/https://doi.org/10.1006/jabr.1998.7558}{\emph{Journal of Algebra} {\bfseries 209} (1998) 692--707}.

\bibitem{Robbins:2025puq}
D.~Robbins and S.~Roy, \emph{{SymTFT actions, Condensable algebras and categorical anomaly resolutions}},  \href{https://arxiv.org/abs/2509.05408}{{\ttfamily 2509.05408}}.

\bibitem{Yu:2023nyn}
X.~Yu, \emph{{Noninvertible symmetries in 2D from type IIB string theory}}, \href{https://doi.org/10.1103/PhysRevD.110.065008}{\emph{Phys. Rev. D} {\bfseries 110} (2024) 065008}, [\href{https://arxiv.org/abs/2310.15339}{{\ttfamily 2310.15339}}].

\bibitem{Kitaev:2005hzj}
A.~Kitaev, \emph{{Anyons in an exactly solved model and beyond}}, \href{https://doi.org/10.1016/j.aop.2005.10.005}{\emph{Annals Phys.} {\bfseries 321} (2006) 2--111}, [\href{https://arxiv.org/abs/cond-mat/0506438}{{\ttfamily cond-mat/0506438}}].

\bibitem{Rowell:2007dge}
E.~Rowell, R.~Stong and Z.~Wang, \emph{{On Classification of Modular Tensor Categories}}, \href{https://doi.org/10.1007/s00220-009-0908-z}{\emph{Commun. Math. Phys.} {\bfseries 292} (2009) 343--389}, [\href{https://arxiv.org/abs/0712.1377}{{\ttfamily 0712.1377}}].

\bibitem{davydov2011wittgroupnondegeneratebraided}
A.~Davydov, M.~Mueger, D.~Nikshych and V.~Ostrik, \emph{{The Witt group of non-degenerate braided fusion categories}},  \href{https://arxiv.org/abs/1009.2117}{{\ttfamily 1009.2117}}.

\bibitem{Thorngren:2021yso}
R.~Thorngren and Y.~Wang, \emph{{Fusion category symmetry. Part II. Categoriosities at c = 1 and beyond}}, \href{https://doi.org/10.1007/JHEP07(2024)051}{\emph{JHEP} {\bfseries 07} (2024) 051}, [\href{https://arxiv.org/abs/2106.12577}{{\ttfamily 2106.12577}}].

\bibitem{Antinucci:2025fjp}
A.~Antinucci, C.~Copetti, Y.~Gai and S.~Schafer-Nameki, \emph{{Categorical anomaly matching}},  \href{https://arxiv.org/abs/2508.00982}{{\ttfamily 2508.00982}}.

\bibitem{Lu:2025yru}
D.-C. Lu and Z.~Sun, \emph{{Intrinsic NISPT phases, igNISPT phases, and mixed anomalies of non-invertible symmetries}},  \href{https://arxiv.org/abs/2511.01965}{{\ttfamily 2511.01965}}.

\bibitem{Gaiotto:2012np}
D.~Gaiotto, \emph{{Domain walls for two-dimensional renormalization group flows}}, \href{https://doi.org/10.1007/JHEP12(2012)103}{\emph{JHEP} {\bfseries 12} (2012) 103}, [\href{https://arxiv.org/abs/1201.0767}{{\ttfamily 1201.0767}}].

\bibitem{Wen:2022tkg}
R.~Wen and A.~C. Potter, \emph{{Bulk-boundary correspondence for intrinsically gapless symmetry-protected topological phases from group cohomology}}, \href{https://doi.org/10.1103/PhysRevB.107.245127}{\emph{Phys. Rev. B} {\bfseries 107} (2023) 245127}, [\href{https://arxiv.org/abs/2208.09001}{{\ttfamily 2208.09001}}].

\bibitem{Bhardwaj:2024qrf}
L.~Bhardwaj, D.~Pajer, S.~Schafer-Nameki and A.~Warman, \emph{{Hasse diagrams for gapless SPT and SSB phases with non-invertible symmetries}}, \href{https://doi.org/10.21468/SciPostPhys.19.4.113}{\emph{SciPost Phys.} {\bfseries 19} (2025) 113}, [\href{https://arxiv.org/abs/2403.00905}{{\ttfamily 2403.00905}}].

\bibitem{Li:2023knf}
L.~Li, M.~Oshikawa and Y.~Zheng, \emph{{Intrinsically/purely gapless-SPT from non-invertible duality transformations}}, \href{https://doi.org/10.21468/SciPostPhys.18.5.153}{\emph{SciPost Phys.} {\bfseries 18} (2025) 153}, [\href{https://arxiv.org/abs/2307.04788}{{\ttfamily 2307.04788}}].

\bibitem{Li:2022jbf}
L.~Li, M.~Oshikawa and Y.~Zheng, \emph{{Decorated defect construction of gapless-SPT states}}, \href{https://doi.org/10.21468/SciPostPhys.17.1.013}{\emph{SciPost Phys.} {\bfseries 17} (2024) 013}, [\href{https://arxiv.org/abs/2204.03131}{{\ttfamily 2204.03131}}].

\bibitem{He:2016xpi}
H.~He, Y.~Zheng and C.~von Keyserlingk, \emph{{Field theories for gauged symmetry-protected topological phases: Non-Abelian anyons with Abelian gauge group $\mathbb Z_2^3$}}, \href{https://doi.org/10.1103/PhysRevB.95.035131}{\emph{Phys. Rev. B} {\bfseries 95} (2017) 035131}, [\href{https://arxiv.org/abs/1608.05393}{{\ttfamily 1608.05393}}].

\bibitem{Franco:2024mxa}
S.~Franco and X.~Yu, \emph{{Generalized symmetries in 2D from string theory: SymTFTs, intrinsic relativeness, and anomalies of non-invertible symmetries}}, \href{https://doi.org/10.1007/JHEP11(2024)004}{\emph{JHEP} {\bfseries 11} (2024) 004}, [\href{https://arxiv.org/abs/2404.19761}{{\ttfamily 2404.19761}}].

\bibitem{Kaidi:2023maf}
J.~Kaidi, E.~Nardoni, G.~Zafrir and Y.~Zheng, \emph{{Symmetry TFTs and anomalies of non-invertible symmetries}}, \href{https://doi.org/10.1007/JHEP10(2023)053}{\emph{JHEP} {\bfseries 10} (2023) 053}, [\href{https://arxiv.org/abs/2301.07112}{{\ttfamily 2301.07112}}].

\bibitem{deWildPropitius:1995cf}
M.~D.~F. de~Wild~Propitius, \emph{{Topological interactions in broken gauge theories}}, Ph.D. thesis, Amsterdam U., 1995.
\newblock \href{https://arxiv.org/abs/hep-th/9511195}{{\ttfamily hep-th/9511195}}.

\bibitem{Xue:2026hji}
Y.~Xue and E.~Y. Yang, \emph{{On Lagrangians of non-abelian Dijkgraaf-Witten theories}},  \href{https://arxiv.org/abs/2604.02414}{{\ttfamily 2604.02414}}.

\bibitem{Cordova:2024mqg}
C.~Cordova, D.~B. Costa and P.-S. Hsin, \emph{{Non-invertible symmetries as condensation defects in finite-group gauge theories}},  \href{https://arxiv.org/abs/2412.16681}{{\ttfamily 2412.16681}}.

\bibitem{Barkeshli:2014cna}
M.~Barkeshli, P.~Bonderson, M.~Cheng and Z.~Wang, \emph{{Symmetry fractionalization, defects, and gauging of topological phases}}, \href{https://doi.org/10.1103/PhysRevB.100.115147}{\emph{Phys. Rev. B} {\bfseries 100} (2019) 115147}, [\href{https://arxiv.org/abs/1410.4540}{{\ttfamily 1410.4540}}].

\bibitem{Tantivasadakarn:2022ceu}
N.~Tantivasadakarn, R.~Verresen and A.~Vishwanath, \emph{{Shortest route to non-Abelian topological order on a quantum processor}}, \href{https://doi.org/10.1103/PhysRevLett.131.060405}{\emph{Phys. Rev. Lett.} {\bfseries 131} (2023) 060405}, [\href{https://arxiv.org/abs/2209.03964}{{\ttfamily 2209.03964}}].

\bibitem{Guarino:2015jca}
A.~Guarino, D.~L. Jafferis and O.~Varela, \emph{{String theory origin of dyonic N=8 supergravity and its Chern-Simons duals}}, \href{https://doi.org/10.1103/PhysRevLett.115.091601}{\emph{Phys. Rev. Lett.} {\bfseries 115} (2015) 091601}, [\href{https://arxiv.org/abs/1504.08009}{{\ttfamily 1504.08009}}].

\bibitem{Gaiotto:2009mv}
D.~Gaiotto and A.~Tomasiello, \emph{{The gauge dual of Romans mass}}, \href{https://doi.org/10.1007/JHEP01(2010)015}{\emph{JHEP} {\bfseries 01} (2010) 015}, [\href{https://arxiv.org/abs/0901.0969}{{\ttfamily 0901.0969}}].

\bibitem{Witten:1995gf}
E.~Witten, \emph{{On $S$-duality in Abelian gauge theory}}, \href{https://doi.org/10.1007/BF01671570}{\emph{Selecta Math.} {\bfseries 1} (1995) 383}, [\href{https://arxiv.org/abs/hep-th/9505186}{{\ttfamily hep-th/9505186}}].

\bibitem{Brennan:2024fgj}
T.~D. Brennan and Z.~Sun, \emph{{A SymTFT for continuous symmetries}}, \href{https://doi.org/10.1007/JHEP12(2024)100}{\emph{JHEP} {\bfseries 12} (2024) 100}, [\href{https://arxiv.org/abs/2401.06128}{{\ttfamily 2401.06128}}].

\bibitem{Hsin:2018vcg}
P.-S. Hsin, H.~T. Lam and N.~Seiberg, \emph{{Comments on one-form global symmetries and their gauging in 3d and 4d}}, \href{https://doi.org/10.21468/SciPostPhys.6.3.039}{\emph{SciPost Phys.} {\bfseries 6} (2019) 039}, [\href{https://arxiv.org/abs/1812.04716}{{\ttfamily 1812.04716}}].

\bibitem{Acharya:2023bth}
B.~S. Acharya, M.~Del~Zotto, J.~J. Heckman, M.~Hubner and E.~Torres, \emph{{Junctions, edge modes, and $G_2$-holonomy orbifolds}}, \href{https://doi.org/10.4310/bpam.2024.v1.n1.a5}{\emph{Beijing J. Pure Appl. Math.} {\bfseries 1} (2024) 273--371}, [\href{https://arxiv.org/abs/2304.03300}{{\ttfamily 2304.03300}}].

\bibitem{Heckman:2017uxe}
J.~J. Heckman and L.~Tizzano, \emph{{6D Fractional Quantum Hall Effect}}, \href{https://doi.org/10.1007/JHEP05(2018)120}{\emph{JHEP} {\bfseries 05} (2018) 120}, [\href{https://arxiv.org/abs/1708.02250}{{\ttfamily 1708.02250}}].

\bibitem{Kapustin:2009av}
A.~Kapustin and M.~Tikhonov, \emph{{Abelian duality, walls and boundary conditions in diverse dimensions}}, \href{https://doi.org/10.1088/1126-6708/2009/11/006}{\emph{JHEP} {\bfseries 11} (2009) 006}, [\href{https://arxiv.org/abs/0904.0840}{{\ttfamily 0904.0840}}].

\bibitem{Choi:2023xjw}
Y.~Choi, B.~C. Rayhaun, Y.~Sanghavi and S.-H. Shao, \emph{{Remarks on boundaries, anomalies, and noninvertible symmetries}}, \href{https://doi.org/10.1103/PhysRevD.108.125005}{\emph{Phys. Rev. D} {\bfseries 108} (2023) 125005}, [\href{https://arxiv.org/abs/2305.09713}{{\ttfamily 2305.09713}}].

\bibitem{Cordova:2023ent}
C.~Cordova and K.~Ohmori, \emph{{Quantum duality in electromagnetism and the fine structure constant}}, \href{https://doi.org/10.1103/PhysRevD.109.105019}{\emph{Phys. Rev. D} {\bfseries 109} (2024) 105019}, [\href{https://arxiv.org/abs/2307.12927}{{\ttfamily 2307.12927}}].

\bibitem{Copetti:2024onh}
C.~Copetti, \emph{{Defect charges, gapped boundary conditions, and the symmetry TFT}}, \href{https://doi.org/10.1007/JHEP04(2026)055}{\emph{JHEP} {\bfseries 04} (2026) 055}, [\href{https://arxiv.org/abs/2408.01490}{{\ttfamily 2408.01490}}].

\bibitem{Copetti:2024dcz}
C.~Copetti, L.~Cordova and S.~Komatsu, \emph{{S-matrix bootstrap and non-invertible symmetries}}, \href{https://doi.org/10.1007/JHEP03(2025)204}{\emph{JHEP} {\bfseries 03} (2025) 204}, [\href{https://arxiv.org/abs/2408.13132}{{\ttfamily 2408.13132}}].

\bibitem{Bhardwaj:2024igy}
L.~Bhardwaj, C.~Copetti, D.~Pajer and S.~Schafer-Nameki, \emph{{Boundary SymTFT}}, \href{https://doi.org/10.21468/SciPostPhys.19.2.061}{\emph{SciPost Phys.} {\bfseries 19} (2025) 061}, [\href{https://arxiv.org/abs/2409.02166}{{\ttfamily 2409.02166}}].

\bibitem{Vandermeulen:2022edk}
T.~Vandermeulen, \emph{{Lower-Form Symmetries}},  \href{https://arxiv.org/abs/2211.04461}{{\ttfamily 2211.04461}}.

\bibitem{Sati:2013rxa}
H.~Sati, \emph{{Framed M-branes, corners, and topological invariants}}, \href{https://doi.org/10.1063/1.5007185}{\emph{J. Math. Phys.} {\bfseries 59} (2018) 062304}, [\href{https://arxiv.org/abs/1310.1060}{{\ttfamily 1310.1060}}].

\bibitem{Giotopoulos:2024xcg}
G.~Giotopoulos, H.~Sati and U.~Schreiber, \emph{{Flux quantization on 11-dimensional superspace}}, \href{https://doi.org/10.1007/JHEP07(2024)082}{\emph{JHEP} {\bfseries 07} (2024) 082}, [\href{https://arxiv.org/abs/2403.16456}{{\ttfamily 2403.16456}}].

\bibitem{Giotopoulos:2024sit}
G.~Giotopoulos, H.~Sati and U.~Schreiber, \emph{{Flux quantization on M5-branes}}, \href{https://doi.org/10.1007/JHEP10(2024)140}{\emph{JHEP} {\bfseries 10} (2024) 140}, [\href{https://arxiv.org/abs/2406.11304}{{\ttfamily 2406.11304}}].

\bibitem{Sati:2025ucv}
H.~Sati and U.~Schreiber, \emph{{Engineering of anyons on M5-probes via flux quantization}}, \href{https://doi.org/10.21468/SciPostPhysLectNotes.107}{\emph{SciPost Phys. Lect. Notes} {\bfseries 107} (2025) 1}, [\href{https://arxiv.org/abs/2501.17927}{{\ttfamily 2501.17927}}].

\bibitem{Bhardwaj:2024xcx}
L.~Bhardwaj, T.~D{\'e}coppet, S.~Schafer-Nameki and M.~Yu, \emph{{Fusion 3-categories for duality defects}}, \href{https://doi.org/10.1007/s00220-025-05388-1}{\emph{Commun. Math. Phys.} {\bfseries 406} (2025) 208}, [\href{https://arxiv.org/abs/2408.13302}{{\ttfamily 2408.13302}}].

\bibitem{Johnson-Freyd:2015sza}
T.~Johnson-Freyd and C.~Scheimbauer, \emph{{(Op)lax natural transformations, twisted quantum field theories, and {\textquotedblleft}even higher{\textquotedblright} Morita categories}}, \href{https://doi.org/10.1016/j.aim.2016.11.014}{\emph{Adv. Math.} {\bfseries 307} (2017) 147--223}, [\href{https://arxiv.org/abs/1502.06526}{{\ttfamily 1502.06526}}].

\bibitem{Brochier:2018kxc}
A.~Brochier, D.~Jordan and N.~Snyder, \emph{{On dualizability of braided tensor categories}}, \href{https://doi.org/10.1112/S0010437X20007630}{\emph{Compos. Math.} {\bfseries 157} (2021) 435--483}, [\href{https://arxiv.org/abs/1804.07538}{{\ttfamily 1804.07538}}].

\bibitem{Reshetikhin:1991tc}
N.~Reshetikhin and V.~G. Turaev, \emph{{Invariants of three manifolds via link polynomials and quantum groups}}, \href{https://doi.org/10.1007/BF01239527}{\emph{Invent. Math.} {\bfseries 103} (1991) 547--597}.

\bibitem{Turaev+2016}
V.~G. Turaev, \emph{Quantum invariants of knots and 3-manifolds}, vol.~18 of \emph{De Gruyter Studies in Mathematics}.
\newblock Walter de Gruyter \& Co., Berlin, revised~ed., 2010.

\end{thebibliography}
\end{document}